\documentclass[5p,times]{elsarticle}
\usepackage{mathtools}
\usepackage[switch]{lineno}

\usepackage{xcolor,colortbl}

\usepackage[flushleft]{threeparttable} 
\usepackage{makecell}
\usepackage{booktabs}
\usepackage{tabularx}
\usepackage{changepage}
\usepackage[%pdfusetitle,
]{hyperref}

\renewcommand{\thesection}{\Roman{section}}

\usepackage{natbib}
\setcitestyle{square, comma, numbers,sort&compress, super}

\newcommand{\beginsupplement}{%
	\setcounter{table}{0}
	\renewcommand{\thetable}{S\arabic{table}}%
	\setcounter{figure}{0}
	\renewcommand{\thefigure}{S\arabic{figure}}%
	\setcounter{section}{0}
	\renewcommand{\thesection}{S\arabic{section}}%
	\setcounter{equation}{0}
	\renewcommand{\theequation}{S\arabic{equation}}
}

\definecolor{Gray}{gray}{0.85}

\newcolumntype{a}{>{\columncolor{Gray}}c}

\usepackage[utf8]{inputenc}
\usepackage{lmodern}

\usepackage{caption}
\usepackage{subcaption}

\journal{Nature Communications}

\begin{document}
	
\begin{frontmatter}

\title{Designing a sector-coupled European energy system  \\ robust to 60 years of historical weather data}

\author{Ebbe Kyhl G\o{}tske$^{a,b,*}$}
\author{Gorm Bruun Andresen$^{a,b}$}
\author{Fabian Neumann$^{d}$}
\author{Marta Victoria$^{a,b,c}$}

\begin{abstract}
As energy systems transform to rely on renewable energy and electrification, they encounter stronger year-to-year variability in energy supply and demand. However, most infrastructure planning is based on a single weather year, resulting in a lack of robustness. In this paper, we optimize energy infrastructure for a European energy system designed for net-zero CO$_2$ emissions in 62 different weather years. Subsequently, we fix the capacity layouts and simulate their operation in every weather year, to evaluate resource adequacy and CO$_2$ emissions abatement. We show that interannual weather variability causes variation of $\pm$10\% in total system cost. The most expensive capacity layout obtains the lowest net CO$_2$ emissions but not the highest resource adequacy. Instead, capacity layouts designed with years including compound weather events result in a more robust and cost-effective design. Deploying CO$_2$-emitting backup generation is a cost-effective robustness measure, which only increase CO$_2$ emissions marginally as the average CO$_2$ emissions remain less than 1\% of 1990 levels. Our findings highlight how extreme weather years drive investments in robustness measures, making them compatible with all weather conditions within six decades of historical weather data.
\end{abstract}

\end{frontmatter}

\begin{minipage}{\textwidth}
	{\small \itshape 
		\noindent 
		$^{a}$ Department of Mechanical and Production Engineering, Katrinebjergvej 89F, 8200 Aarhus N, Denmark\\
		$^{b}$ iCLIMATE Interdisciplinary Centre for Climate Change, Aarhus University, Denmark \\
		$^{c}$ Novo Nordisk Foundation CO$_2$ Research Center, Aarhus University, Gustav Wieds Vej 10, 8000 Aarhus C, Denmark \\
		$^{d}$ Department of Digital Transformation in Energy Systems, Institute of Energy Technology, Technische Universität Berlin, Fakultät III, Einsteinufer 25 (TA 8), 10587 Berlin, Germany \\
		$^{*}$ Lead contact and corresponding author, Email: ekg@mpe.au.dk
	}
\end{minipage}

\clearpage

Decarbonizing power systems by rebuilding them around wind and solar photovoltaic (PV) generation has shown to be a cost-effective strategy to reduce greenhouse gas emissions in Europe \cite{Brown_2018, Victoria_2022} and globally \cite{Luderer_2022}. This strategy can also help decarbonize other sectors using direct electrification (via heat pumps or electric vehicles), and indirect electrification (using electrolytic hydrogen and other synthetic fuels)\cite{Brown_2018, Victoria_2022}. Future energy systems with high wind and solar penetration might be more sensitive to interannual weather variation and extreme weather events, which should be considered in the planning of infrastructure for the energy transition. However, considering decades of weather data in a single capacity expansion optimization is computationally expensive at the high resolution that infrastructure planning requires. Hence, most previous studies use only a single weather year in their analysis. 

Recent efforts to investigate interannual weather variability impacts on energy systems include modeling system dispatch \cite{Collins_2018, Grochowicz_2024} or estimating optimal capacities \cite{Pfenninger_2017, Schlachtberger_2018, Zeyringer_2018, Schyska_2021, Grochowicz_2023} for several weather years, estimating the storage capacity required to deal with \textit{dunkeflauten}, \textit{i.e.} a period of low wind and solar generation \cite{Ruhnau_2021}, determining the overlapping of near-optimal solutions spaces calculated for different weather years \cite{Grochowicz_2023}, or attempting to classify the weather regimes that are particularly challenging for the European power system \cite{Mockert_2023, Bloomfield_2020, Grochowicz_2024, Gruber2022, RAYNAUD2018578}. Assessments have also been conducted on the impact of extreme weather events on conventional infrastructure, including investigations of persistent cold spells in Texas \cite{Gruber2022} and heat waves in North America \cite{KE2016504, Bartusek2022}. However, previous research either considers one country in isolation \cite{Pfenninger_2017, Zeyringer_2018, Ruhnau_2021, Mockert_2023, JIMENEZ2024122695} (omitting the impact of cross-border energy flows, transmission bottlenecks and regional variability), includes only the power system \cite{Grochowicz_2023, Grochowicz_2024, Pfenninger_2017, Schlachtberger_2018, Collins_2018, Schyska_2021, Ruhnau_2021, Klimaraadet} (omitting the synergetic benefits from sector-coupling \cite{Brown_2018}), or consider only interannual variability of wind and solar capacity factors, and electricity demand \cite{Pfenninger_2017, Schlachtberger_2018, Collins_2018, Schyska_2021, Ruhnau_2021, Klimaraadet} (disregarding the impact of large interannual variation of hydropower inflow \cite{VICTORIA2019674, GOTSKE2021102999}). Furthermore, interannual weather variability does not only impact the energy supply but also has large impacts on the demand \cite{Ruhnau2023,PEACOCK2023120885, STAFFELL201865}.

Our analysis employs PyPSA-Eur, an open networked model of the European sector-coupled energy system \cite{Victoria_2022, Neumann_2023}, which performs regional capacity expansion planning of energy infrastructure, covering the energy demands and CO$_2$ emissions in the electricity, heating, land transport, shipping and aviation, and industry (including industrial feedstock) sectors with comprehensive carbon management. Using 62 years of weather data (1960-2021), we investigate the impacts of different solar PV, onshore and offshore wind capacity factors, hydropower inflow, heating demand, and coefficient of performance (COP) of heat pumps time series.  We first perform a greenfield joint capacity and dispatch optimization for a European system constrained to net-zero CO$_2$ emissions for every weather year (design year) to determine 62 alternative capacity layouts. Throughout this paper, we use the term \textit{design year} to refer to the weather year used as input in the joint capacity and dispatch optimization. Subsequently, we keep the capacities fixed and analyze their robustness to interannual weather variations by simulating the dispatch in each of the remaining 61 weather years (operational years) different from the design year. Here, we refer to the weather years in the dispatch optimization as \textit{operational years}. To accommodate extreme events, we allow the activation of CO$_2$-emitting backup generation, which is subjected to a CO$_2$ tax derived from the shadow price calculated on the design year. Focusing on two key criteria, avoiding unserved energy and mitigating CO$_2$ emissions, we address how interannual weather variations impact a highly renewable European energy system. Furthermore, we shed light on the characteristics that a robust energy system exhibits. 

\vspace{-0.25cm}
\section*{\Large Results}
\vspace{-0.25cm}

\begin{figure*}[!h]
	\centering
	\includegraphics[width=0.75\textwidth]{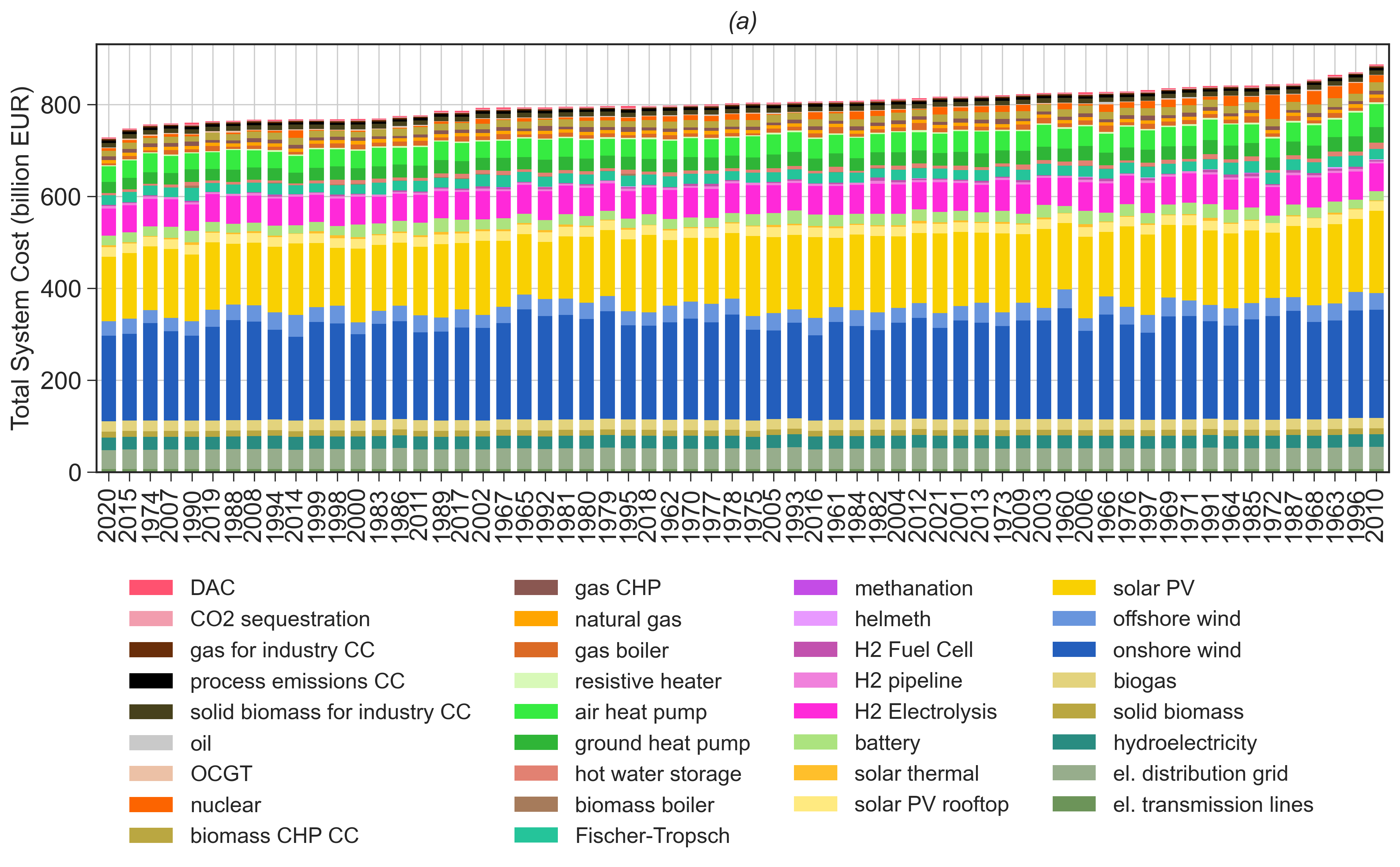}
	\includegraphics[width=0.75\textwidth]{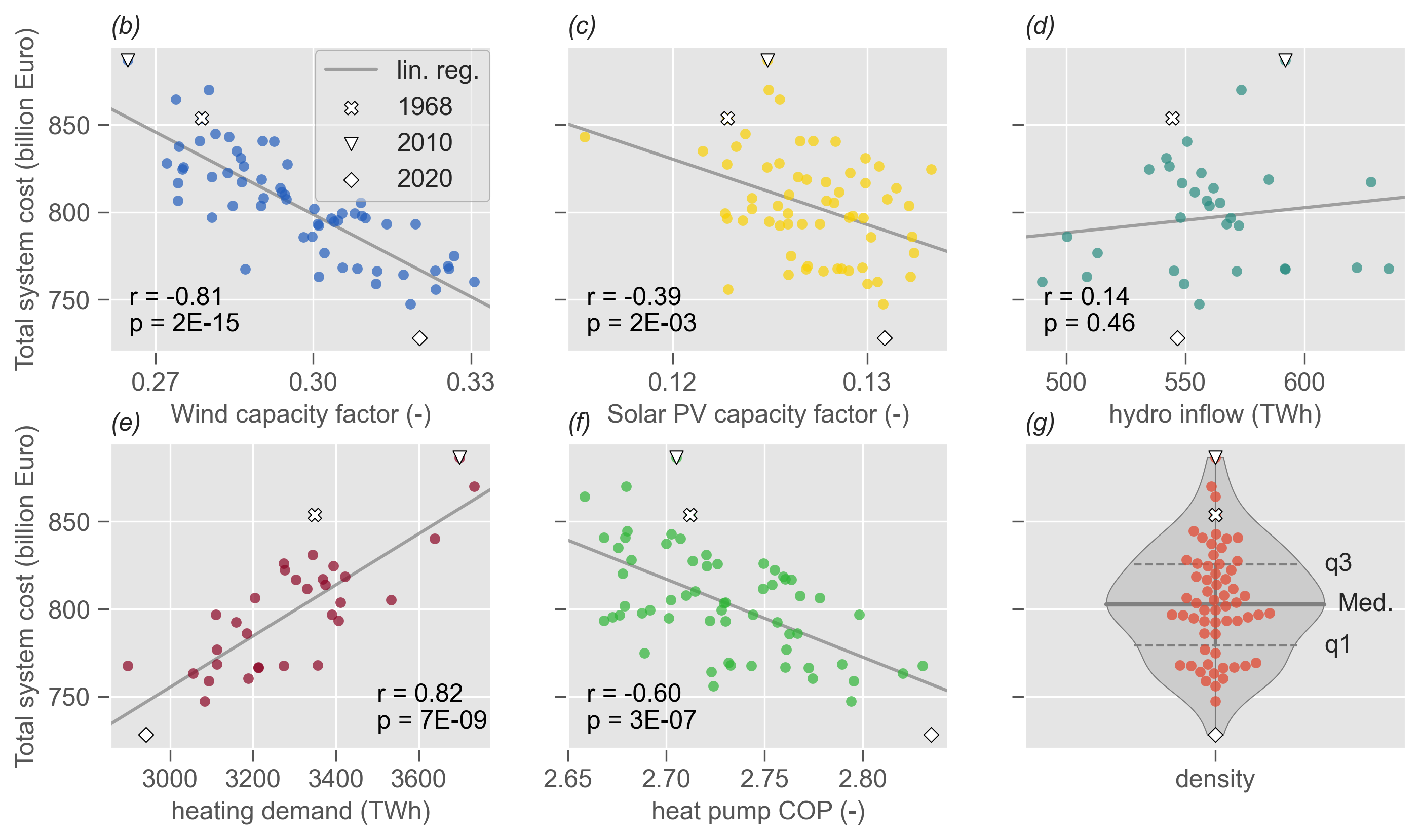}
	\caption{\textbf{Total annual system costs for the capacity optimization} using as input weather years from 1960 to 2021. The figure shows (a) the Europe-aggregate system costs split by technology, (b-f) the total system costs correlated with the Europe-aggregate weather-dependent variables for the corresponding years, and (g) violin plot of the total system cost with annotations of the 25\% (q1), 50\% (Med.), and 75\% (q3) quantiles. In a-f, the solid lines depict the linear regression, while the coefficient of determination r and p-values are shown in the lower left corner. See Supplemental Fig. \ref{sfig:co2_emissions_price} for the distribution of the CO$_2$ emissions prices derived from the capacity optimization.}
	\label{fig:system_cost_all_years}
\end{figure*}

\noindent
\textbf{Capacity optimization for different weather years}. The optimization obtained with all design years (1960 - 2021) in Fig. \ref{fig:system_cost_all_years}a yields an average system cost of €803 billion per year with a variation of $\pm$10\% in the most expensive and the cheapest capacity layouts. The distribution of system costs exhibits long tails (Fig. \ref{fig:system_cost_all_years}g), suggesting that rare extreme weather years drive the range of system costs. In searching for the main causes of system cost variation, Fig. \ref{fig:system_cost_all_years}b-d present the annual available renewable resources from wind energy (aggregate of offshore and onshore), solar PV (utility and rooftop), and hydro inflow (reservoir and run-of-river) that can be utilized for electricity generation in every design year, plotted against total system cost. Wind and solar PV resources are shown as normalized power output (annual capacity factors): Stronger wind speeds result in larger capacity factors, hence, fewer wind turbines are needed to produce the same amount of electricity. For this reason, we see a strong anticorrelation between annual wind resource and total system cost ($r$=-0.81). The same effect applies to solar PV but with a weaker correlation coefficient ($r$=-0.39). Conversely, while the relative interannual variation is highest for hydro inflow on a country level (see Supplemental Material \ref{smat:interannual_variability}), its impact on the capacity optimization is small (correlation coefficient is not significant) because the volume of wind and solar PV generation is much larger. Fig. \ref{fig:system_cost_all_years}e-f show the heating demand and the COP of heat pumps. Persistent cold weather leads to high heating demand requiring more generation capacity for the provision of heating. For this reason, a strong correlation appears between annual heating demand and total system cost ($r$=0.82). Simultaneously, cold weather leads to low COP for the heat pumps converting electricity to heat which has similar impact on the total system cost ($r$=-0.60).

As all capacity optimizations fulfill the net-zero CO$_2$ emissions constraint, the predominant investments are consistently placed in wind and solar PV electricity generation capacity, accounting for approximately 50\% of total system costs. As consequence, they comprise on average 93\% of the electricity generation, and when including hydropower, renewable electricity account for 98\% (Supplemental Fig. \ref{sfig:vre_share}). The remaining share primarily relies on firm generation from Combined Heat and Power (CHP) plants fueled with solid biomass and incorporating point-source carbon capture (operated at 3,864 full load hours/year), as well as flexible generation from gas CHP (611 full load hours/year) and Open Cycle Gas Turbine (OCGT; 152 full load hours/year). A small fraction of firm generation is supplied by nuclear power which operates as base load (8,293 full load hours/year). In certain design years, reserve capacity is deployed, with synthetic oil-fired power plants being activated for only a few hours (40 hours/year). Here, the oil is produced synthetically in Fischer-Tropsch plants combining electrolytic H$_2$ and captured CO$_2$. The gas usage in OCGT and CHP plants can be of fossil origin (the system relies on an annual production or import of 121~TWh/year), from biogas (336~TWh/year), and, in some design years, produced synthetically with Sabatier or Helmeth \cite{helmeth} reaction (see Supplemental Figs. \ref{sfig:capacity_mining} and \ref{sfig:methane_production}).

Design years with less favorable renewable resources compensate by increasing their dependence on firm generation, in this case nuclear, which operates at almost full capacity. As consequence, during the subsequent simulation of every design year across 61 operational years, the firm generation may not alleviate challenging events since it can not increase further its power production (see Supplemental Figures \ref{sfig:firm_and_flexible_generation}, \ref{sfig:capacity_firm_generation}, and \ref{sfig:capacity_CHP_el}).

Decarbonization of other sectors is obtained via direct electrification (heat pumps and electric vehicles) and indirect electrification (production of H$_2$ and synthetic fuels) that are used in land transport, shipping, aviation and industry (see Supplemental Figs. \ref{sfig:capacity_VRE} - \ref{sfig:capacity_other} for the deployed capacities in different sectors). Lastly, reaching net-zero emissions requires investment into point-source and air capture of CO$_2$, and underground CO$_2$ sequestration. All design years fully utilize the sequestration limit of 200 MtCO$_2$/year assumed in our analysis. See previous works for detailed discussions on the interplay of different technologies \cite{Brown_2018, Victoria_2022, victoria_early_2020, Neumann_2023}.\\

\begin{figure*}[!h]
	\centering
	 \includegraphics[width=0.95\textwidth]{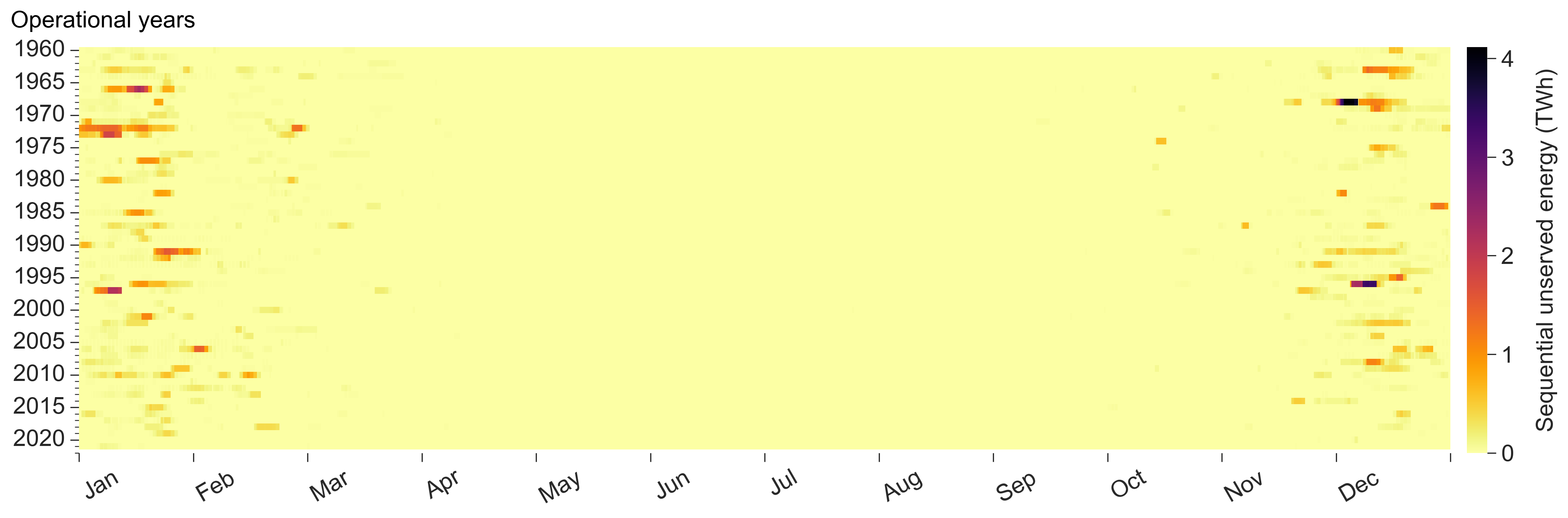}
	\includegraphics[width=0.95\textwidth]{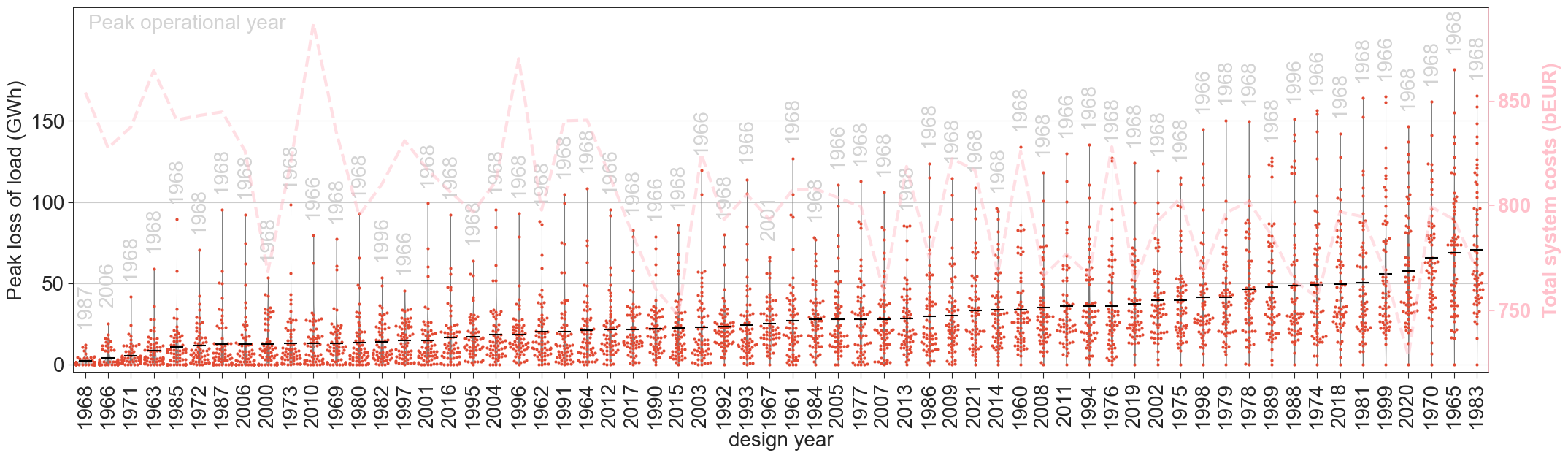}				\includegraphics[width=0.95\textwidth]{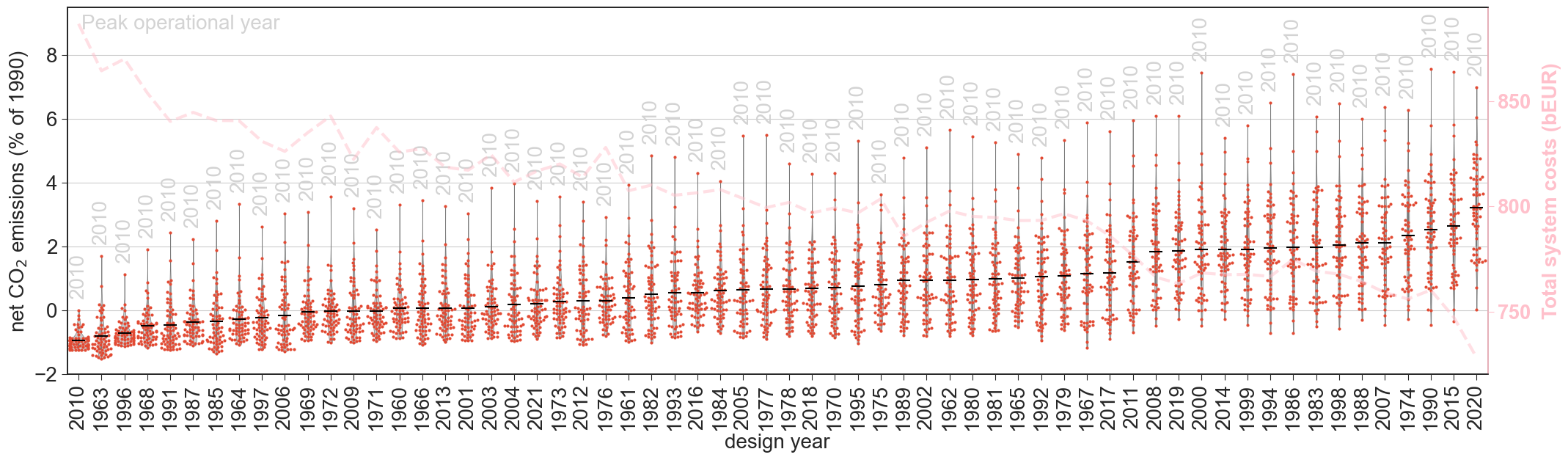}
	
	\caption{\textbf{European aggregate robustness metrics} summarized by (a) loss of load aggregated for every coherent hours with unserved energy (sequential unserved energy) for Europe in all operational years averaged for all capacity layouts, (b) peak loss of load across all 3 hour time steps and (c) net CO$_2$ emissions for all operational years (red markers) run for the different capacity layouts (design year). The black lines in (b) and (c) indicate the average of all operational years, and gray annotations indicate the operational year causing the highest peak loss of load and net CO$_2$ emissions. See Supplemental Fig. \ref{sfig:unserved_energy_avg} for the loss of load in every 3 hour timestep, Supplemental Fig. \ref{sfig:CO2_emissions_t} for the net-CO$_2$ emissions, and Supplemental Fig. \ref{sfig:dispatch_optimization_all} for distribution plots of cumulative and sequential unserved energy.}
	\label{fig:dispatch_optimization_dy2013}
\end{figure*}

\noindent
\textbf{Unserved energy happens in winter periods}. The capacity layouts were determined while assuming no precautionary robustness measure, such as a safety reserve capacity margin targeted at interannual weather variation. Consequently, during operational years, instances may arise where not all energy demands can be met. While imposing the CO$_2$ tax resulting from the design year on the dispatch optimization, shortages in supply can be alleviated by gas and oil-fired backup capacities if built in the design year. In more severe cases, where available capacities do not suffice to bridge the gap between supply and demand, the system resorts to load shedding in either heating or electricity demand. We use \textit{loss of load} or \textit{unserved energy} interchangably to describe this phenomena. See Table \ref{tab:metrics} for definitions of resource adequacy metrics. Fig. \ref{fig:dispatch_optimization_dy2013}a shows the loss of load aggregated for coherent hours with unserved energy (for the disaggregated loss of load in every 3 hour time step, we refer to Supplemental Fig. \ref{sfig:unserved_energy_avg}). From  Fig. \ref{fig:dispatch_optimization_dy2013}a, unserved energy is most frequently observed during winter, in agreement with previous studies \cite{Brown_2018, Grochowicz_2024, Mockert_2023, Bloomfield_2020}. In addition, we also observe variation from year to year: Some operational years entail almost full resource adequacy (e.g., 2020), while others display periods with unserved energy (the highest sequential unserved energy is encountered in Dec 1968, Dec 1996, and Jan 1966, ordered by magnitude) of different duration. The longest period with unserved energy happens in January during the 1972 operational year.\\

\noindent
\textbf{Extreme operational years drive the range of loss of load and CO$_2$ emissions}. As a next step, we explore all the combinations of operational years based on different capacity layouts (design years). We focus on evaluating the peak loss of load (i.e. the 3-hour period with the largest unserved energy) and the annual net CO$_2$ emissions obtained for different combinations of design and operational years. Fig. \ref{fig:dispatch_optimization_dy2013}b reveals that, in most cases, the highest peak loss of load are driven by operational years with extreme weather conditions (such as 1966, 1968, and 1996), as indicated by the long tail of the distributions. Consequently, when simulating the capacity layouts obtained with those years as design years, the system demonstrates greater robustness as they are capable of withstanding all operational years with loss of load lower than 25~GWh (2\% of the average hourly electricity demand of Europe; see also Supplemental Fig. \ref{sfig:LOL_duration_curves}).

\begin{figure}[!h]
	\centering
	\includegraphics[width=0.5\textwidth]{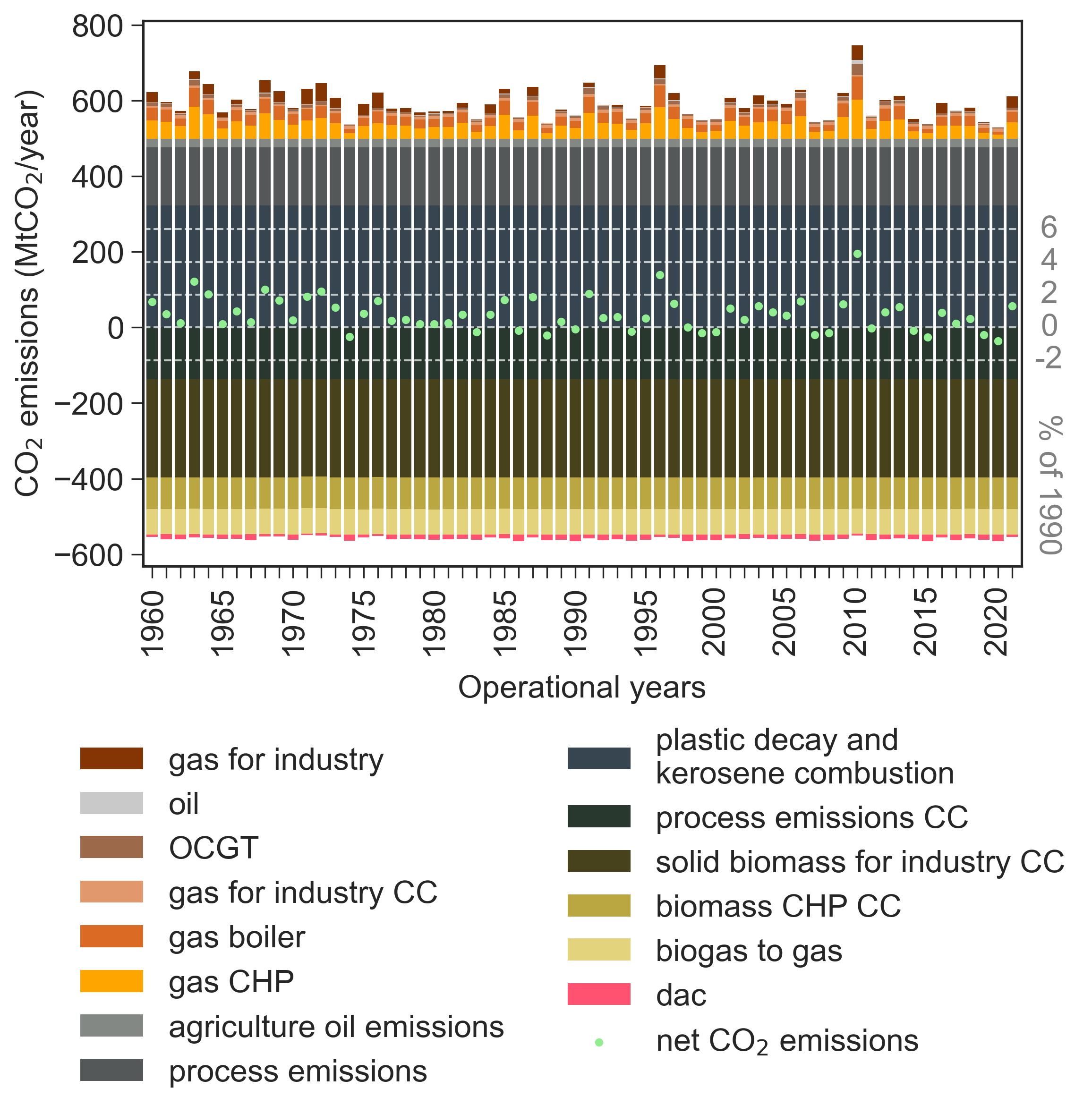}
	\caption{\textbf{CO$_2$ emissions split by technology for every operational year}, averaged over all capacity layouts. For the CO$_2$ emissions in every capacity layout, see Supplemental Fig. \ref{sfig:CO2_emissions}.}
	\label{fig:CO2_emissions}
\end{figure}

\begin{figure*}[!h]
	\centering
	\includegraphics[width=\textwidth]{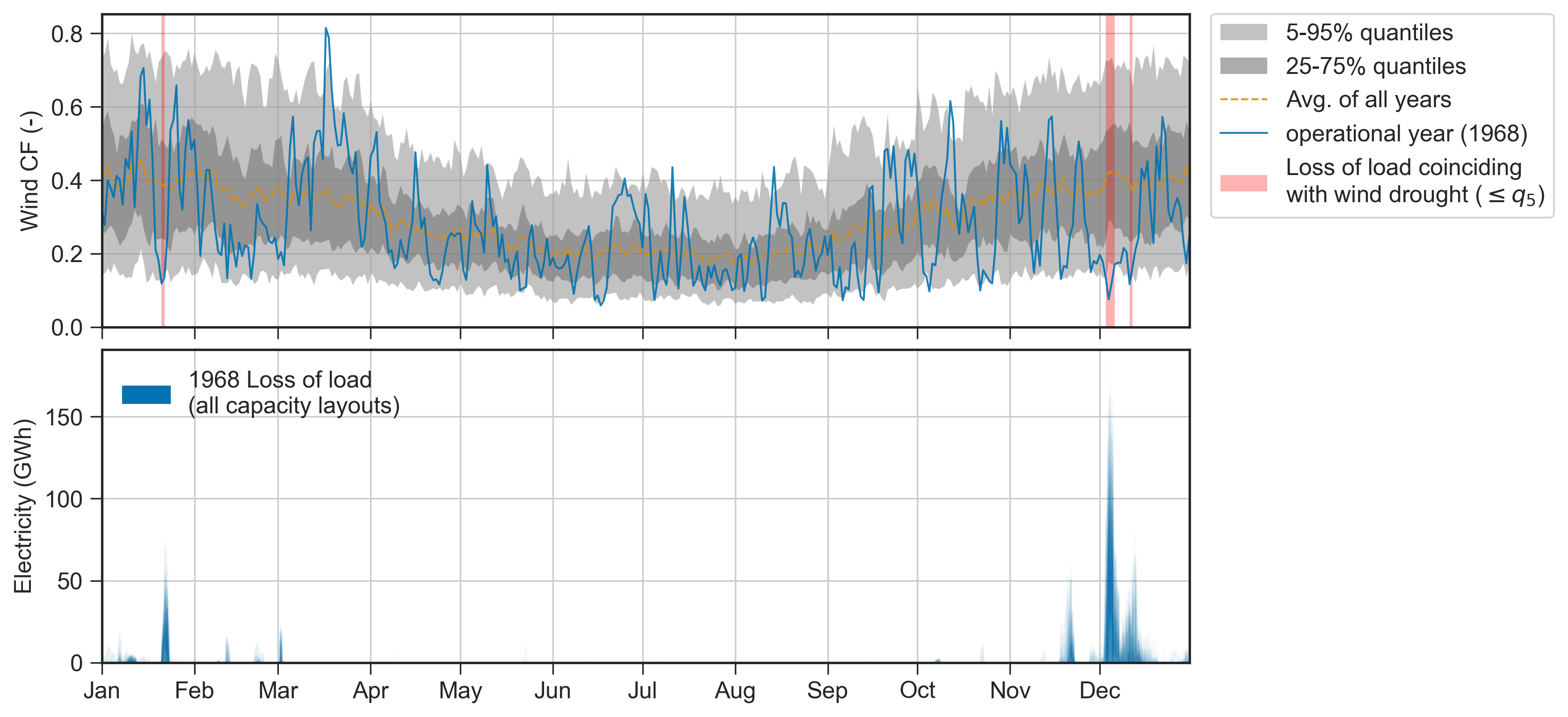}
	\caption{\textbf{Wind resources in the operational year (1968) causing the highest peak loss of load}. See Supplemental Fig. \ref{sfig:wind_December} for the nodal wind resources in the loss of load event occuring in December. See Supplemental Figs. \ref{sfig:unserved_energy_and_renwable_droughts_EU_1966}-\ref{sfig:unserved_energy_and_renwable_droughts_EU_2010} for the same depiction with wind and solar PV resources, hydro inflow, and heating demand in 1966, 1968, 1972, 1996, and 2010.}
	\label{fig:wind_most_extreme_year}
\end{figure*}

For the CO$_2$ emissions (Fig. \ref{fig:dispatch_optimization_dy2013}c), we observe a similar phenomenon to that of the peak loss of load, but in this case, we see consistently the same operational year (2010) causing the highest CO$_2$ emissions regardless of the design year. In the extreme case, an overshoot of 7.6\% relative to 1990 emissions levels is observed, while the best case (2020 operational year) shows an undershoot of -1.5\%. The overshoot observed during the 2010 operational years can be largely attributed to the characteristics of the weather, which leads to the second highest annual heating demand and the lowest annual wind resources of all operational years (Fig. \ref{fig:system_cost_all_years}). As a response (Fig. \ref{fig:CO2_emissions}), the system increases the heating generation from the fixed capacity of gas boilers, electricity generation from OCGT, co-generation from gas CHP plants, and, in few cases, synthetic oil-fired power plants, which leads to higher CO$_2$ emissions. In some cases, the system is forced to shut off the direct air capture (DAC). Thus, preventing to offset the increased emissions, which consequently leads to a higher net CO$_2$ emissions balance. On average for the different capacity layouts, the operational years create net CO$_2$ emissions of 0.8\% relative to 1990 historical levels, i.e., a slight increase compared to the net-zero design requirement.\\

\noindent
\textbf{Compound weather events trigger periods with unserved energy}. In our analysis, we identify distinct compound weather events triggering loss of load events. 

The 1968 operational year entails the highest peak loss of load (and on average, the highest sequential unserved energy). While we observe low hydro inflow and high heating demand (both on the border of the 5th and 95th percentiles), the loss of load is mostly driven by a pronounced scarcity of wind resources (below the 5th percentile) in early December (Fig. \ref{fig:wind_most_extreme_year}). Here, wind lulls are observed in countries such as Germany, Netherlands, Poland, and Denmark, lasting approximately one week (see Supplemental Fig. \ref{sfig:wind_December}). Previous analyses have made similar observations on the high impacts on energy systems of wind energy droughts \cite{Brown_2018, Grochowicz_2024, Mockert_2023, Bloomfield_2020}. The 1996 operational year encounters a similar wind energy drop causing the second highest sequential unserved energy. However, the latter coincides with an average heating demand and a period with abundant hydropower resources, which suggests that the unserved energy is caused solely by the wind scarcity
(see Supplemental Fig. \ref{sfig:unserved_energy_and_renwable_droughts_EU_1996}).

While the unserved energy of the 1968 operational year also shows some correlation with other phenomena than the wind scarcity, we find that other years face more clear evidence of compound weather events driving loss of load. The second highest peak loss of load across all years is encountered in the 1966 operational year. This is caused by a widespread wind scarcity, see supplemental Fig. \ref{sfig:wind_middle_January}, and a concurrent abnormal heating demand due to low temperatures (Supplemental Fig. \ref{sfig:unserved_energy_and_renwable_droughts_EU_1966}), both transcending the 5th and 95th percentiles. By closer inspection, the 1966 European heating demand is highly driven by unusually high demands in France and Germany during that period (Supplemental Fig. \ref{sfig:heating_demand}). This coincides with a significant drop in wind resources, reaching a stagnation in those countries for two days (from January 16 to 18; Supplemental Fig. \ref{sfig:wind_middle_January}). For the longer lasting loss of load event occurring in the 1972 operational year, we observe a prolonged hydro reservoir drought throughout January. This is uniformly distributed over Europe (see Supplemental Fig. \ref{sfig:hydro}) and is caused by a seasonal shift (from winter to summer) in the hydropower inflow. Additionally, in the same period, Europe also encounters a scarcity of wind energy (lasting four days) and solar PV resources (lasting two days), see Supplemental Fig. \ref{sfig:unserved_energy_and_renwable_droughts_EU_1972}. Solar PV resources are generally less variable from year to year, and for this reason, we do not find similar cases with solar PV droughts causing significant load shedding. The 2010 operational year has the lowest annual wind resources and the second highest annual heating demand, but does not cause a similarly large impact as the above-mentioned conditions. This is because the wind lulls are less severe and do not coincide with other weather impacts (Supplemental Fig. \ref{sfig:unserved_energy_and_renwable_droughts_EU_2010}).

\begin{figure}[!t]
	\centering
	\includegraphics[width=0.5\textwidth]{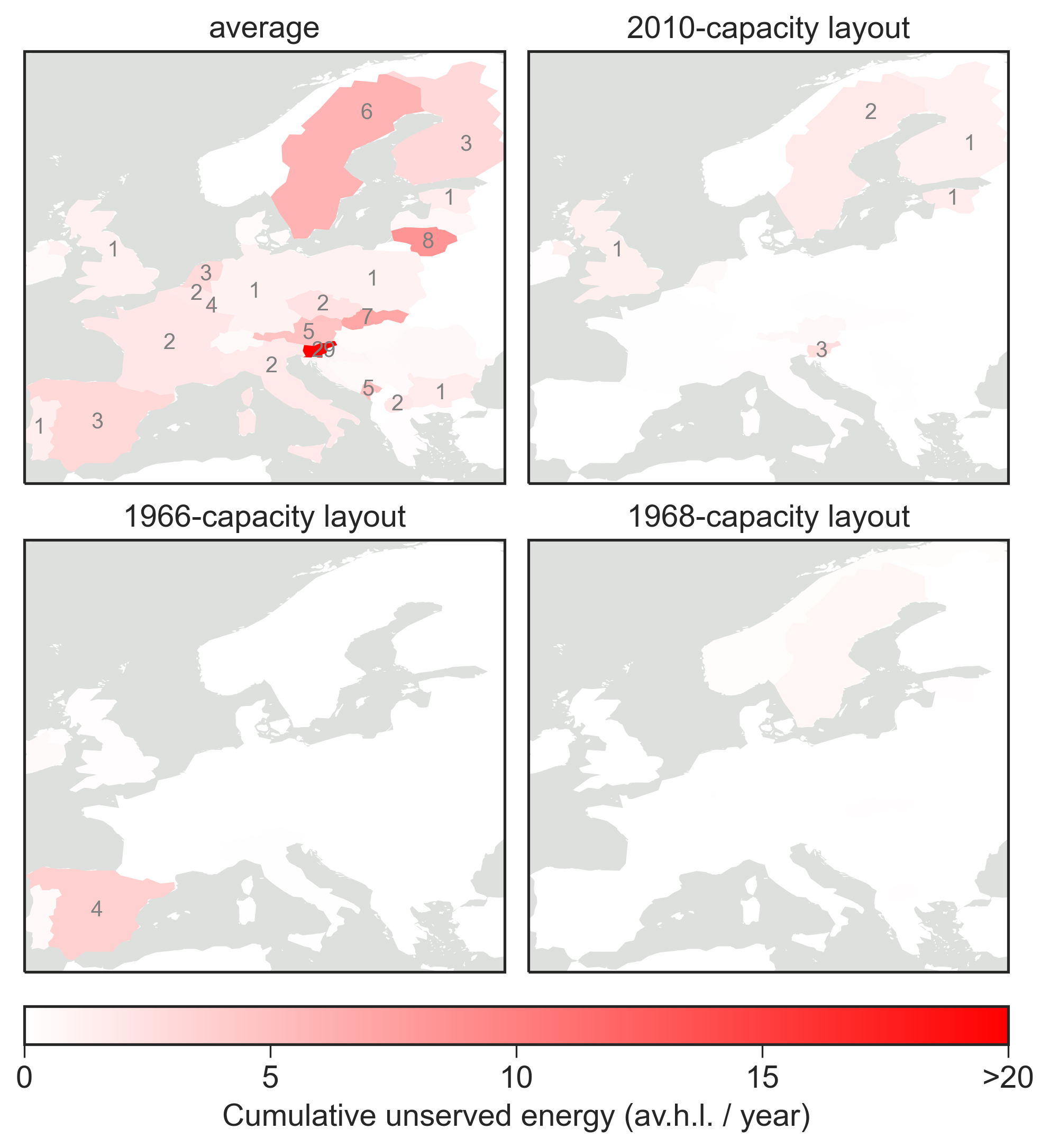}
	\caption{\textbf{Geographical distribution of the annual unserved energy} obtained with different capacity layouts (design years) simulated with all operational years. The top-left subfigure depicts the average unserved energy in all capacity layouts for every operational years. Furthermore, we show the two capacity layouts obtaining the lowest peak loss of load (1966 and 1968), and the most expensive system (2010). The annotations indicate the number of hours with average electricity demand (average hourly load; av.h.l.) uncovered per year in every country. See Supplemental Figs. \ref{sfig:map_unserved_energy_all_designs} and  \ref{sfig:map_unserved_energy_all_operations} for all capacity layouts and all operational years.}
	\label{fig:map_unserved_energy}
\end{figure}

Fig. \ref{fig:map_unserved_energy} depicts the average cumulative unserved energy of all capacity layouts simulated in every operational year. The figure indicates how well the capacity layout prevents load shedding in each country, and how the impacts of the compound weather events are generally distributed among countries. For the latter, we observe a shift between Northeast (NE) and Southwest (SW) encountering high levels of unserved energy in both regions, varying for different operational years. See Supplemental Fig. \ref{sfig:map_unserved_energy_all_operations} and \ref{sfig:map_unserved_energy_all_designs} for the geospatial distribution of unserved energy in every operational year, averaged for all capacity layouts and vice versa. Most design years show load shedding in either NE or SW but is more commonly observed in the NE of Europe. For the simulation of the most expensive capacity layout (2010) and the two capacity layouts obtaining the lowest peak loss of load on European level (1966 and 1968), we notice that energy deficits are also concentrated in either NE or SW. However, the 1968 capacity layout shows almost zero unserved energy for all operational years.\\

\noindent
\textbf{System cost impact on robustness}. Our investigation leads us to the question of whether choosing the most expensive year in the capacity optimization is the most effective measure towards mitigating loss of load events. From Fig. \ref{fig:dispatch_optimization_dy2013}b, we notice some relation between total system cost and avoiding loss of load. However, the most expensive design year (2010) does not show the lowest peak loss of load, while the lowest system cost (2020) does not bring the highest peak loss of load. See Supplemental Fig. \ref{sfig:LOL_duration_curves} for the loss of load duration curves in every design year, showing eight capacity layouts exhibiting higher robustness than 2010, of which the 1968 design year has the highest robustness.

\begin{figure}[!t]
	\centering
	\includegraphics[width=0.5\textwidth]{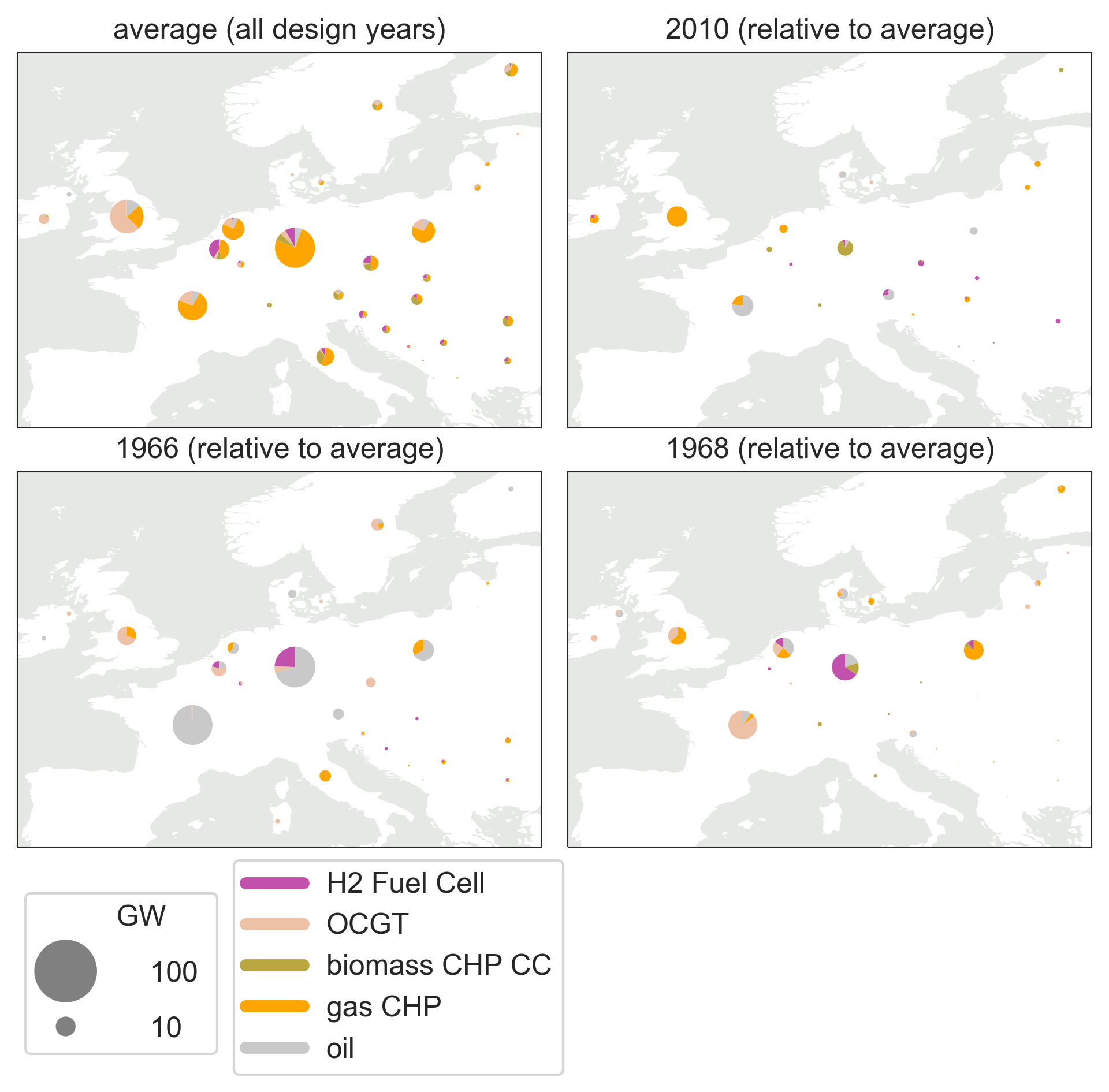}
	\includegraphics[width=0.5\textwidth]{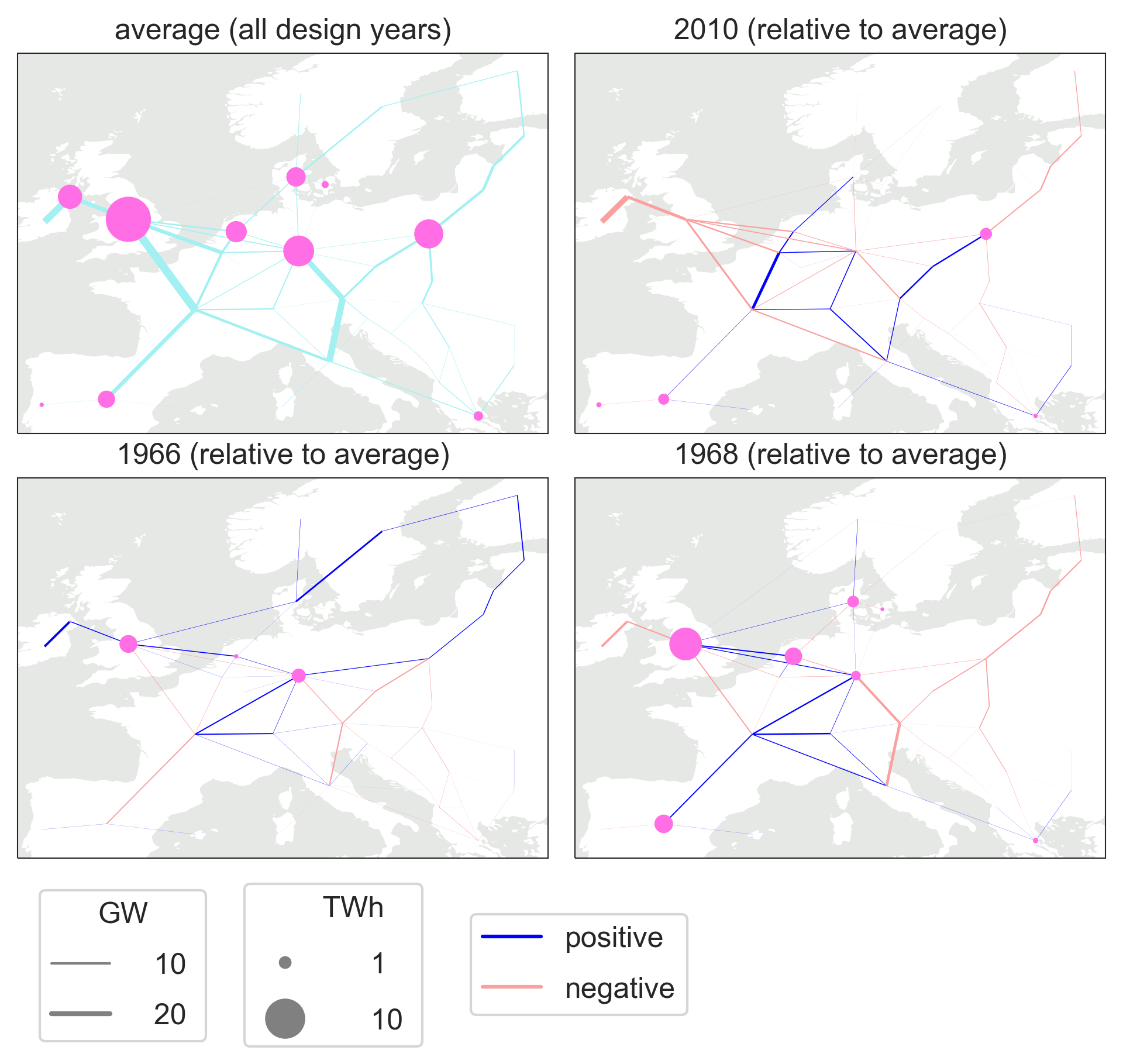}
	\caption{\textbf{Nodal electricity generation capacity and H$_2$ infrastructure}. The figure depicts (top) the capacity of flexible generation and H$_2$ fuel cell and reserve capacity, and (bottom) the H$_2$ network and storage, illustrated for the average of every design year, the two systems obtaining the lowest peak loss of load (1966 and 1968), and the most expensive system (2010). See Supplemental Figs. \ref{sfig:renewable_generation_capacity}, \ref{sfig:H2_electrolysis_maps}, \ref{sfig:Heating_maps} for renewable capacity, H$_2$ electrolysis, and decentral heating options. See Supplemental Fig. \ref{sfig:Eurupe_aggregate_capacity} for the aggregated European capacities.}
	\label{fig:map_robustness}
\end{figure}

% This can be explained by the 2010 weather, characterized by significant low annual wind and high annual heating demand.
For the net-CO$_2$ emissions, we see a more clear linear relation with total system cost (Fig. \ref{fig:dispatch_optimization_dy2013}c). The most expensive design year also entails the lowest net CO$_2$ emissions (net-negative on average). In the capacity optimization, this drives more investments in heating generation and storage capacities, electricity generation capacity, and DAC (Supplemental Figs. \ref{sfig:Eurupe_aggregate_capacity}, \ref{sfig:Eurupe_aggregate_capacity_H2_storage}, \ref{sfig:Eurupe_aggregate_capacity_thermal_storage} \ref{sfig:Eurupe_aggregate_capacity_DAC}). As a result, the system becomes the most prepared for mitigating CO$_2$ emissions.\\

\noindent
\textbf{Strategies to increase robustness}. As a last step, we investigate which properties are needed for the energy system to be robust. First, we describe the characteristics of robust capacity layouts. From Fig. \ref{fig:map_robustness}, the most robust systems, in terms of avoiding loss of load (1966 and 1968), have more flexible generation (OCGT and gas CHP) in France and Italy, and reserve capacity (synthetic oil-fired power plants) in countries dependent on wind (GB, France, Germany, Netherlands, and Denmark), indicating that it compensates for the observed wind droughts. Spain, which has a solar-dominated electricity generation, does not install backup generation but increases the capacity of solar PV and batteries for compatibility with the 1968 operational year. 

Regarding H$_2$ pipelines, we observe distinct topology layouts: The 1968 design year features a more distributed network in the Southwest, whereas the 1966 layout exhibits more reinforcements in the Northeast. This pattern aligns well with the regional distribution of the loss of load for the 1966 and 1968 capacity layouts (Fig. \ref{fig:map_unserved_energy}), wherein the loss of load occur in the region that lacks reinforcements.

The reason for the system preferring CO$_2$-intense synthetic oil-fueled power plants over OCGT in some regions is explained by the intended operation scheme. The synthetic oil-fired power plant has higher CO$_2$ emissions, lower energy efficiency, and higher running costs, but it has lower fixed cost. I.e., it is built to run only for few hours per year, targeting periods with severe wind droughts. This is more pronounced for the 1966 design year, and in combination with a higher reliance on gas boilers in the heating supply (Supplemental Fig. \ref{sfig:Heating_maps}), this design year is suffering a higher CO$_2$ emissions (on average 0.1\% net CO$_2$ emissions; Fig. \ref{fig:dispatch_optimization_dy2013}). Conversely, the 1968 design year relies more on renewable electrified technologies, with additional deployment of heat pumps and resistive heating, and a higher capacity of H$_2$ production, storage, and fuel cells, compared to the average. For this reason, the 1968 design year emits less CO$_2$ (on average, it becomes net negative with a -0.5\% undershoot) and deploys less DAC (Supplemental Fig. \ref{sfig:Eurupe_aggregate_capacity_DAC}).\\

Second, we estimate the safety reserve capacity margin in every design year necessary for achieving a fully robust system. The regional distribution of peak loss of load is different across operational years. For this reason, we determine the reserve capacity margin from the sum of the peak loss of load in every country across all operational years as a conservative estimate (see Supplemental Fig. \ref{sfig:cap_deficits}). For the 1968 design year, which has the most promising tradeoff of avoiding unserved energy and mitigating CO$_2$ emissions, our approach leads to a European backup capacity requirement of 42~GW (0.6\% of the total electricity generation capacity), with 2/3 located in Great Britain, Sweden, and Italy (see Supplemental Fig. \ref{sfig:map_nodal_loss_of_load}). Furthermore, the deficit lasts approximately 20 hours/year (0.27\% of time), see Supplemental Fig. \ref{sfig:LOL_duration_curves}.\\

\noindent
\textbf{Sensitivity to electricity transmission expansion and seasonal hydropower constraint}. In this paper, we assume a fixed electricity transmission equivalent to today's capacity. In Fig. \ref{fig:map_robustness}, we showed reinforcements in the H$_2$ pipelines for robust design years, indicating a beneficial impact from a larger volume of transmission, in line with the study by Grochowicz et al. \cite{Grochowicz_2023}. In Supplemental Material \ref{smat:senstivity}, we assess the sensitivity to an expansion of the electricity transmission. We show an improved robustness from electricity transmission for an average design year. Since transmission expansion entails more wind generation, the sensitivity to wind variability increases. Concurrently, due to a smaller fleet of backup generation, we observe higher peak loss of load during compound weather events in the capacity layouts with electricity transmission expansion. 

To avoid unrealistic operation of hydropower reservoirs, conflicting with environmental aspects and water accessibility, we included a seasonal constraint on the reservoir filling level according to historical observations. From a sensitivity study in Supplemental Material \ref{smat:senstivity}, we show that the reservoir operation constraint has negligible impact on total system cost. It furthermore leads to a small increase in unserved energy, but this is negligible compared to the variation across all operational years.

\vspace{-0.25cm}
\section*{\Large Discussion}
\vspace{-0.25cm}

\noindent
In this paper, we show that challenging weather conditions affecting the energy system predominantly occur during winter. However, capacity layouts with increased backup generation, H$_2$ fuel cells, and a more distributed H$_2$ network, show capable of alleviating many of the instances.

We identified wind droughts as one of the main causes of unserved energy. In particular when they are reinforced by cold spells that increase heating demand. These compound events have also been identified in previous literature \cite{Brown_2018, Bloomfield_2020, Mockert_2023, Otero2022, Grochowicz_2023} where they are explained by the existence of certain weather regimes across Europe. In particular, European blocking and Greenland blocking weather regimes are known to create high pressure over central Europe causing a period of prolonged low wind and low ambient temperature \cite{Bloomfield_2020, Mockert_2023, Grochowicz_2023}. In our study, the impacts of cold spells are captured better since they entail higher heating demand and lower coefficient of performance of heat pumps, reducing the available electric heating supply. Moreover, we identified an additional type of compound events that results in load shedding. They are caused by a combination of wind drought and low availability of hydropower generation.

The literature underscores the importance of examining longer lasting energy droughts \cite{Grochowicz_2024, Ruhnau_2021}, as they often exert greater system impacts than isolated extreme hours. Our findings align with this perspective, as the high-impact operational years we identified (1966 and 1968) are characterized by compound events lasting multiple days. In addition, our  study emphasizes the pitfalls of relying solely on average renewable resources and demand of a certain year to assess system reliability. The identified high-impact operational years do not show extreme values for the annual wind resources, hydro inflow, and heating demand. In fact, they show average annual hydro inflow and heating demand, while the annual wind energy is higher than eight other operational years.

Across all design years examined, we observe a variation in total system costs of $\pm$10\% relative to the mean, largely determined by the annual wind resources and the demand for heating. Grochowicz et al. \cite{Grochowicz_2023} observed a comparable variation in total system costs using 40 design years, and Schlachtberger et al. \cite{Schlachtberger_2018} showed similar correlations between annual wind resources and system costs using four design years. Both studies investigate only the power system. It is relevant to notice that interannual system cost variability remain similar in our case where the power system is coupled with heating, transport and industry. The core compensation to compound weather events is still located in the power sector (more backup generation from gas OCGT and synthetic oil-fired power plants). Synthetic oil is produced by Fischer-Tropsch plants that cover the demand for oil in industry and aviation. In some countries, oil power plants are preferred over OCGT since they are only used a dozen hours per year. The H$_2$ network built also enhance robustness by interconnecting countries in Europe.

Although our system can use CO$_2$-emitting generators (subject to a high CO$_2$ tax) to supply demand in challenging periods, we still observe deficits in the electricity supply. For the most robust capacity layout (1968), we show that an additional reserve capacity of 42~GW, operated 20 hours/year, is needed to fully avoid unserved energy in every operational year. Assuming that this capacity is provided by OCGT power plants fueled with fossil gas, this adds annualized capacity costs of €777 million and fuel costs of €3 million (assuming a gas price of 20~€/MWh), corresponding to a cost increase of 0.1\%, relative to the average of all design years. Furthermore, this adds emissions of 26 ktCO$_2$/year, corresponding an overshoot of less than 0.001\% relative to 1990 historical levels. Previous literature has shown the benefit of integrating green methanol in the backup generation, since methanol storage has a low capacity cost, and it can be stored and dispatched with interannual fluctuations \cite{BROWN20232414}. Choosing OCGT fueled with green methanol (with a market price of 100~€/MWh \cite{BROWN20232414}) instead changes the picture marginally (same capital costs + €13 million in fuel costs).

Our study addresses also the interannual variation of CO$_2$ emissions. We show how years with persistent low wind energy and high heating demand (e.g., 2010) causes the highest net CO$_2$ emissions, while other operational years show capable of reaching net negative. Over the six decades of weather data, the capacity layouts reach an average net CO$_2$ emissions of 0.8\% relative to 1990, equivalent to 34 MtCO$_2$/year. This could be offset by installing DAC with a capacity of 4 ktCO$_2$/h (which is almost equivalent to the capacity built in the design year with the largest DAC installments). This would add annualized capital costs of €3 million for the DAC, but would require additionally 17\% of the assumed underground sequestration potential to be exploited.\\

In the following paragraphs, we discuss the main limitations of our analysis.

Deploying an additional capacity for backup generation is the conservative option to avoid load shedding. In reality, fluctuating electricity prices drive the incentive for demand-side management which would shift some energy services to a later instance when electricity prices are lower. While charging of electric vehicles and power-to-X technologies operate flexibly, the remaining residential electricity demand is assumed to be perfectly inelastic. For this reason, the system does not represent the consumer behavior as would be expected in reality, which could further prevent instances with unserved energy. Demand elasticities could also be utilized by reducing demand with an appropriate monetary compensation. Another aspect is that the model is constrained by its assumption on Europe being autarkic as we do not allow imports of electricity and green fuels from outside of Europe. In reality, intercontinental energy trades could also increase the robustness under compound events, e.g., by storing imported green fuels as a strategic long-term reserve.

Our assessment relies on a joint capacity and dispatch optimization, considering the operation throughout a calendar year, repeated for different design years. The model assumes long-term market equilibrium, and for this reason, assets are only deployed if they fully recover their costs. The cost recovery is, however, only guaranteed for the considered design year. For instance, the increased share of backup generation in robust capacity layouts is a result of a difficult design year, but in more favorable design years, less backup generation is needed. This might challenge the financial aspect of the investments associated with the proposed backup infrastructure in the robust capacity layouts, assuming this happens on a free market. In the model, this could be addressed by extending the horizon, and thus the perfect foresight, from one year to multiple years, similar to previous research \cite{BROWN20232414,Ruhnau_2021,DOWLING20201907, Grochowicz_2024, Zeyringer_2018} while including one of the extreme weather years identified in our work. This could potentially replace some of the backup generation with interannual energy storage \cite{BROWN20232414, DOWLING20201907}. This approach is, however, much more computationally expensive, and hos not yet been done with the level of spatial, temporal, and technological resolution used in our analysis.   

In our work, we assume that the 62 weather reanalysis years encapsulate the realm of likely weather events that may interfere with a future fully decarbonized energy system. Already, from the weather reanalysis, we were able to see a long-term trend in the heating demand and hydro inflow. The aim of our work was, however, to analyze the impact of the natural variability in a steady-state climate. In the future, climate change signal might be even more pronounced and could have more complex implications impacting not only the annual inflow and heating demand. Increasing ambient temperatures leads to a negative trend in the heating demand, while for cooling demand, it is the opposite \cite{KOZARCANIN2020111386}. This necessitates the deployment of air-condition units in more locations, imposing more stress on the electricity grid during heat waves \cite{WECC_heatwave_report}. Concurrently, droughts in hydro reservoirs is expected to be more severe in Southern regions of Europe \cite{GOTSKE2021102999}. Moreover, wind energy \cite{Hahmann2022} and solar PV \cite{Hou2021} resources might also be impacted. Future studies could address this in a similar assessment, but with more focus on the climate change signal instead of the natural variability of renewable resources and energy demand as highlighted in our work \cite{eurocordex}.\\
  
In conclusion, we show variations in total system cost from weather years of 10\%, with the largest determinants being the annual wind resources and the demand for heating. A net-zero CO$_2$ emission European energy system comprising high renewable penetration shows robust to historical weather years if having built backup generation to compensate for week-long wind droughts. Due to higher utilization of CO$_2$-emitting technologies, extreme weather years drive higher CO$_2$ emissions, which leads to a maximum overshoot of 7\% relative to 1990 levels for a particularly challenging operational year, but reaches an average level of 1\% CO$_2$ emissions for all operational years. Our results provide significant insights into the design of robust highly renewable energy systems but further research is needed to understand the impacts of compound weather events on highly renewable energy systems and the best strategy to make them robust against a changing climate.

\section*{\Large Methods} % max 3000 words
\vspace{-0.25cm}

We investigate 62 alternative future European energy systems, determined for different weather years, using the open sector-coupled energy system model \textit{PyPSA-Eur}. The model performs capacity expansion and dispatch optimization of generation, transmission, storage, and conversion technologies. The model assumes an idealized market with perfect competition, and it has perfect operational foresight with some limitations for hydropower explained below. All energy storage (including hydropower reservoirs) is modeled assuming a cyclic operation, ensuring that the storage filling level at the end of the year is equivalent to the initial filling level. Our investigation is split in two steps: 1) determining 62 alternative capacity layouts for different weather years (design years), and 2) simulating the 62 capacity layouts in every other weather year (operational year) different than the design year, leading to a total of 3,844 optimization runs. The capacity layouts are determined from an overnight optimization (with key assumptions summarized in Table \ref{tab:assumptions}) based on a network resolution of 37 nodes, covering 33 ENTSO-E member countries, and is solved with a 3-hourly resolution to reduce computation time. Previous research \cite{Schlachtberger_2018, Neumann_2023} shows negligible impact of downsampling from hourly to 3-hourly resolution for long-term planning studies, in particular for studies including multiple weather years. The model includes energy and non-energy feedstock demand and CO$_2$ emissions (energy-related and from industrial processes) in the electricity, heating, land transport, shipping, aviation, and industry sectors and energy demand in agriculture. For heating, we assume district heating is expanded to cover 60\% of urban demand. For land transport, we assume 85\% of the heavy- and light-duty vehicle fleet to be battery electric vehicles, while the remaining heavy-duty transport uses H$_2$ fuel cells. The shipping sector is fueled with H$_2$. For steel manufacturing, 30\% is assumed to originate from the primary route with H$_2$-based Direct Reduced Iron, while the remaining share is from the secondary route (i.e., from recycling). For aluminum, we assume 20\% is produced via primary route and 80\% is recycled, while we assume no recycling of plastics and high-value chemicals. The overnight optimization relies on a greenfield investment approach, except for hydropower and electricity transmission, for which we assume the capacity of today (disregarding planned lines in the Ten Year Network Development Plan (TYNDP 2018) by ENTSO-E \cite{entsoe_tyndp}; see Supplemental Fig. \ref{sfig:transmission_capacity}). We fix the hydropower capacity as we assume that the cost-effective potential in Europe has already been fully exploited, while we assume a fixed electricity transmission to represent rising network expansion costs in recent years, due to a supplier scarcity, and social acceptance constraints. For the electricity transmission assumption, we provide a sensitivity analysis in Supplemental Material \ref{smat:senstivity}. We include interannual variability for wind energy, solar PV, hydropower inflow, heating demand, and coefficient of performance (COP) of heat pumps. We include the historical national electricity demand from ENTSO-E (deducting the estimated share for heating) as an inflexible electricity demand, assuming zero price elasticity and no interannual variability. Additional electricity demand from electrification in the heating sector determined endogenously by the model can have interannual variation due to the included temperature dependence of the heating demand and COP. Cooling demand is included in the inflexible electricity load which is kept constant in all design and operational years. For the mathematical formulation, we refer to Supplemental Material \ref{smat:math_formulation}.\\ 

\begin{table}[!h]
	\centering
	\caption{Key assumptions in the capacity optimization.}
	\label{tab:assumptions}
	\begin{tabularx}{0.45\textwidth}
		{>{\hsize=.7\hsize\linewidth=\hsize}X
			>{\hsize=1.3\hsize\linewidth=\hsize}X}
		\toprule
		\textbf{Assumption} & \textbf{Description} \\ \midrule
		\textit{Net-zero emissions} & Net-zero CO$_2$ emissions constraint, corresponding to the target for Europe 2050\\
		\textit{Overnight} & We do not model the transition. \\
		\textit{Greenfield} & We do not consider existing power plants and infrastructure, except hydropower and electricity transmission.\\
		\textit{Transmission} & We assume that electricity transmission is fixed to today's capacity. For this assumption, we perform a sensitivity study, found in Supplemental Material \ref{smat:senstivity}.\\
		\textit{Hydropower} & We assume that hydropower is fixed to today's capacity.\\
		\textit{Network} & Network of 37 nodes covering 33 ENTSO-E member countries.\\
		\textit{Renewables} & The renewable resources are estimated from the weather reanalysis in 370 nodes, connected to the 37 network nodes.\\
		\textit{Resolution} & 3-hourly.\\
		\textit{Technology costs} & To reduce uncertainties in technology cost evolution, we use assumptions for 2030. \\
		\textit{Sectors} & Electricity, heating, land transport, aviation, shipping, industry including industrial feedstock and comprehensive carbon management.\\
		\textit{Flexibility strategies present} & Batteries, H$_2$ electrolysis and fuel cells, EV batteries, thermal energy storage, production of synthetic fuel, gas- and synthetic oil-fired power plants, co-generation of heat and electricity (CHP) fueled with gas and biomass.\\ 
		\textit{Interannual variability} & Wind energy, solar PV, hydropower inflow, heating demand, Coefficient of Performance (COP) of heat pumps.\\
		\bottomrule
	\end{tabularx}%
\end{table}

Following the capacity optimization, we fix the capacity layout and perform a dispatch optimization for every other operational year than the design year. We allow the system to shed electricity and heat load, thus violating the energy balance imposed in the capacity optimization (Eq. \ref{eq:energy_balance}). This should be considered a last resort since it is a very expensive solution, as we assume a social cost of unserved energy of 100,000~€/MWh, similar to the social cost (87,000\$/MWh) estimated for the 2021 blackout in Texas \cite{Gruber2022}. To evaluate the loss of load events, we identify the metrics summarized in Table \ref{tab:metrics}. Similarly, while the capacity optimization was subject to a net-zero CO$_2$ emissions constraint, in the dispatch optimization, we allow the model to deviate from this constraint as a measure to avoid load shedding. This requires the payment of a CO$_2$ tax, derived from the CO$_2$ shadow price $\mu_{CO_2}$ in the capacity optimization. See Eq. \ref{eq:co2_constraint} and Supplemental Fig. \ref{sfig:co2_emissions_price} for the derived CO$_2$ tax in every design year. We assume a fixed level of CO$_2$ emissions from industrial processes (153~MtCO$_2$/year) and plastic decay and kerosene combustion (323~MtCO$_2$) which can be offset by carbon capture on industrial sites, biomass CHP with carbon capture, or direct air capture (DAC).\\

\begin{table}[!ht]
	\caption{Metrics used in the assessment of resource adequacy.}
	\label{tab:metrics}
	\begin{tabularx}{0.45\textwidth}
		{>{\hsize=.7\hsize\linewidth=\hsize}X
			>{\hsize=1.3\hsize\linewidth=\hsize}X}
		\toprule 
		\textbf{Metric} & \textbf{Description} \\ \midrule
		\textit{Loss of load} or \textit{Unserved energy} & 3-hourly deficits of electricity in units of GWh. In describing this phenomena, we use loss of load and unserved energy interchangeably. The number of hours in which loss of load happens divided by 8760 hours in a year corresponds to the more common metric, Loss of Load Expectation (LOLE).\\
		
		\textit{Peak loss of load} & The maximum 3-hourly loss of load, indicating the most difficult time step during an operational year. In units of GWh.\\
		
		\makecell[cl]{\textit{Sequential} \\ \textit{unserved energy}} & Sum of loss of load during a period with consecutive hours of unserved energy. In units of TWh.\\
		
		\makecell[cl]{\textit{Cumulative} \\ \textit{unserved energy}} & Sum of loss of load during a full operational year. In units of TWh when discussing European aggregates, and units of average electricity demand (average hourly load; av.h.l.) for countries. The average electricity demand includes both the exogenous and the endogenous electricity demand from the electrification in the energy sectors.\\
	\end{tabularx}%
\end{table}

Using Atlite \cite{Hofmann2021}, we convert weather reanalysis from ERA5 \cite{era5}, acquired for the years 1960 - 2021, to estimate the available renewable time series for wind energy, solar PV, and hydropower. From the same source, we also estimate the heating demand and the coefficient of performance (COP) of heat pumps. For wind energy and solar PV resources (Supplemental Fig. \ref{sfig:aggregated_wind_and_solarPV}), and the COP (Supplemental Fig. \ref{sfig:average_COP}), this is based on the available data for all weather years, without further data processing. Hydro inflow is estimated with Atlite based on the surface runoff data from ERA5, data on hydropower plant locations \cite{jrc} and upstream basins \cite{Lehner2013}, and is calibrated with historical annual production from IRENA \cite{IRENA}. Despite being a conventional electricity generation technology, the historical hydropower capacity has increased from 1960 and forward (either by exploring new sites or expanding the capacity at an existing site). This has led to a positive trend in the historical generation. This is accounted for by normalizing the historical generation data by the installed capacity for the corresponding year. The heating demand is estimated based on the heating degree days (HDD) computed from the ERA5 temperature and fitted with historical intraday profiles from BDEW \cite{oemof}. From the JRC IDEES \cite{JRCIDEES2015}, this is calibrated with the 2015 reported annual data. For the hydro inflow and heating demand, previous work has shown a historical long-term trend in hydro resources \cite{HADDELAND2022625} and heating demand \cite{Deroubaix2021}. For this reason, we apply a method to detrend the data. To omit the historical climate change signal, we impose a cutoff at the 1990 weather year, dividing the dataset into two 30-year climate periods. The first 30 years are then scaled to match the average of the subsequent 30-year period, while retaining the intraannual variation according to the ERA5 data. See Supplemental Figs. \ref{sfig:annual_hydropower_compare} and \ref{sfig:annual_heating_demand_compare}.

When comparing the annual renewable resources (average capacity factors) and heating demand for all weather years, we aggregate on both a technological and geospatial level. Solar PV resources include utility scale and rooftop panels, while wind energy includes onshore and offshore turbines. The geospatial aggregation is capacity-weighted, based on the optimal capacity layout of the capacity optimization. The sensitivity of the European aggregation of wind and solar PV resources to the capacity layout was tested in Supplemental Fig. \ref{sfig:aggregated_wind_and_solarPV} which shows negligible impact.\\

\begin{figure}[!h]
	\centering
	\includegraphics[width=0.45\textwidth]{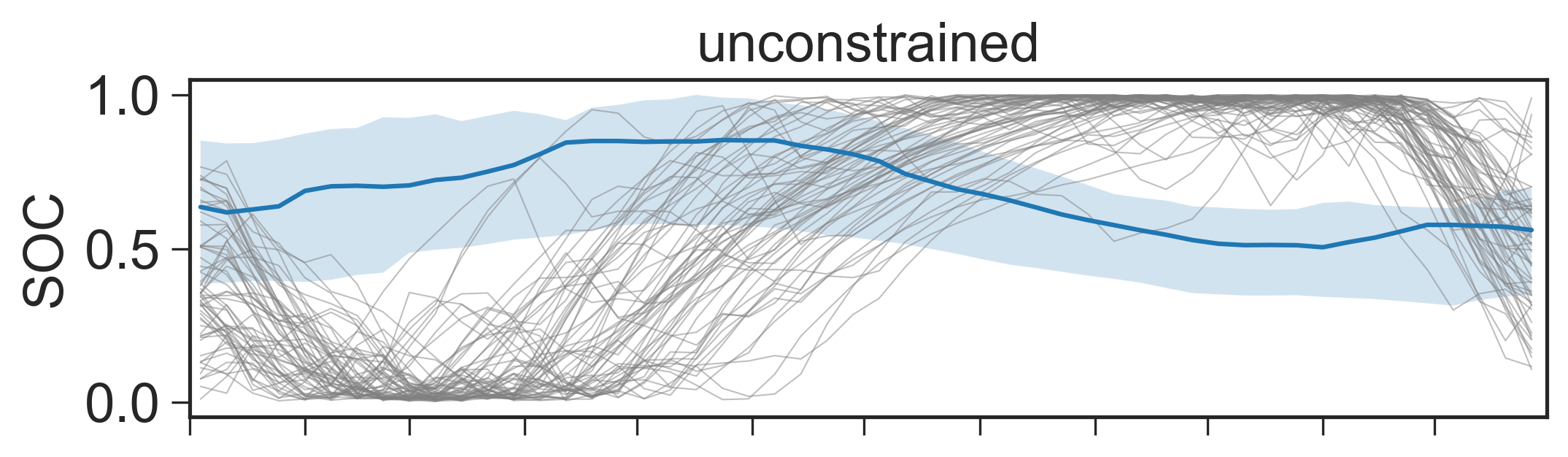}
	\includegraphics[width=0.45\textwidth]{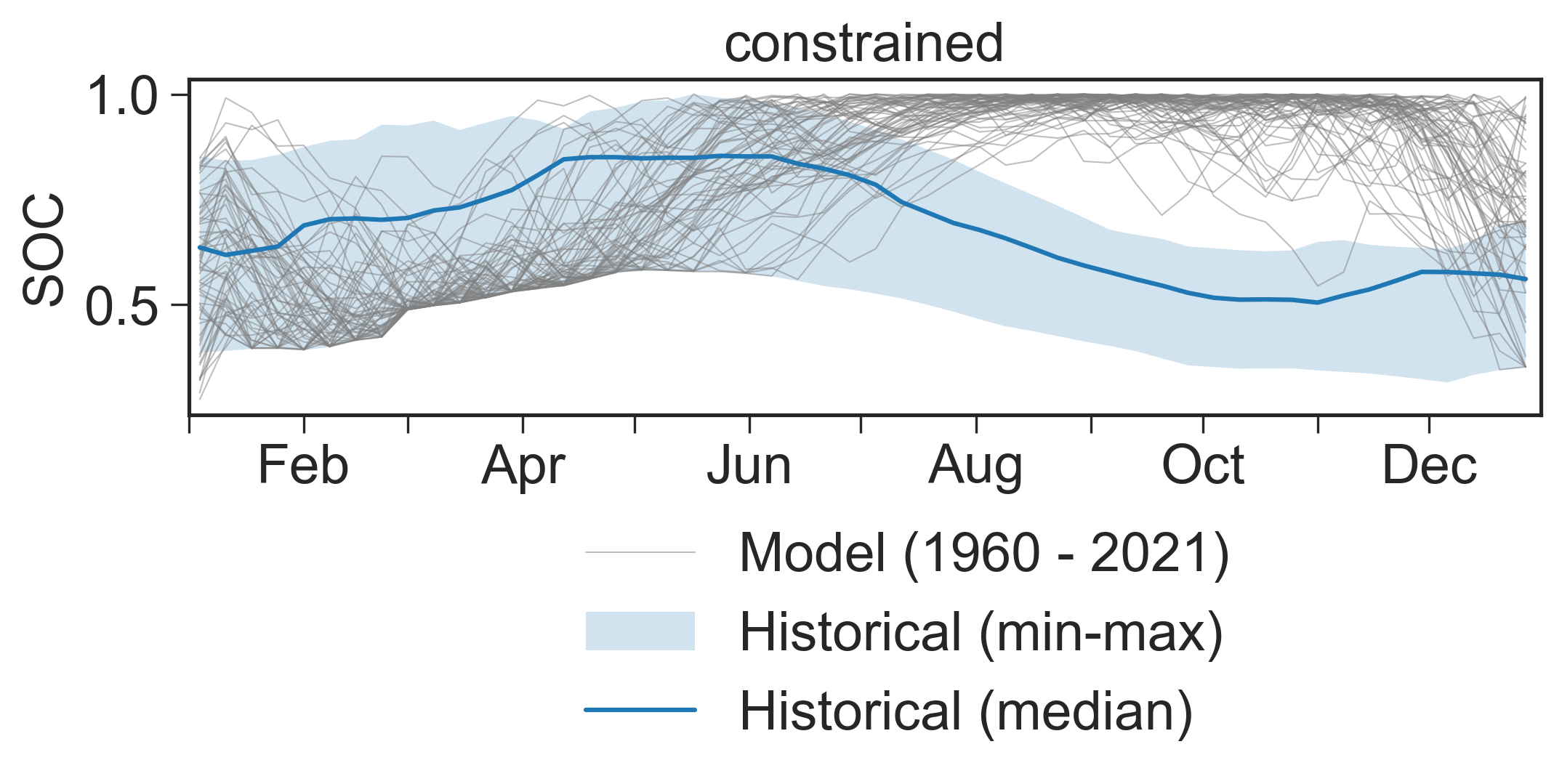}
	\caption{\textbf{State-of-charge (SOC) of country-aggregate hydro reservoirs} for Spain where (top) the hydropower operation is unconstrained and (bottom) where the filling level is bounded by the range of historical operation. The gray lines indicate the modeled PyPSA-Eur capacity-optimization scenarios with weather data from 1960 to 2021. See Supplemental Figs. \ref{sfig:hydro_constraint} and \ref{sfig:hydro_constraint_result} for the unconstrained and constrained reservoir operation for all countries with hydropower.}
	\label{fig:hydro_constraint_ES}
\end{figure}

To emulate a water-energy nexus model that accounts for the access to water for other purposes than energy, we consider historical data to delineate the range of realistic operations of hydropower. This is important for the cost-optimal operation of hydropower to harmonize with other natural and human requirements. Previous work has shown that without this, the model tends to operate hydro reservoir in a more aggressive manner compared to the historical seasonal trend, mostly driven by the perfect foresight assumption \cite{GOTSKE2021102999}. In our model, we also observe how countries such as Spain and Greece exert substantial deviation from the historical data, when we do not impose any operational constraint on the reservoir filling level (Fig. \ref{fig:hydro_constraint_ES}). This could cause unintended problems in water accessibility and river flow continuity. Here, we apply the ENTSO-E data on historical reservoir filling level to impose a lower bound of the modeled weekly average state-of-charge (SOC). See Supplemental Material \ref{smat:math_formulation} for the mathematical formulation of the constraint. The resulting SOC is found in Fig. \ref{fig:hydro_constraint_ES} which shows that this is a conservative approach. 

\subsubsection*{Data availability}

Datasets including results from the capacity optimization and from the robustness assessment can be accessed from the public repository: \href{https://doi.org/10.5281/zenodo.10891263}{10.5281/zenodo.10891263}.

\subsubsection*{Code availability}

\begin{sloppypar}
	Our assessment relies on the open energy modeling framework PyPSA and makes use of the model \href{https://github.com/PyPSA/pypsa-eur-sec}{PyPSA-Eur-Sec} v0.6.0. Costs and technology assumptions are obtained with \href{https://github.com/PyPSA/technology-data}{technology-data} v0.4.0. The code used in the robustness assessment can be accessed from the public repository:
\end{sloppypar}

\noindent
\href{https://github.com/ebbekyhl/multi-weather-year-assessment/tree/master}{github.com/ebbekyhl/multi-weather-year-assessment/}.\\

\noindent
Scripts produced for the data visualization can be accessed from the public repository: \href{https://doi.org/10.5281/zenodo.10891263}{10.5281/zenodo.10891263}.

%\bibliography{Bibliography.bib}

\subsubsection*{Acknowledgement}

E.K.G., M.V., and G.B.A. are partially funded by the GridScale project supported by the Danish Energy Technology Development and Demonstration Program
under Grant No. 64020-2120. 
This research has received funding from DFF Sapere Aude - EXTREMES project (2067-00009B).

\subsubsection*{Author contributions}

E.K.G. designed the analysis, drafted the manuscript and contributed to the analysis and interpretation of data. G.B.A., F.N. and M.V. contributed to the initial idea and made substantial revisions of the manuscript. M.V. contributed to writing the introduction and supervised the investigation.

\subsubsection*{Competing interests}
The authors declare no competing interests.

\clearpage % <--------------- uncomment to include sup. mat.
\onecolumn % <--------------- uncomment to include sup. mat.
\beginsupplement % <--------------- uncomment to include sup. mat.
\textbf{\textcolor{gray}{Supplemental Material}}\\

\begin{adjustwidth}{50pt}{50pt}
	
\section{Mathematical formulation}\label{smat:math_formulation}

Here, we present the governing equations of our analysis. First, the objective function of the join capacity and dispatch optimization is formulated in Eq. \ref{eq:objective}:
\begin{equation}\label{eq:objective}
	\min_{G_{n,s},E_{n,s},F_{l},g_{n,s,t}} 
	\bigg[ \sum_{n,s} c_{n,s}G_{n,s} + \sum_{n,s} \hat{c}_{n,s}E_{n,s}
	+ \sum_l c_l F_l + \sum_{n,s,t} o_{n,s,t}g_{n,s,t} \bigg]
\end{equation}

\noindent
where $c_{n,s}$ and $\hat{c}_{n,s}$ are the annualized costs for generator power capacity $G_{n,s}$ and storage energy capacity $E_{n,s}$ for technology $s$ in node $n$, $c_l$ are the fixed annualized costs for the capacities $F_l$ of the links $l$, and $o_{n,s,t}$ are the marginal costs of generation and storage dispatch $g_{n,s,t}$ at time $t$.\\

\noindent
In the join capacity and dispatch optimization, the energy balance most be fulfilled in all hours and all locations, formulated in Eq. \ref{eq:energy_balance}, ensuring that the energy generation, the charging and discharging of energy storage, and the import/export are exactly balanced with the demand $d_{n,t}$:
\begin{equation}\label{eq:energy_balance}
	\begin{split}
		\sum_s g_{n,s,t} + \sum_l \alpha_{n,l,t} f_{l,t} = d_{n,t} &\leftrightarrow \lambda_{n,t} \hspace{0.5cm} \forall n,t
	\end{split}
\end{equation}
where $f_{l,t}$ is the power flow through link $l$ at time $t$, $\alpha_{n,l,t}$ represent the direction and efficiency of the power flow through links, and $\lambda_{n,t}$ is the electricity shadow price in node $n$ and time $t$. 

For the join capacity and dispatch optimization, we impose an upper bound of zero on the net CO$_2$ emissions, aggregated for Europe, formulated in Eq. \ref{eq:co2_constraint}:
\begin{equation}\label{eq:co2_constraint}
	\begin{split}
		\sum_{n,s,t} \varepsilon_s \frac{g_{n,s,t}}{\eta_{s}} \leq 0 &\leftrightarrow \mu_{CO_2} \hspace{0.5cm}
	\end{split}
\end{equation}
where, $\varepsilon_s$ is the CO$_2$ intensity in tonne CO$_2$ per MWh$_\text{th}$ for technology $s$, $\eta_{s}$ is the energy efficiency, and $\mu_{CO_2}$ is the resulting CO$_2$ shadow price.\\

In the subsequent dispatch optimization, we modify Eq. \ref{eq:energy_balance} and remove \ref{eq:co2_constraint}. In Eq. \ref{eq:energy_balance}, we include load shedding, assuming a social cost of unserved energy of 100,000 EUR/MWh. For CO$_2$ emissions, we replace the net CO$_2$ emissions constraint with a CO$_2$ tax, derived from the shadow price $\mu_{CO_2}$ obtained in the join capacity and dispatch optimization.\\

\noindent
To avoid that the operation of hydropower reservoirs does not conflict with other non-energy related constraints, we impose a lower bound on the hydropower filling level according to the historical range:
\begin{equation}\label{eq:SOC_hydro}
	SOC_t^{\text{model}} \geq \min SOC_t^{\text{hist}}
\end{equation}

where $SOC_t$ is the state of charge (SOC) in time step $t$. The SOC is determined from the filling level $e_t$ divided by the reservoir energy capacity $E$. The filling level in time step $t$ is determined from the reservoir energy balance:  
\begin{equation}
	e_t = e_{t-1} + \Delta e_t^{+, \text{inflow}} + g_t/\eta_{\text{el}} + \Delta e_t^{-,\text{spill}} \\
\end{equation}
where $e_{t-1}$ is the filling level in the previous time step, $g_t$ is the hydropower generation, $\eta_{\text{el}}$ is the turbine energy efficiency, and $\Delta e_t^{\text{spill}}$ is the water spillage.

\newpage
\section{Supplemental Results}\label{smat:interannual_variability}

\textbf{Interannual variability of renewable resources and heating demand}. 
Fig. \ref{fig:interannual_variability} shows the annual wind, solar PV, and hydro energy that can be utilized for electricity production (hereafter referred to as resources), as well as heating demand for weather years ranging from 1960 to 2021, for the aggregated Europe and Spain (See Supplemental Fig. \ref{sfig:interannual_variability} for other countries). Note that the estimated hydro inflow and heating demand show significantly different values observed in the years before 1990 compared to the subsequent years (see Supplemental Figures \ref{sfig:annual_hydropower_compare} and \ref{sfig:annual_heating_demand_compare}). For this reason, we define a cutoff at 1990, and scale the years earlier than this to match the average of the subsequent years, while maintaining the intraannual variation. Consequently, the values for hydro inflow and heating demand in Fig. \ref{fig:interannual_variability} are only displayed for 1990 - 2021.\\

\noindent
On a country level, the largest source of interannual variability arises from variations in the hydro reservoir inflow, particularly pronounced for Spain. Here, deviations up to 40 \% from the long-term mean are shown, aligning closely with previous reporting \cite{VICTORIA2019674}. The aggregation on a European level reduces the interannual variability of hydro inflow. This observation can be explained by an anticorrelation between the inflow in Northern and Southern reservoirs. For instance, the year with the lowest inflow in Norway and Sweden (1996) coincides with the year featuring the highest inflow in Spain. Those geographical differences are counterbalanced in the aggregation. Wind speeds and temperatures vary greatly due to complex atmospheric dynamics, while solar irradiation is more predictable due to stable orbital dynamics. Wind, solar PV, and heating demand do not show similar sensitivity to the aggregation. For a chronological depiction of wind and solar PV resources across the full range of weather years (1960 to 2021), see Supplemental Fig. \ref{sfig:aggregated_wind_and_solarPV}.\\

% Of the depicted countries, six of them are heavily relying on hydropower, while we also include a country without hydropower (Denmark) for comparison.

\begin{figure}[!h]
	\centering
	\includegraphics[width=0.85\textwidth]{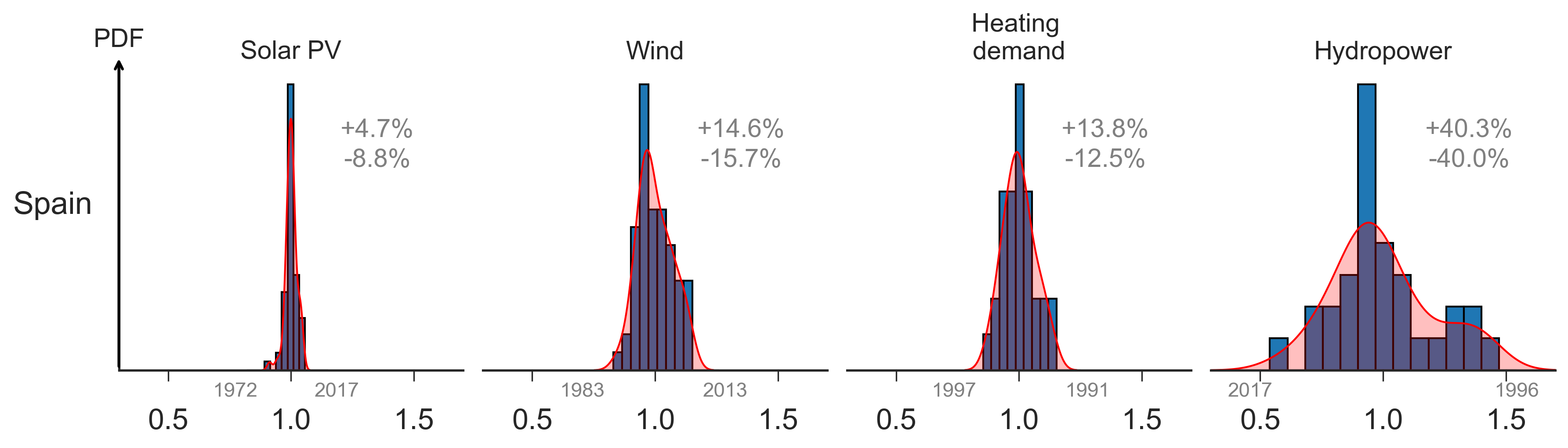}
	\includegraphics[width=0.85\textwidth]{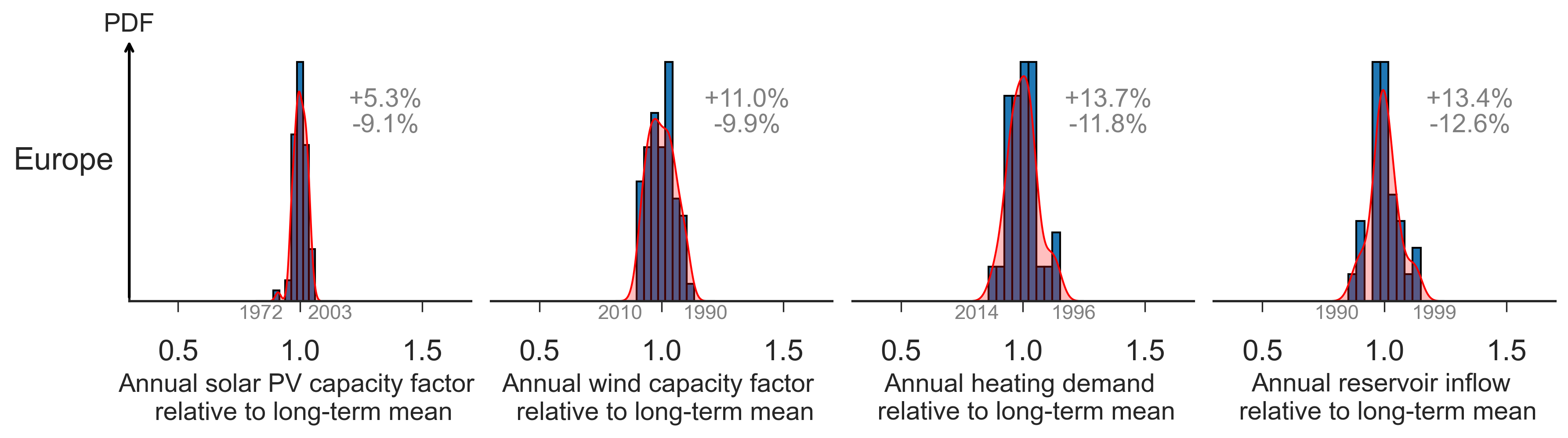}
	\captionsetup{width=14.8cm}
	\caption{\textbf{Histogram of annual solar PV and wind energy resources, heating demand, and hydro inflow} from 1960 to 2021 for wind and solar PV, and 1990 to 2021 for hydro and heating demand for (top) Spain and (bottom) aggregated Europe. The annotations below the first axis show the year with the maximum and minimum resources and demand. The percentage values in grey indicate the maximum and minimum deviation from the average. See Supplementary Fig. \ref{sfig:interannual_variability} for other countries.}
	\label{fig:interannual_variability}
\end{figure}

\newpage
\textbf{Distributions of annual renewable resources and heating demand}
\begin{figure}[!h]
	\centering
	\includegraphics[width=0.85\textwidth]{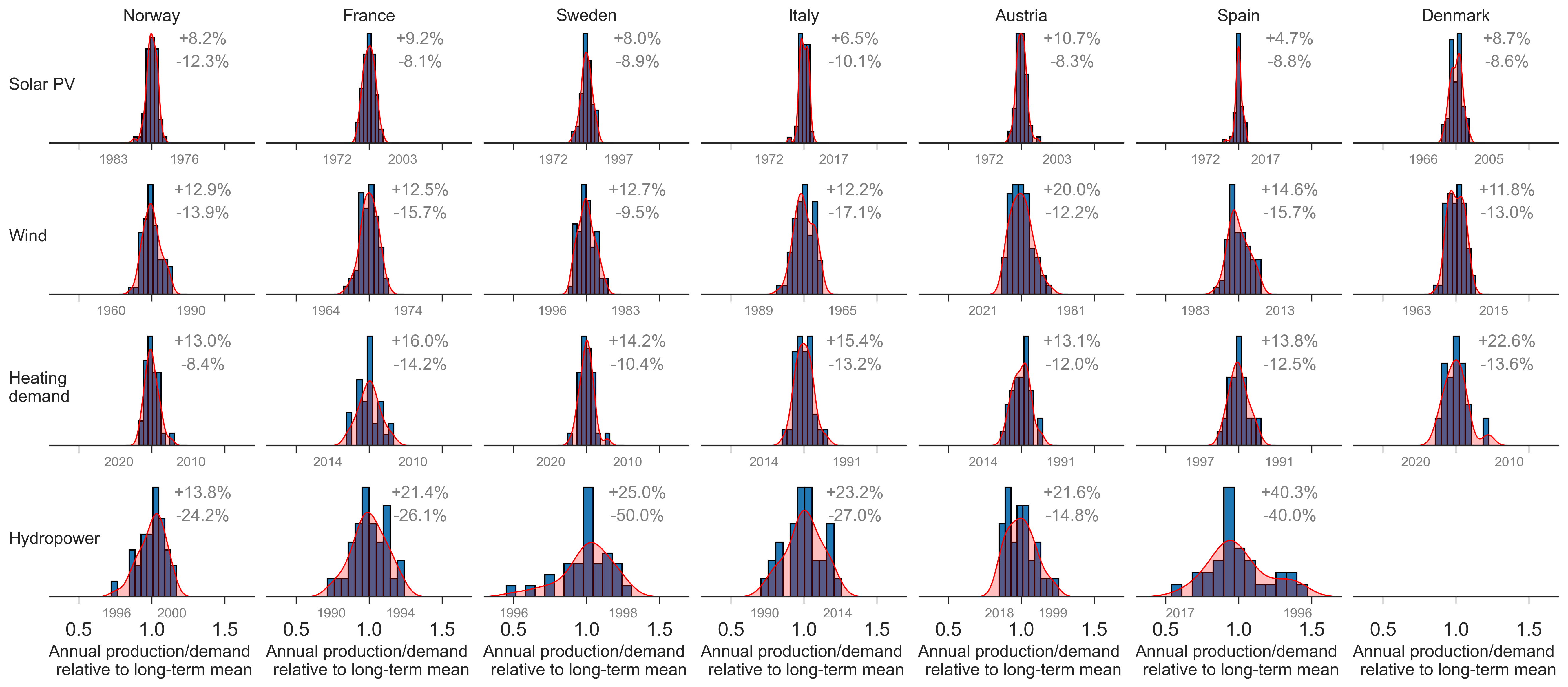}
	\captionsetup{width=14.8cm}
	\caption{Histogram of annual solar PV and wind energy resources, heating demand, and hydro inflow from 1990 to 2021 for Europe (top row) and seven countries (columns) of which six of them have large supply of hydropower. The annotations below the first axis show the year with the maximum and minimum resources and demand, and the percentages show the corresponding deviation from the average.}
	\label{sfig:interannual_variability}
\end{figure}

\newpage
\section{Sensitivity analysis}\label{smat:senstivity}

\textbf{Analysis with transmission expansion}. In our main study, we assumed a fixed electricity transmission capacity equivalent of today's, including existing and planned lines according to the Ten Year Network Development Plan (TYNDP 2018) by ENTSO-E \cite{entsoe_tyndp}. Here, we perform a sensitivity study where we allow the electricity transmission to be extended. We focus on the 1968 design year, since this showed the best tradeoff between avoiding loss of load and mitigating CO$_2$ emissions, and compare with an average design year (2013). Consistent with the process described in the main body of this paper, we first run a join capacity and dispatch optimization for one weather year as input (design year) followed by a dispatch optimization of the capacity layout under every other 61 weather years (operational years). In the join capacity and dispatch optimization, we allow the expansion of the electricity transmission, assuming an upper bound of 10\%, 30\%, 50\%, 70\%, 100\% of today's volume. We also include a scenario where we do not impose any upper bound on the transmission expansion, corresponding to the cost-optimal expansion. Fig. \ref{sfig:sensitivity_transmission_1968}a shows the investment costs in all cases of transmission expansion, and Fig. \ref{sfig:sensitivity_transmission_1968}b shows the changes in capacity deployment relative to the zero-expansion scenario. We observe reductions of up to 5\% in total system costs from transmission expansion. This is because of the stronger interconnection requiring less backup generation (gas CHP, OCGT, and synthetic oil-fired power plants) in every node, making the system less expensive. With an expanded electricity transmission, windy weather regimes can be better distributed, entailing a higher capacity of wind energy, substituting electricity generation from solar PV + battery. The average design year shows similar impact, but chooses to substitute with offshore wind, instead of onshore, while also causing some of the interconnection with H$_2$ pipelines to be replaced by the increased electricity transmission. The difference here is that the 1968 design year still sees the need for the H$_2$ pipeline, to overcome the December wind drought discussed in the main body of this paper. For this reason, it retains the H$_2$ pipeline despite an increased volume of electricity transmission, while this is not the case for the average design year.\\ 

\begin{figure}[!htbp]
	\centering
	\includegraphics[width=0.44\textwidth]{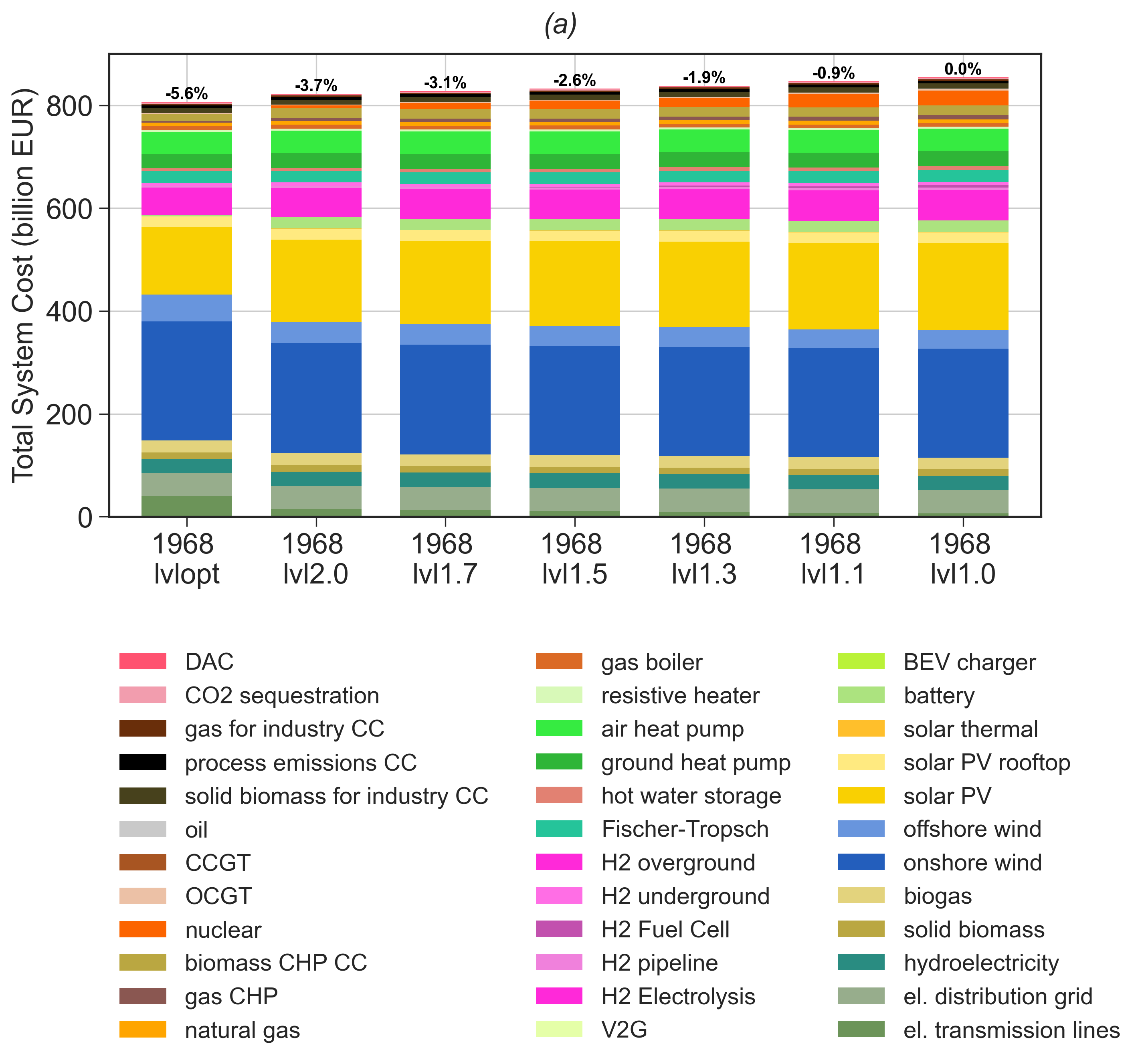}
	\includegraphics[width=0.285\textwidth]{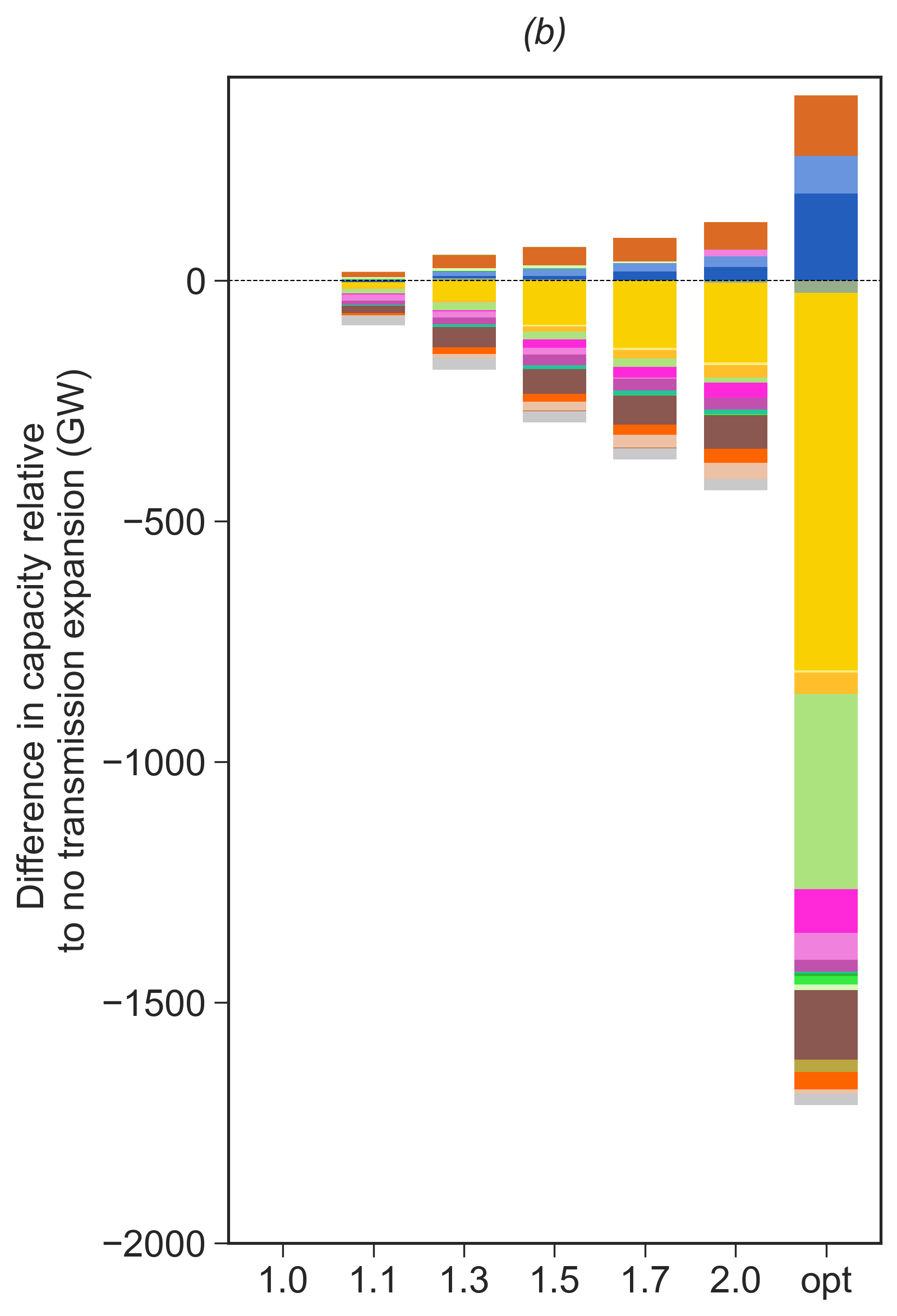}
	\captionsetup{width=14.8cm}
	\caption{Capacity optimization for the 1968 design year, showing (a) total system costs and (b) the deployed capacities in the scenarios with allowed transmission capacity. We impose an upper bound of 10\%, 30\%, 50\%, 70\%, 100\% (1.1, 1.3, 1.5, 1.7, and 2.0) of today's volume (1.0). We also include a scenario without an upper bound (opt).}
	\label{sfig:sensitivity_transmission_1968}
\end{figure}

\begin{figure}[!htbp]
	\centering
	\includegraphics[width=0.44\textwidth]{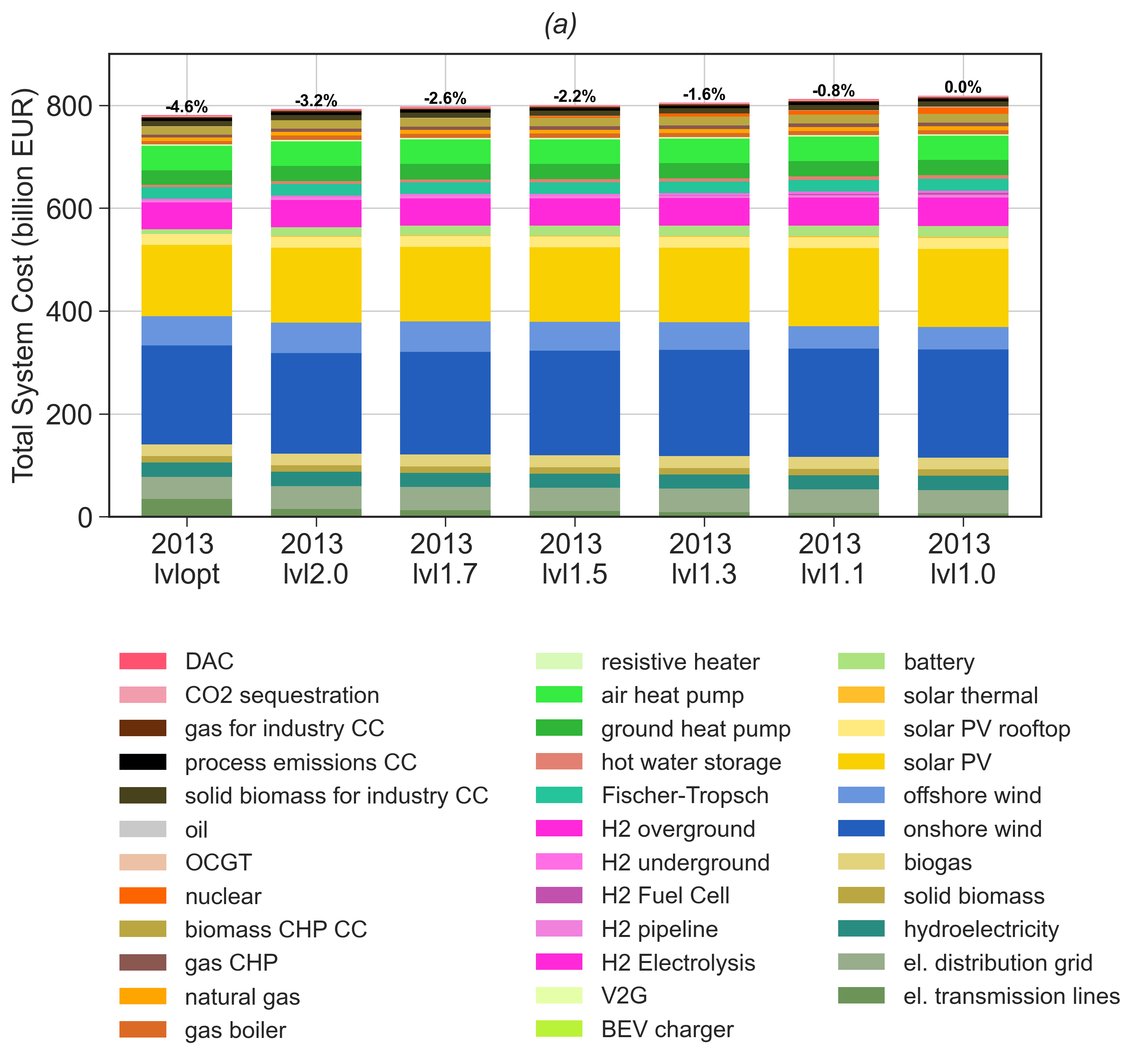}
	\includegraphics[width=0.285\textwidth]{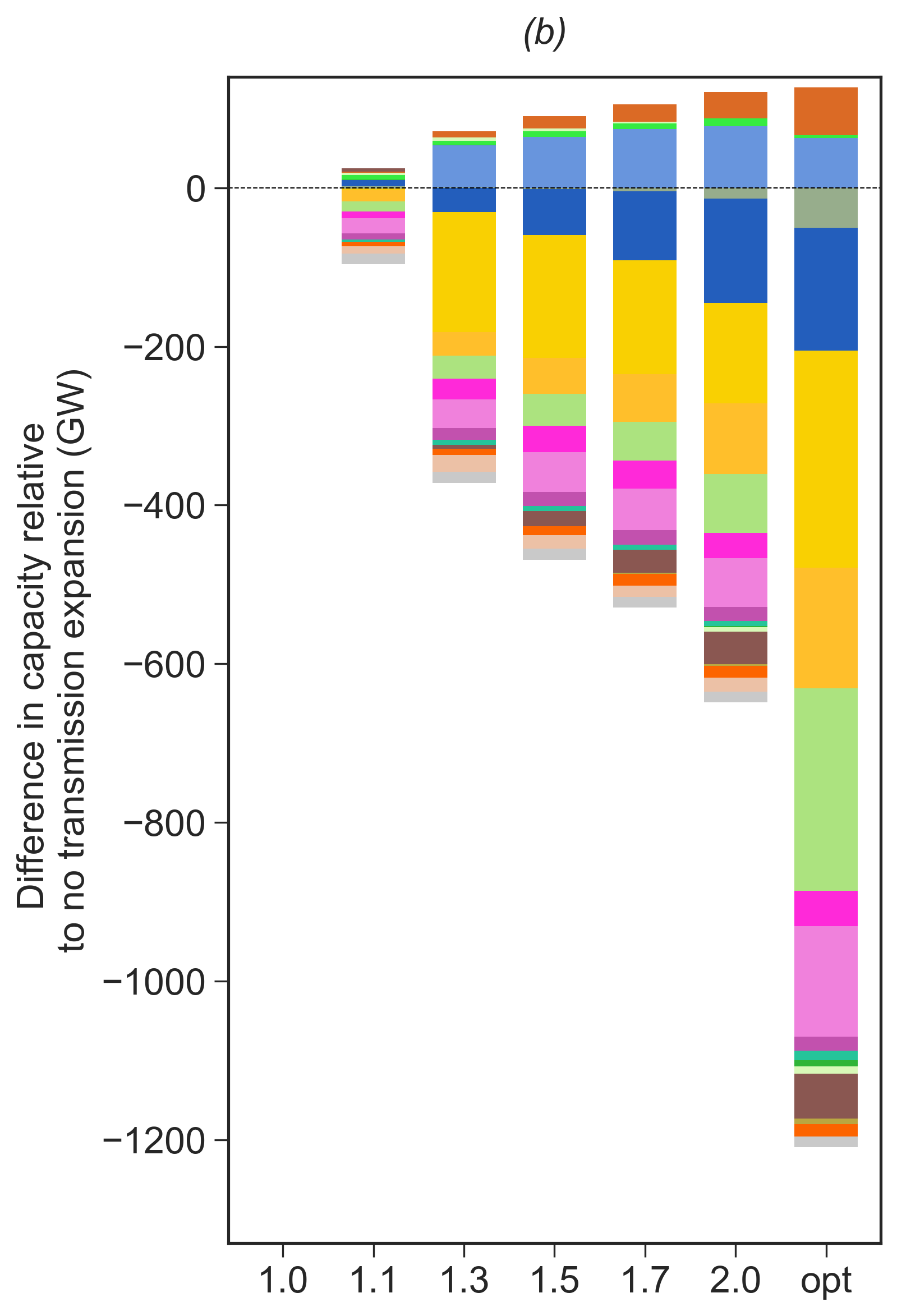}
	\captionsetup{width=14.8cm}
	\caption{Same as Fig. \ref{sfig:sensitivity_transmission_1968} performed for the 2013 design year.}
	\label{sfig:sensitivity_transmission_2013}
\end{figure}

When proceeding with the subsequent dispatch optimization, different outcomes exist. Since we, in the main body of this paper, identified nodal backup generation as one of the key robustness measures, the reduced backup capacity driven by the transmission expansion is expected to have a detrimental impact on the capability of avoiding loss of load. Conversely, the increased transmission enables the transport of energy over a wider area with fewer constraining bottlenecks in the transmission network, which is expected to have a positive impact on the robustness. For this reason, there is a turning point, where the benefit of increased transmission overweights the deficiency of having less local backup generation, as illustrated for the cumulative unserved energy for the 2013 design year (Fig. \ref{sfig:sensitivity_transmission_robustness}a), with transmission expansion driving a reduction in unserved energy when above 50\% volume expansion. However, for the peak loss of load (Fig. \ref{sfig:sensitivity_transmission_robustness}b), we do not see similar impact. This can be explained by the higher rollout of wind energy driven by the electricity transmission, causing a higher vulnerability to wind droughts.\\

\begin{figure}[!htbp]
	\centering
	\includegraphics[width=0.37\textwidth]{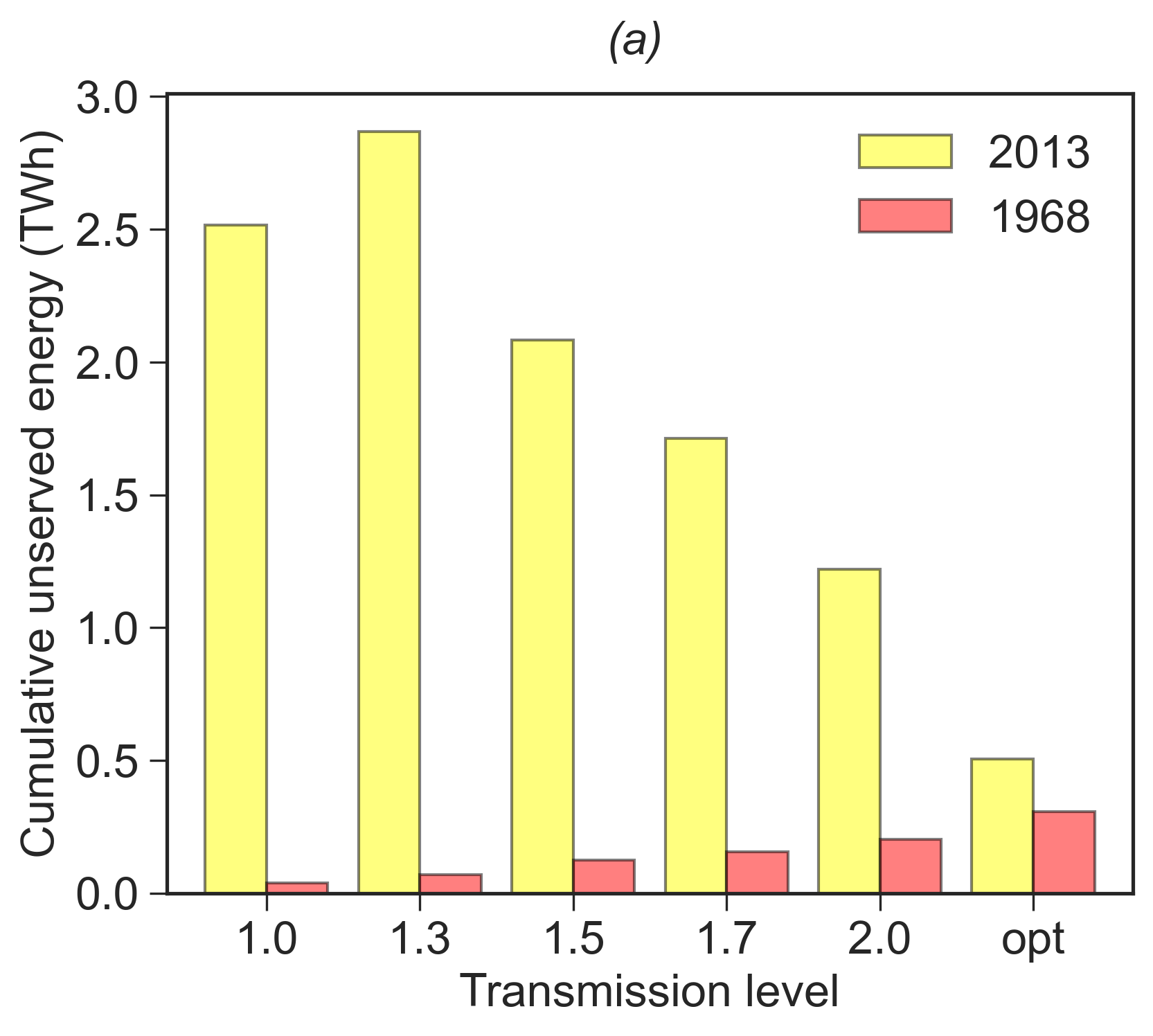}
	\includegraphics[width=0.37\textwidth]{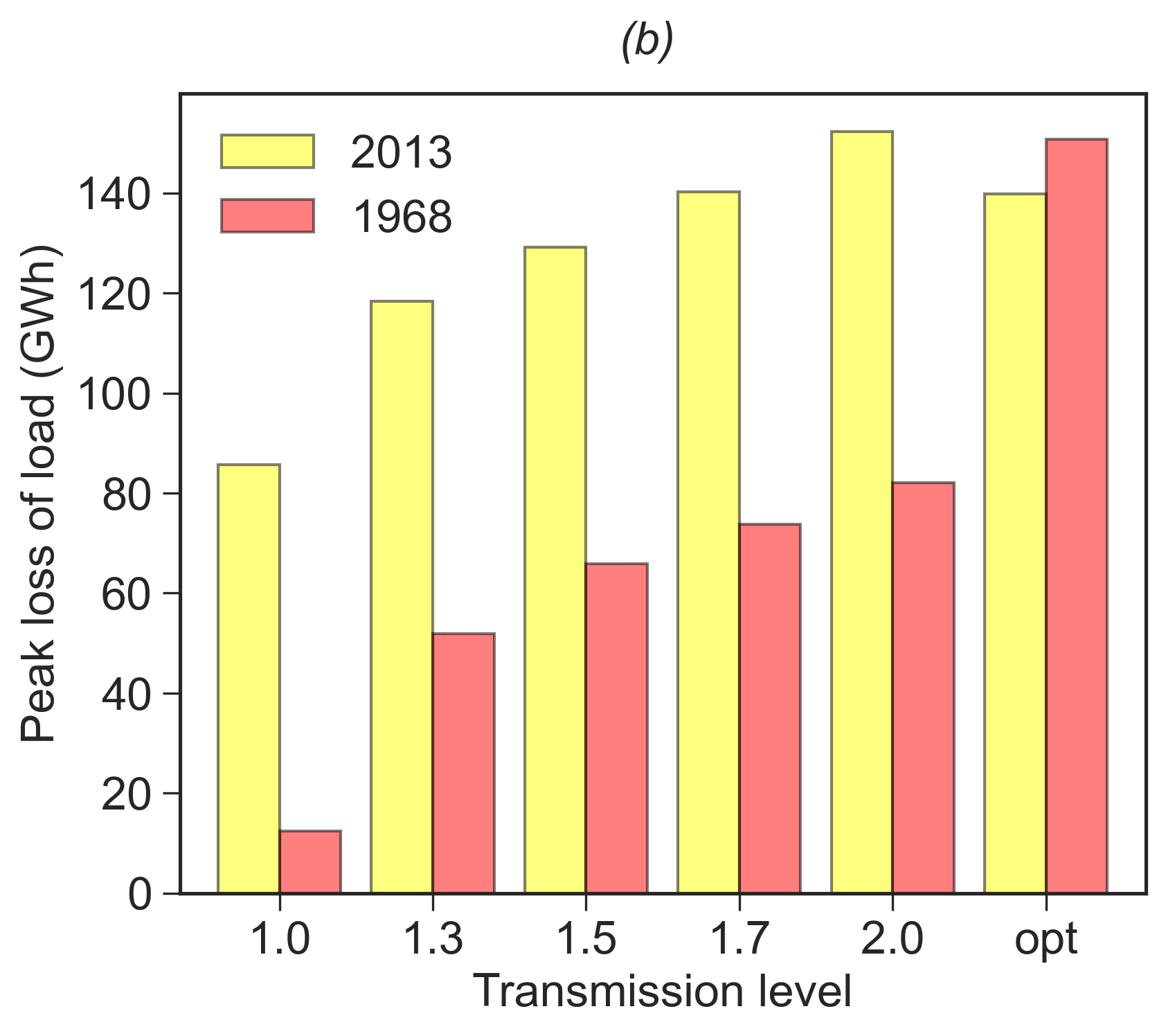}
	\captionsetup{width=14.8cm}
	\caption{Aggregated robustness metrics for the 1968 and 2013 design years, showing (a) the cumulative unserved energy, and (b) the peak loss of load during a 3-hour time step.}
	\label{sfig:sensitivity_transmission_robustness}
\end{figure}

\begin{figure}[!htbp]
	\centering
	\includegraphics[width=0.7\textwidth]{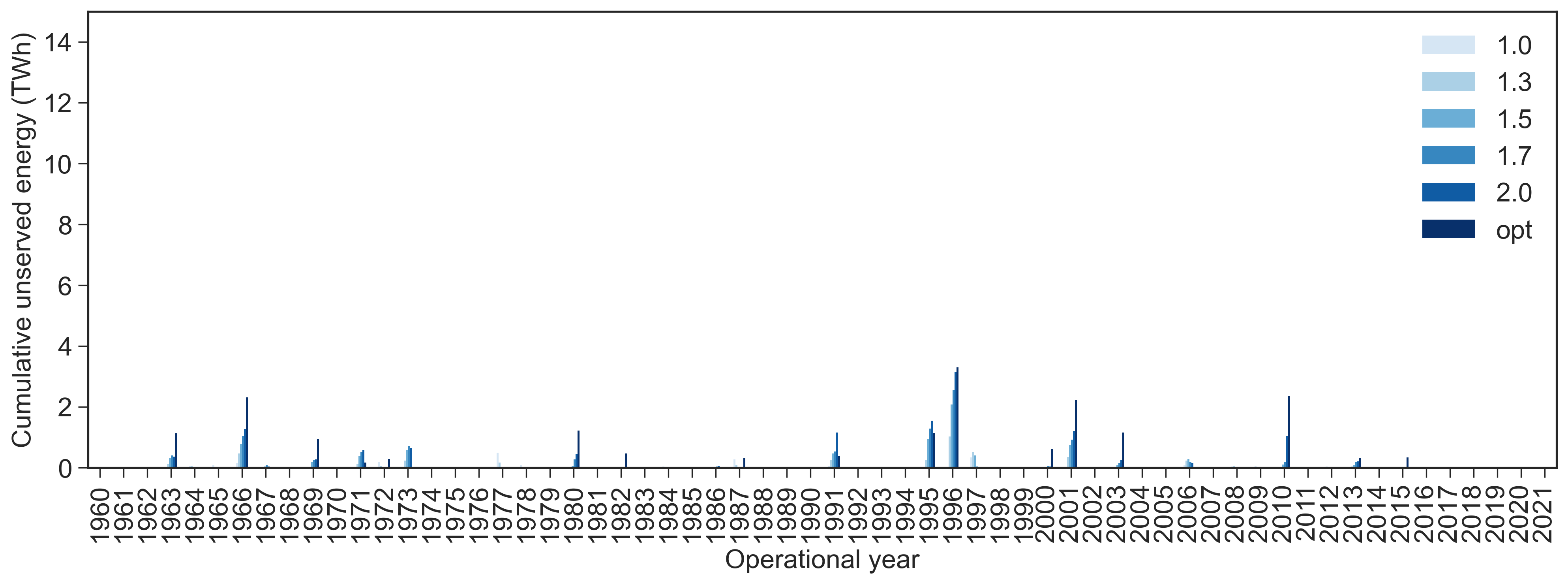}
	\includegraphics[width=0.7\textwidth]{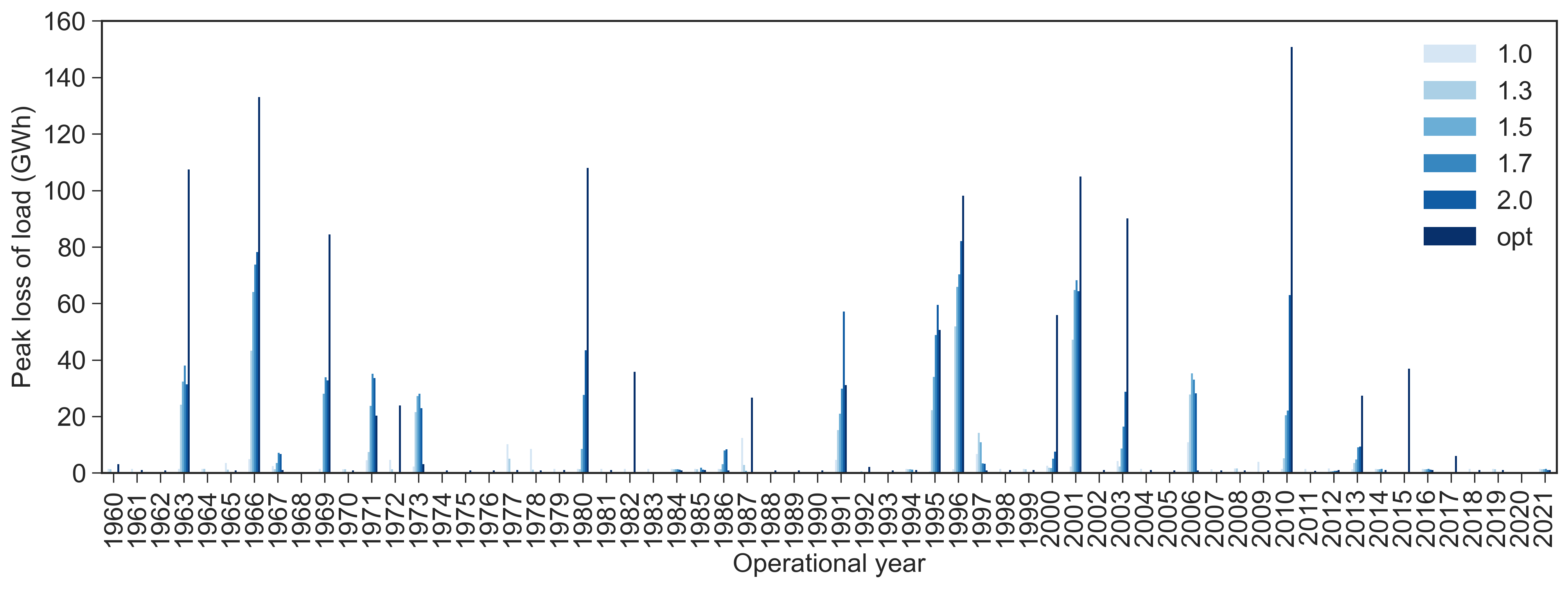}
	\captionsetup{width=14.8cm}
	\caption{Cumulative unserved energy and peak loss of load for the 1968 design year, with different levels of electricity transmission expansion.}
	\label{sfig:sensitivity_transmission_robustness_by_year_1968}
\end{figure}

The analysis is showcased for every operational year in Fig. \ref{sfig:sensitivity_transmission_robustness_by_year_1968} and \ref{sfig:sensitivity_transmission_robustness_by_year_2013} for the 1968 and 2013 design years, respectively. From here, we also see that the impact on the 1968 design year is mostly impacted on the robustness during the hour with peak loss of load, while the impact on the average unserved energy is small because it already possesses robust characteristics. For the 2013 design year, we see a noticeable alleviation of the average loss of load in most of the operational years, driven by the transmission expansion. For the explained reasons, this is not the case for the peak loss of load. 

\begin{figure}[!htbp]
	\centering
	\includegraphics[width=0.7\textwidth]{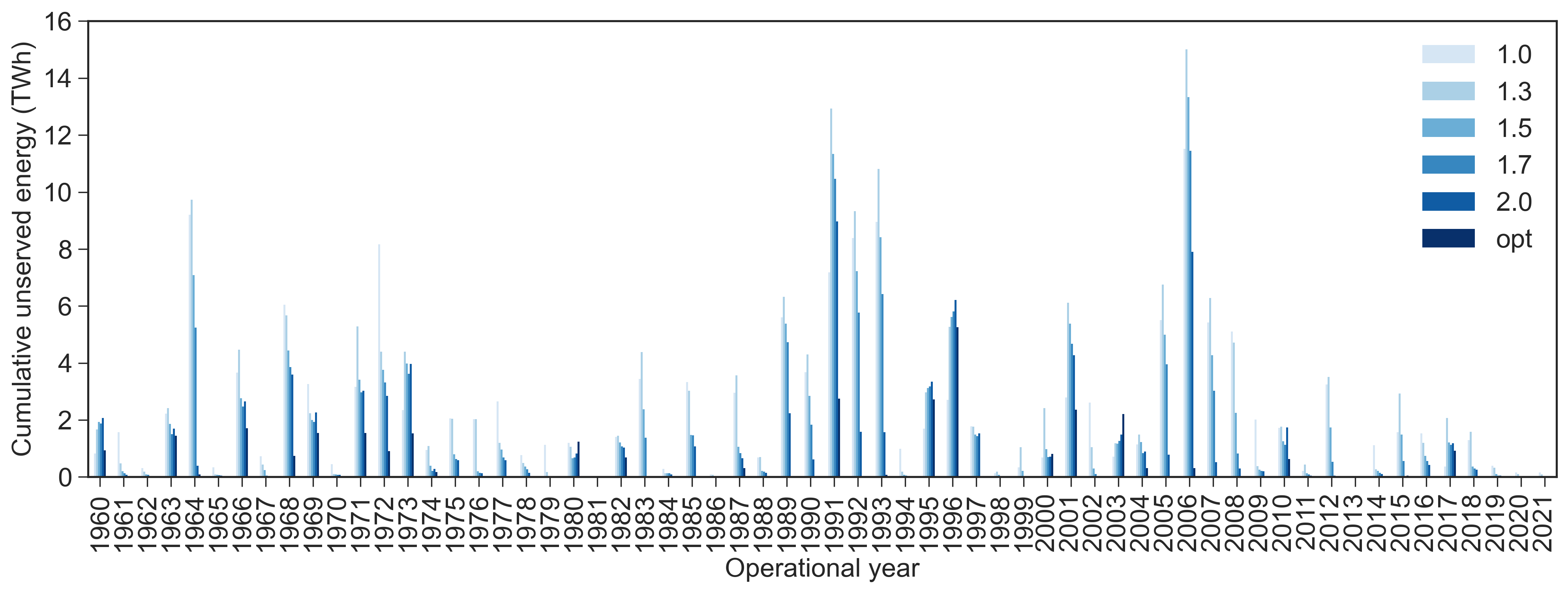}
	\includegraphics[width=0.7\textwidth]{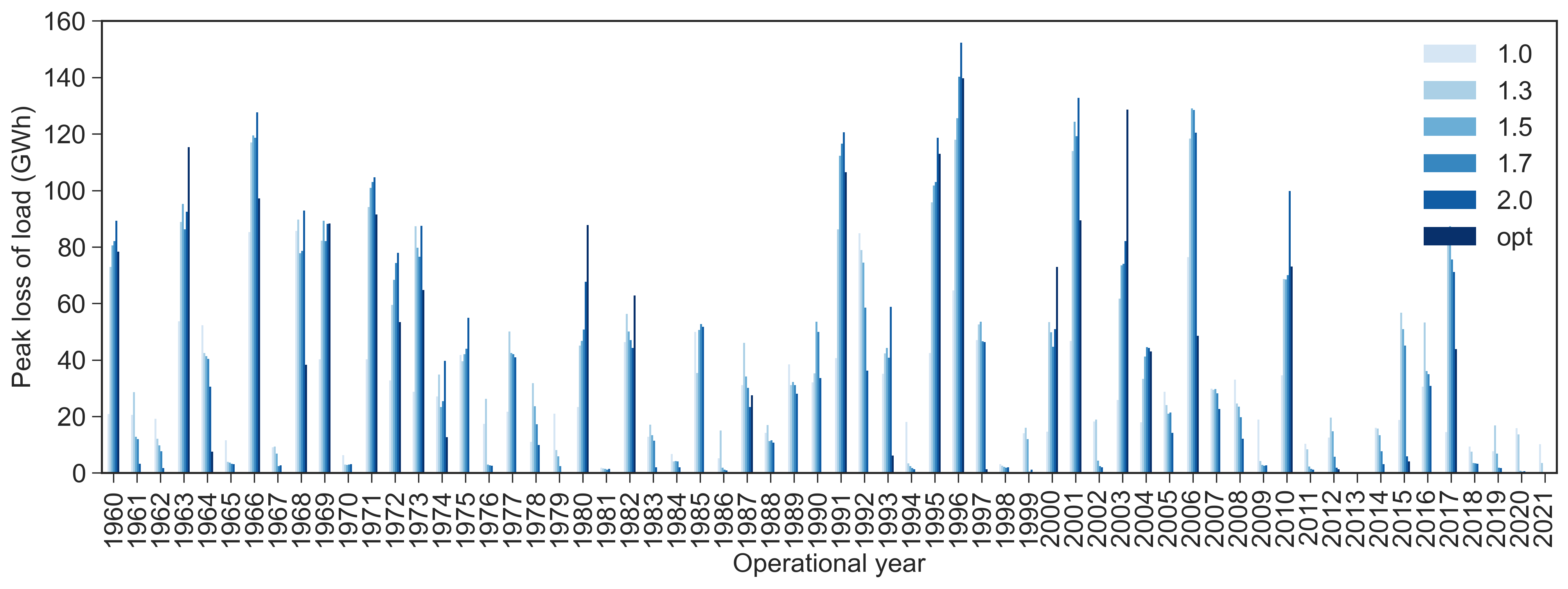}
	\captionsetup{width=14.8cm}
	\caption{Cumulative unserved energy and peak loss of load for the 2013 design year, with different levels of electricity transmission expansion.}
	\label{sfig:sensitivity_transmission_robustness_by_year_2013}
\end{figure}

\newpage
\vspace{\baselineskip}
\noindent
\textbf{Hydro constraint}. In our analysis, we included a constraint on the filling level of hydropower reservoirs, to represent non-energy related constraints such as water access for irrigation and flow continuity to avoid negative impacts on aquatic organisms. The constraint imposed a lower bound on the reservoir filling level normalized by the reservoir energy capacity (Eq. \ref{eq:SOC_hydro}), hereinafter state of charge (SOC). Here, we perform a sensitivity study where we omit the SOC constraint, allowing the reservoirs to be operated fully according to the cost-optimization. We run a dispatch optimization of the 2013 capacity layout, which showed average weather conditions, for which we compare the peak loss of load and unserved energy in the case with and without the hydropower reservoir constraint.\\

Two aspects apply when constraining the hydropower SOC. First, in the capacity optimization, the SOC constraint adds more dispatchable generation (solar thermal, OCGT), and storage (battery and H$_2$ fuel cells). The hydropower SOC limits the possibility of displacing all the hydropower generation from summer to winter. Consequently, the SOC constraints drives more wind generation. For total system cost, we see negligible impact of the SOC constraint. Second, in the subsequent dispatch optimization for every operational year, the balancing from hydropower reservoirs is limited by the SOC constraint which causes a small increase in the cumulative unserved energy, compared to the unconstrained reservoir operation. For the peak loss of load, we observe negligible impact. Considering the two metrics for energy adequacy across the different operational years, we observe that the impact of the SOC constraint is small compared to the year-to-year variation. 

\begin{figure}[!htbp]
	\centering
	\includegraphics[width=0.45\textwidth]{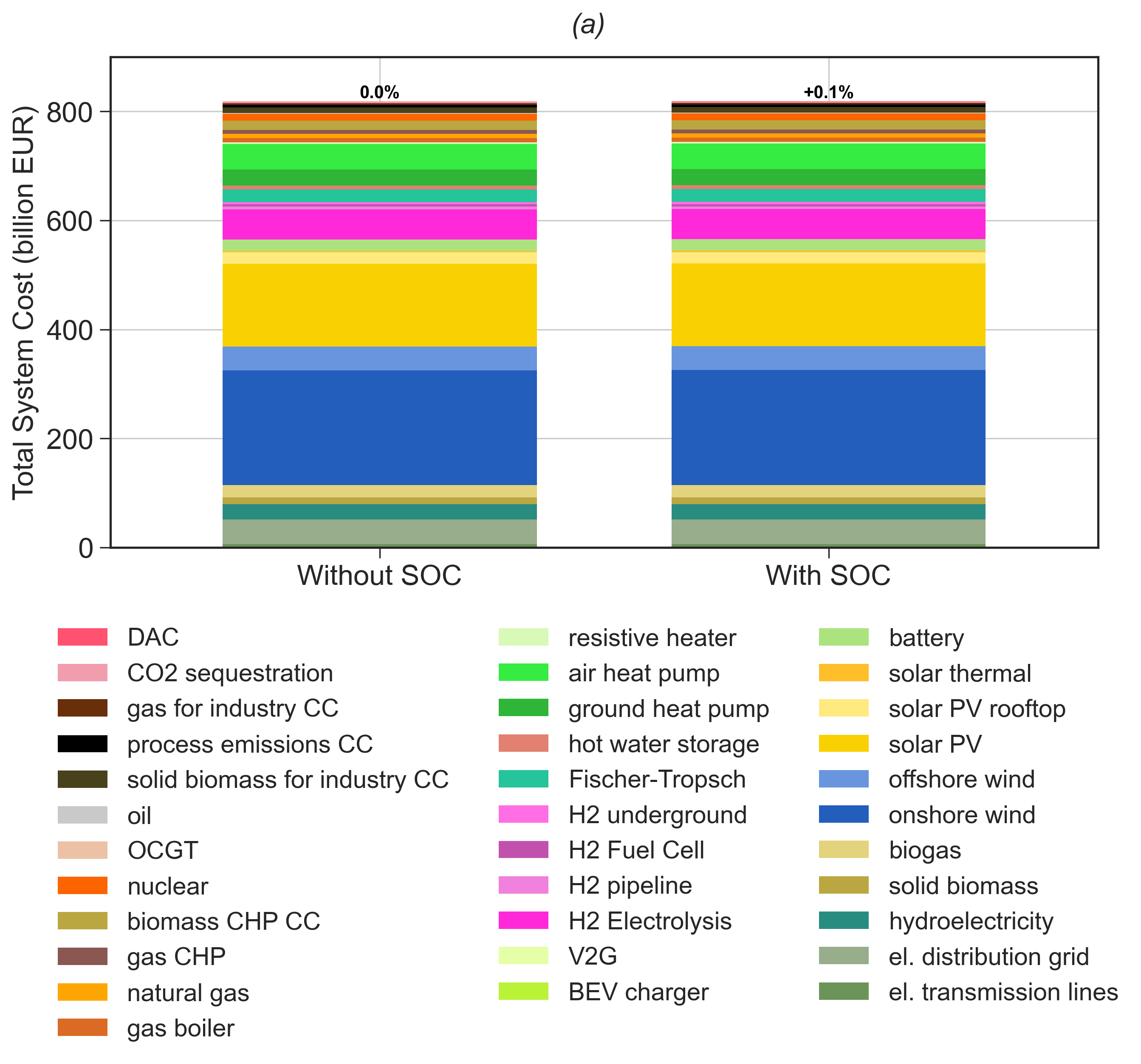}
	\includegraphics[width=0.27\textwidth]{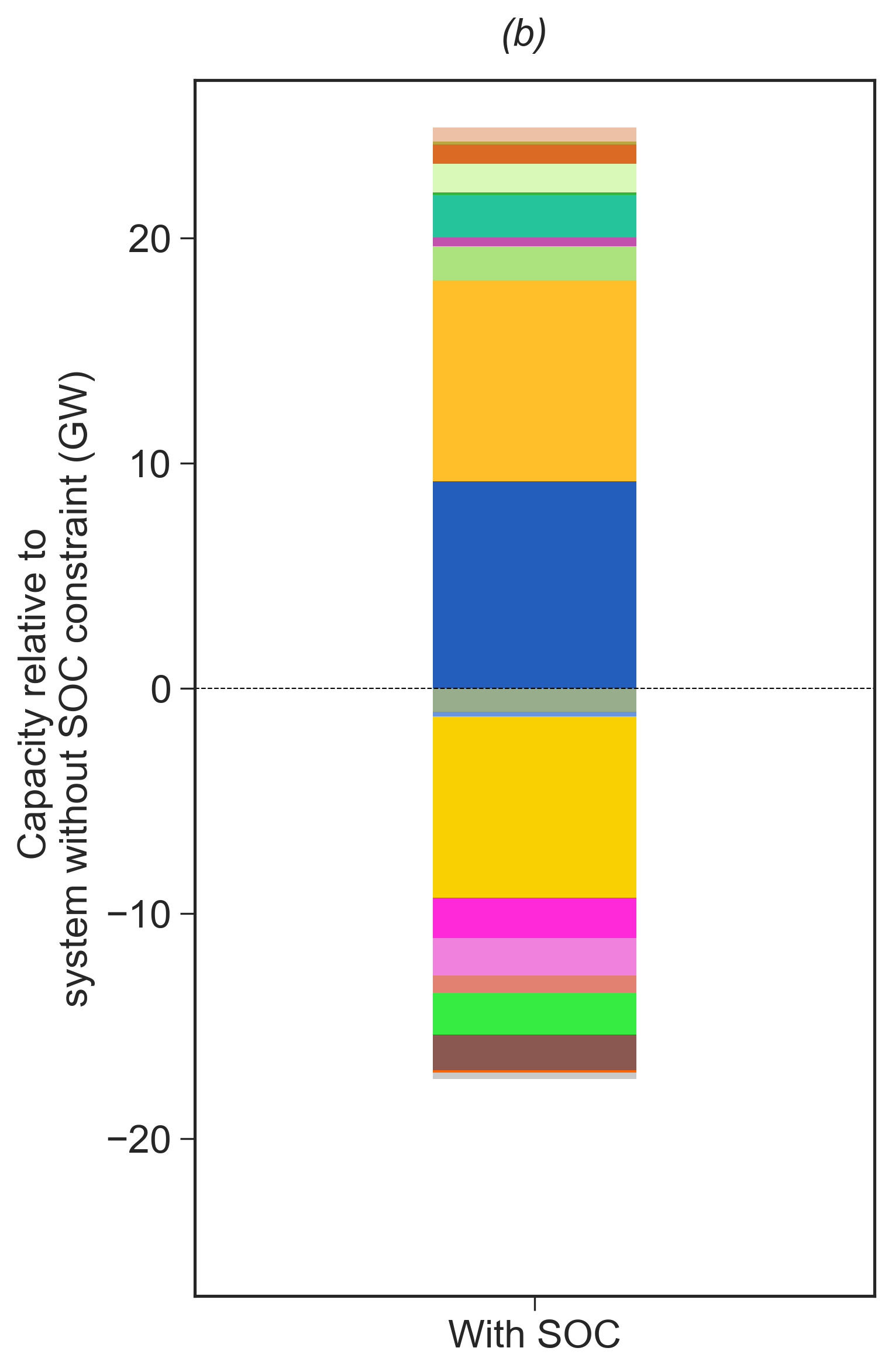}
	\captionsetup{width=14.8cm}
	\caption{Capacity optimization for the scenarios with and without the hydropower filling level (SOC) constraint, showing (a) the total system costs and (b) the deployed capacities. Results are obtained with the 2013 design year}
	\label{sfig:sensitivity_hydro_soc1}
\end{figure}

\begin{figure}[!htbp]
	\centering
	\includegraphics[width=0.37\textwidth]{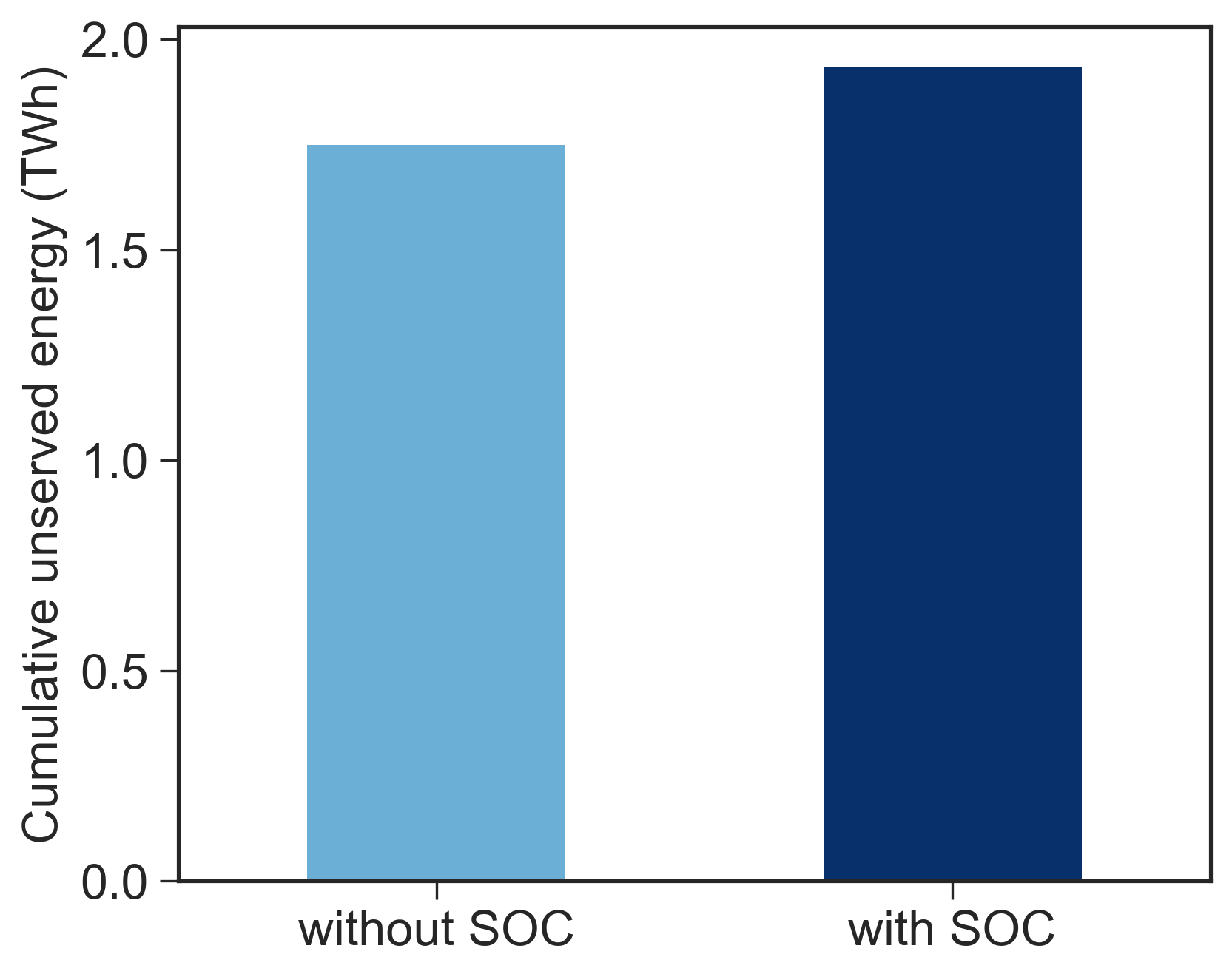}
	\includegraphics[width=0.37\textwidth]{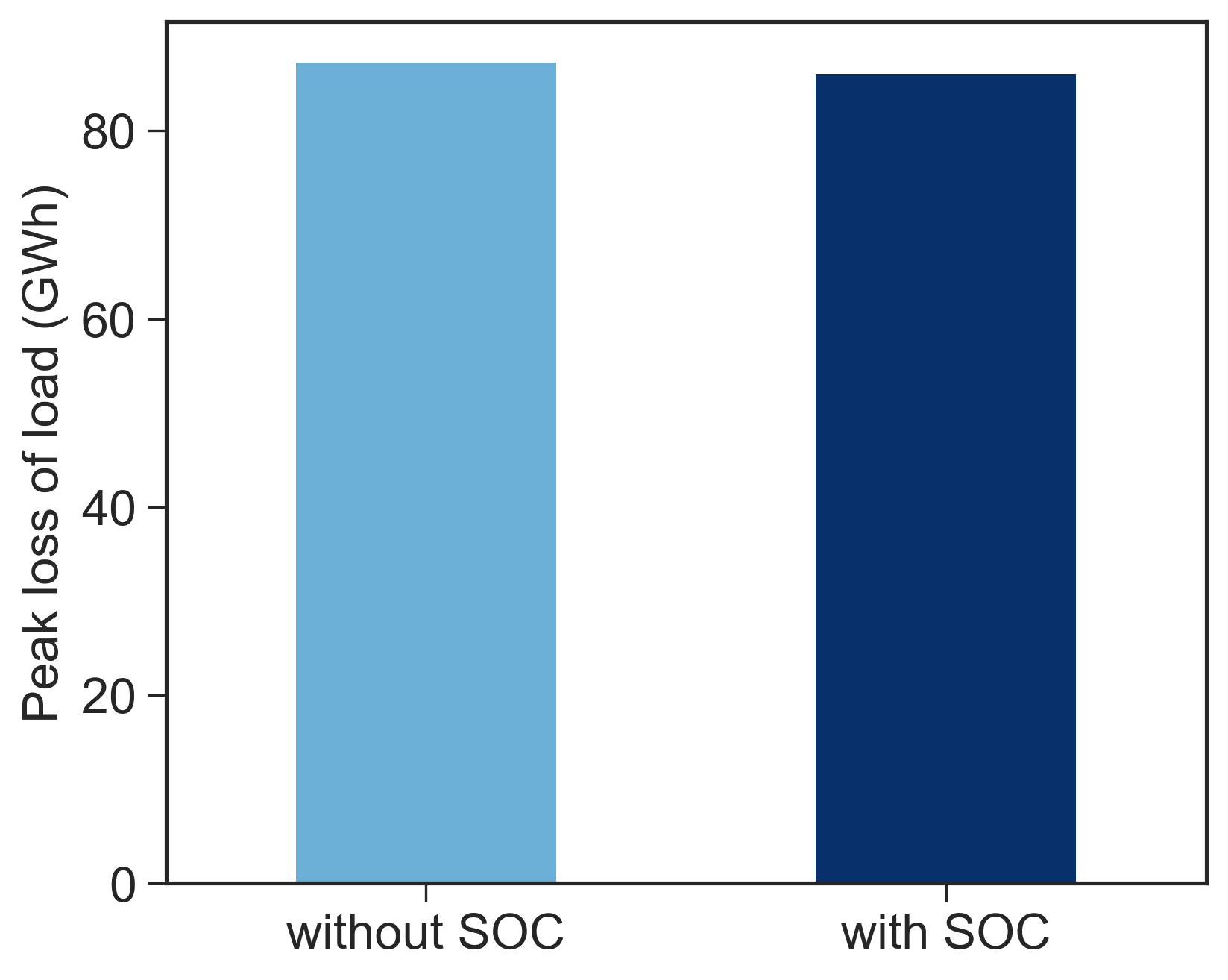}
	\captionsetup{width=14.8cm}
	\caption{Aggregated robustness metrics for the scenarios with and without the SOC constraint, showing (a) the cumulative unserved energy, and (b) the peak loss of load during a 3-hour time step. Results are obtained with the 2013 design year.}
	\label{sfig:sensitivity_hydro_soc2}
\end{figure}

\newpage
\textcolor{white}{Hydro SOC}
\begin{figure}[!htbp]
	\centering
	\includegraphics[width=0.7\textwidth]{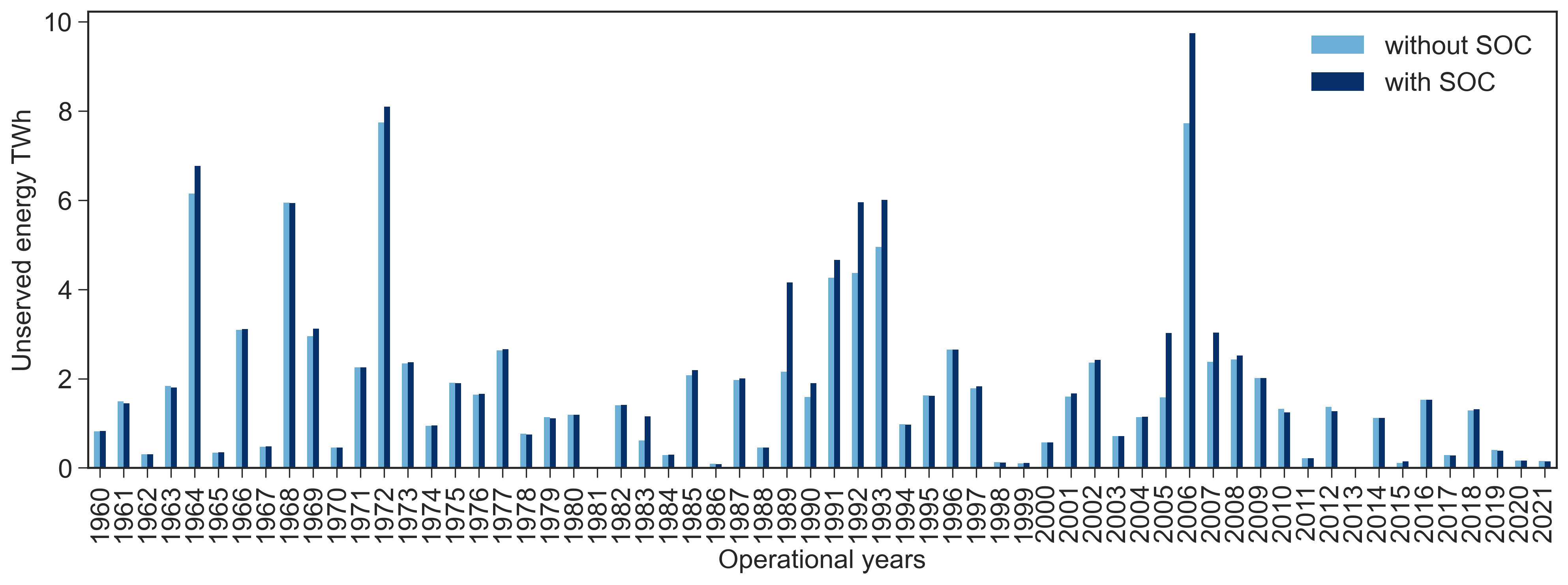}
	\includegraphics[width=0.7\textwidth]{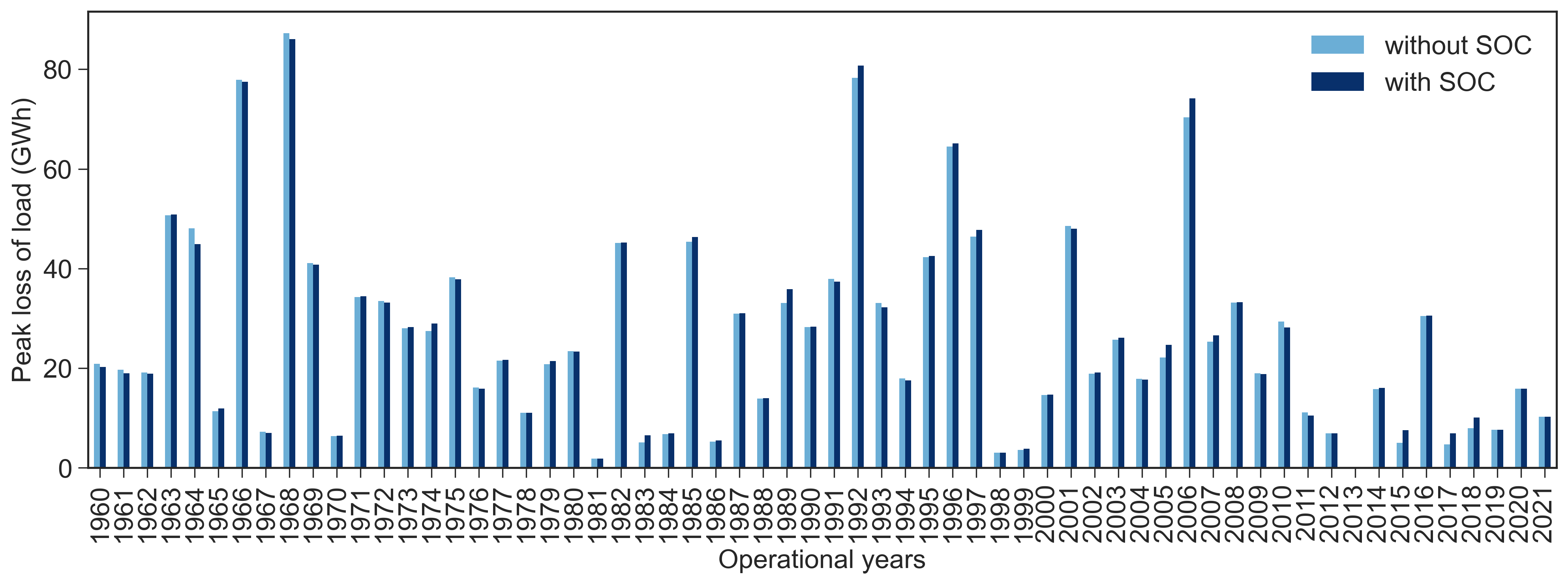}
	\captionsetup{width=14.8cm}
	\caption{Cumulative unserved energy and peak loss of load for the 2013 design year with and without the hydropower filling level (SOC) constraint.}
	\label{sfig:sensitivity_hydro_soc3}
\end{figure}

\newpage
\section{Derivation of capacity deficits}

\textbf{Loss of load duration curves for the aggregated Europe}. We first show the loss of load duration curves for all capacity layouts (Fig. \ref{sfig:LOL_duration_curves}), indicating the peak loss of load and the percentage of time encountering energy deficits.
\begin{figure}[!h]
	\centering
	\includegraphics[width=0.8\textwidth]{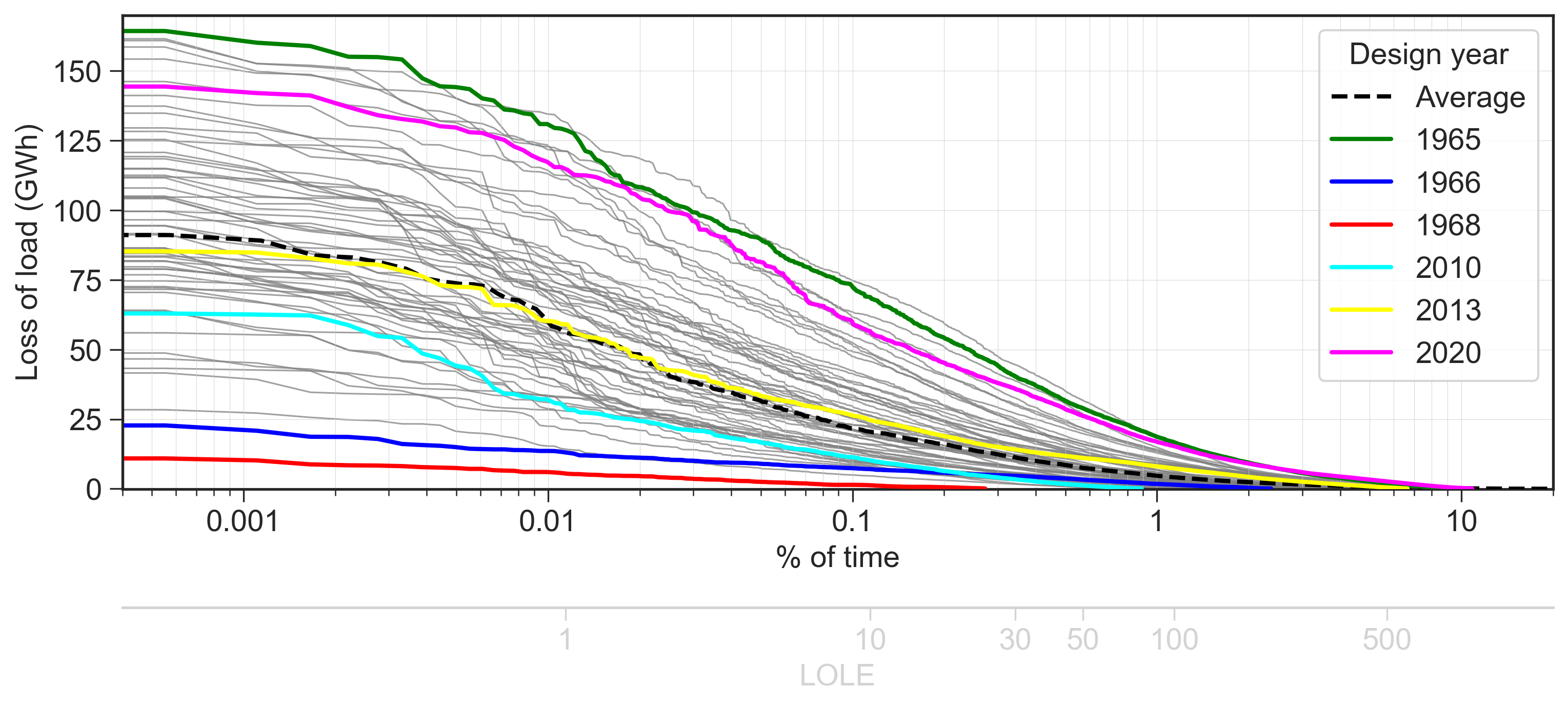}
	\captionsetup{width=14.8cm}
	\caption{Loss of load (aggregated Europe) duration curves for all design years simulated in every operational year. From the data along the x-axis, we derive the more common metric LOLE.}
	\label{sfig:LOL_duration_curves}
\end{figure}

\noindent
\textbf{Capacity deficits derived from peak loss of load}. In a theoretical case where energy could move across countries without any limitations, the peak loss of load in Fig. \ref{sfig:LOL_duration_curves} would correspond to the inadequate power capacity. Since cross-border energy flows are constrained by the capacity of deployed transmission lines, we derive the power generation capacity deficits $\Delta G$ from the peak loss of load in every country across all operational years:
\begin{equation}\label{eq:capacity_deficit}
	\Delta G = \sum_c \max_y \left[ \max_t \Delta e_{c,t,y} \right]	
\end{equation}
where $\Delta e_{c,t,y}$ is the loss of load in the 3-hourly timestep $t$ in country $c$ for operational year $y$. The resulting power capacity deficits, derived for all capacity layouts, are shown in Fig. \ref{sfig:capacity_deficits_duration_curves_all}.

\begin{figure}[!h]
	\centering
	\includegraphics[width=0.8\textwidth]{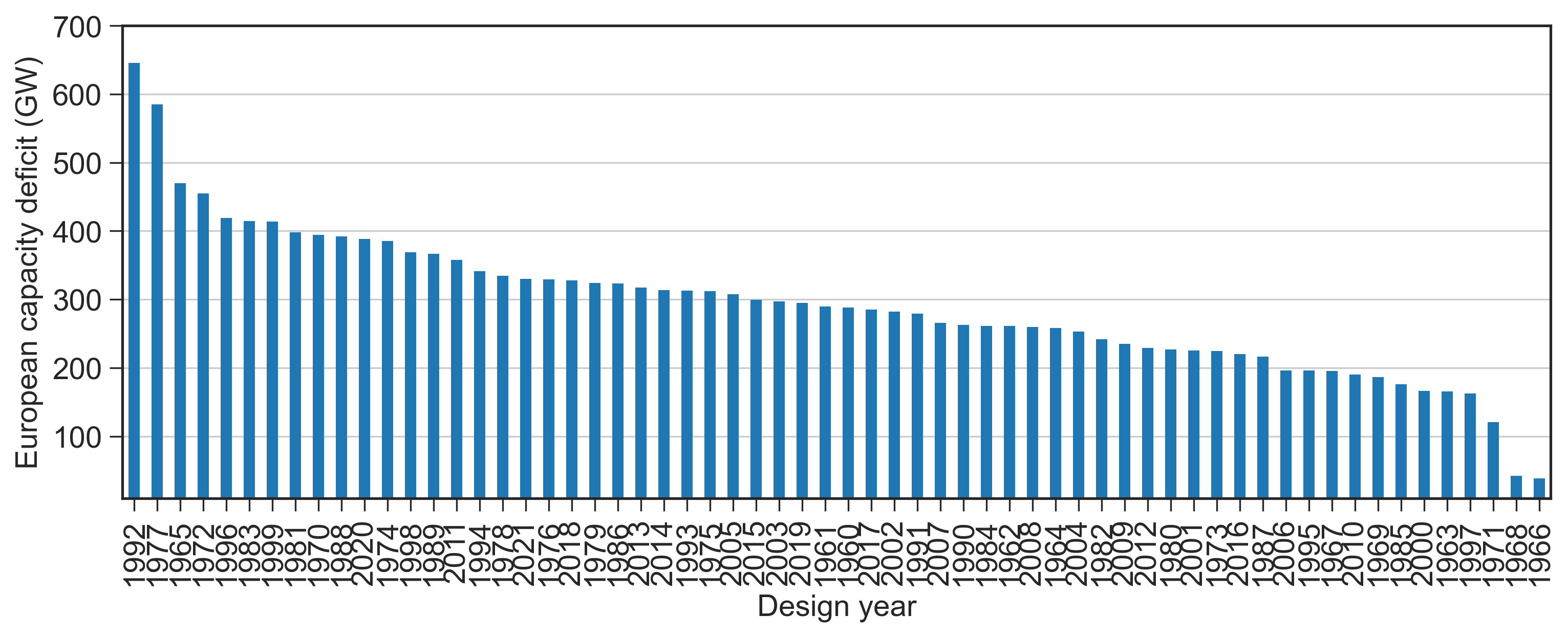}
	\caption{Capacity deficits for every design year, derived with Eq. \ref{eq:capacity_deficit}, with data from Fig. \ref{sfig:map_nodal_loss_of_load}.}
	\label{sfig:capacity_deficits_duration_curves_all}
\end{figure}

\newpage
\noindent
\textbf{Capacity deficit and operation for the 1968 design year}
\begin{figure}[!h]
	\centering
	\includegraphics[width=0.8\textwidth]{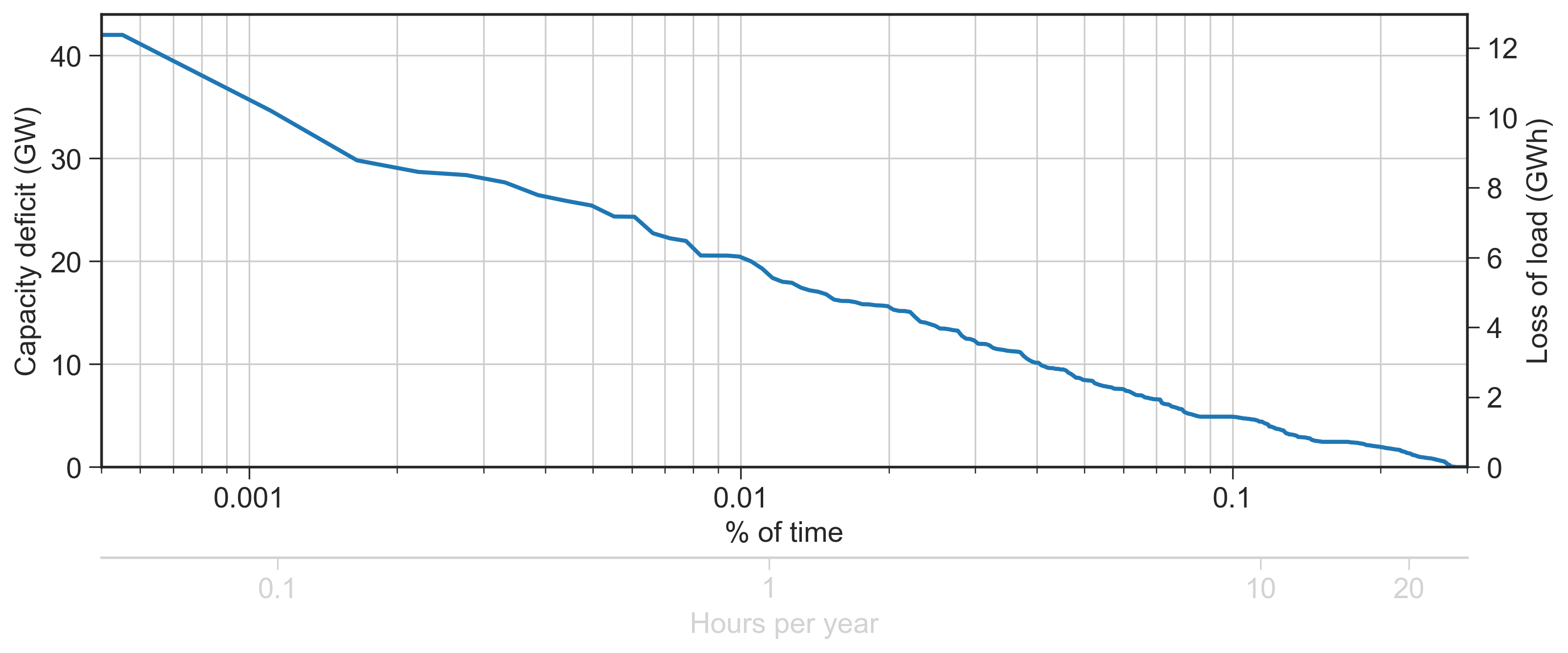}
	\captionsetup{width=14.8cm}
	\caption{Capacity deficit for the 1968 design year and the operational scheme required to remove energy deficits, derived from Fig. \ref{sfig:LOL_duration_curves} and \ref{sfig:capacity_deficits_duration_curves_all}.}
	\label{sfig:cap_deficits}
\end{figure}

\newpage
\section{Supplemental Figures}

\subsection{Electricity transmission capacity}
\begin{figure}[!htbp]
	\centering
	\includegraphics[width=0.75\textwidth]{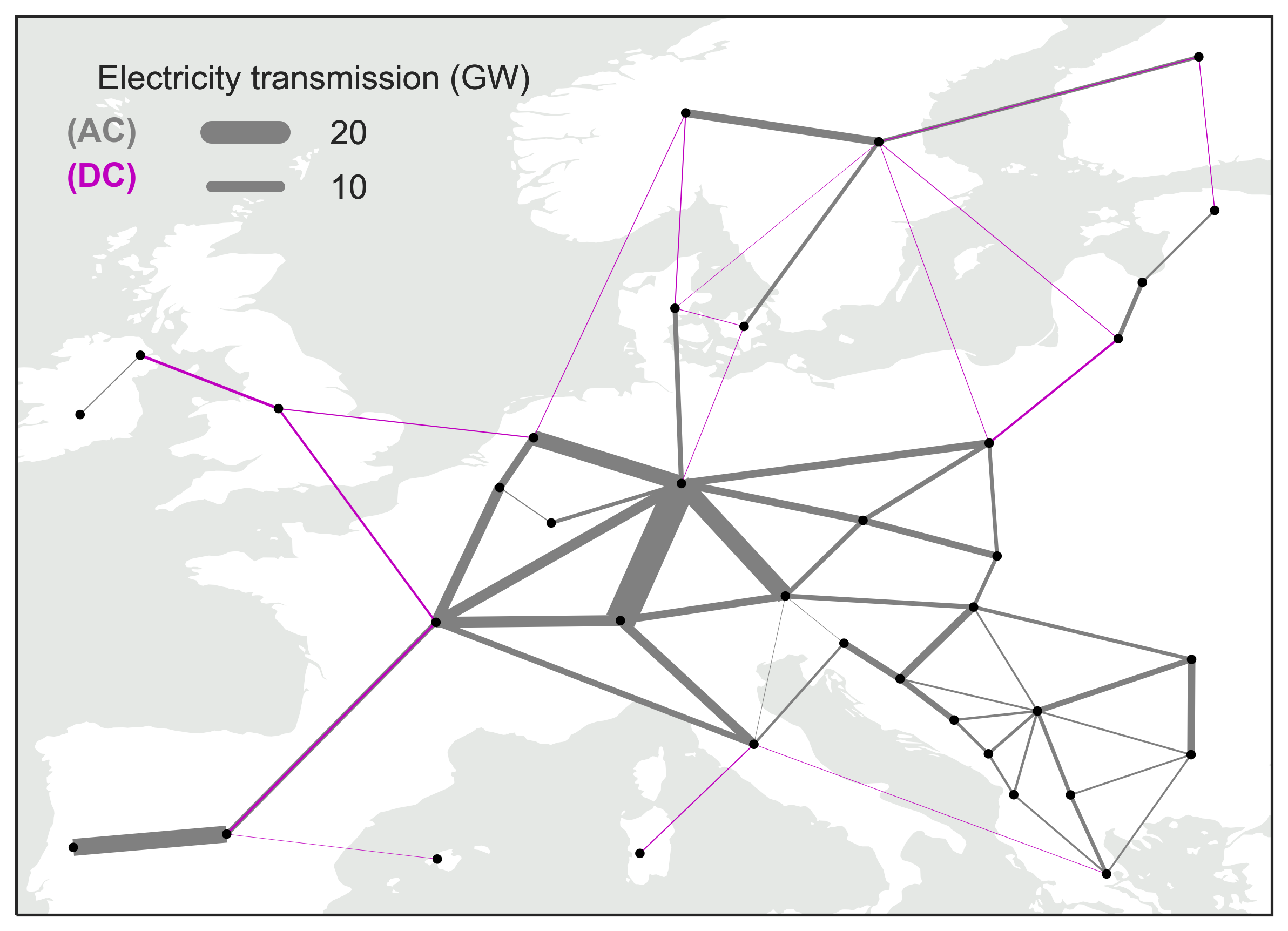}
	\captionsetup{width=14.8cm}
	\caption{Electricity transmission capacity used in our analysis.}
	\label{sfig:transmission_capacity}
\end{figure}

\newpage
\subsection{Long-term historical trend in hydro inflow and generation}
\begin{figure}[!htbp]
	\centering
	\includegraphics[width=0.85\textwidth]{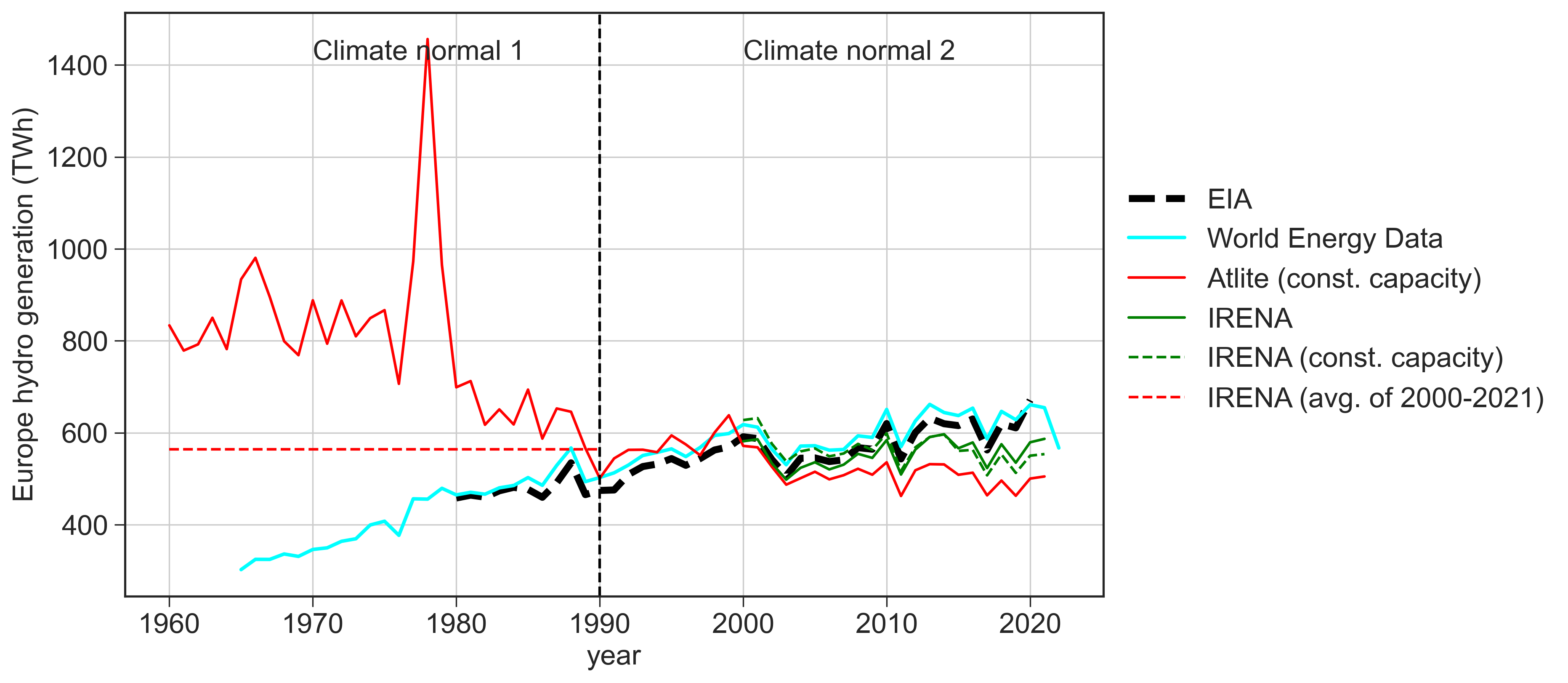}
	\includegraphics[width=0.85\textwidth]{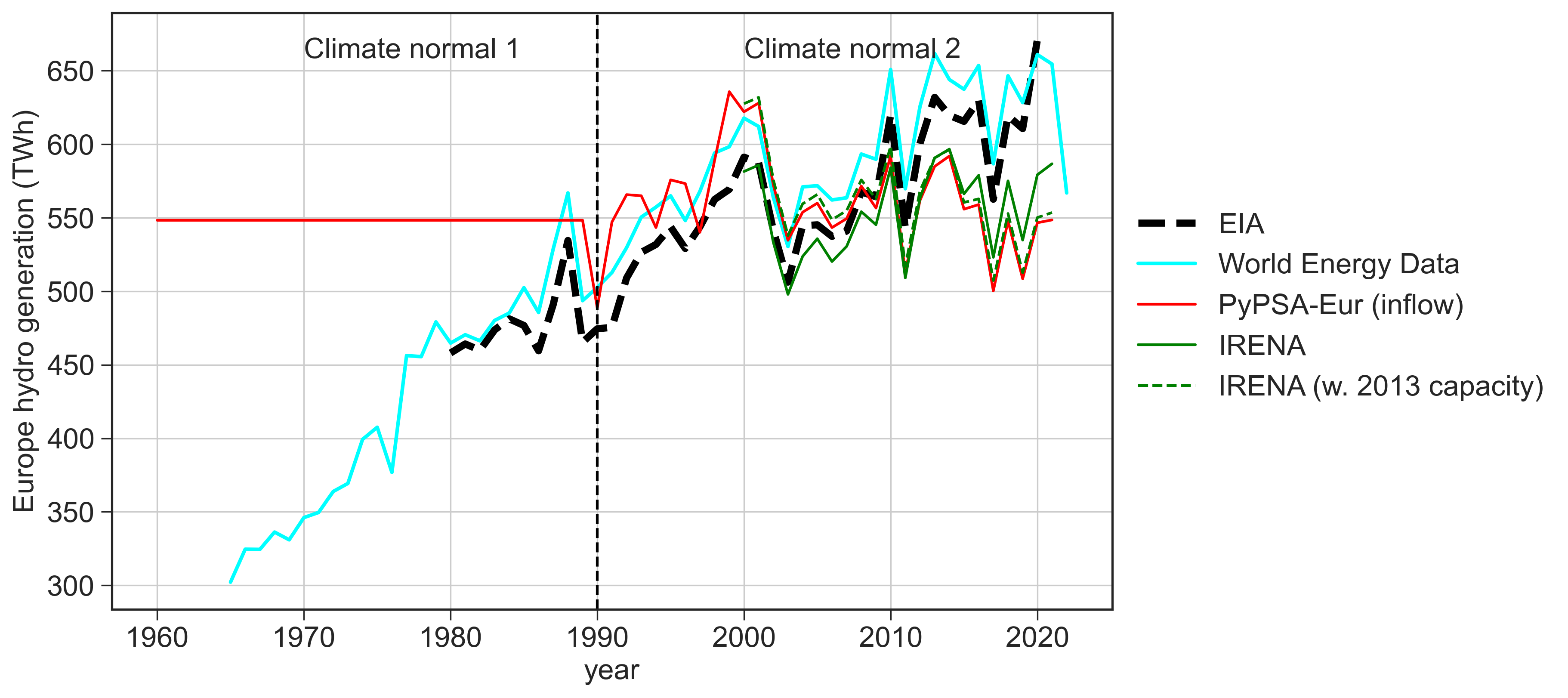}
	\captionsetup{width=14.8cm}
	\caption{Historical hydroelectricity generation from three different sources, compared to the estimated inflow from Atlite using ERA5 reanalysis data. The top figure shows the default estimate for every year, and the bottom figure shows the detrended inflow. The latter was obtained by scaling the first 30 years ("Climate normal 1") according to the average of the last 32 years ("Climate normal 2"). This was done to avoid climate change masking the steady-state natural variability of hydro which was our main focus in this paper, and to avoid including the positive trend in the historical hydropower production due to a graduate capacity expansion of existing and new hydro reservoirs. Intraannual variation according to the ERA5 data is retained.}
	\label{sfig:annual_hydropower_compare}
\end{figure}

\newpage
\subsection{Long-term historical trend in heating demand}
\begin{figure}[!htbp]
	\centering
	\includegraphics[width=0.85\textwidth]{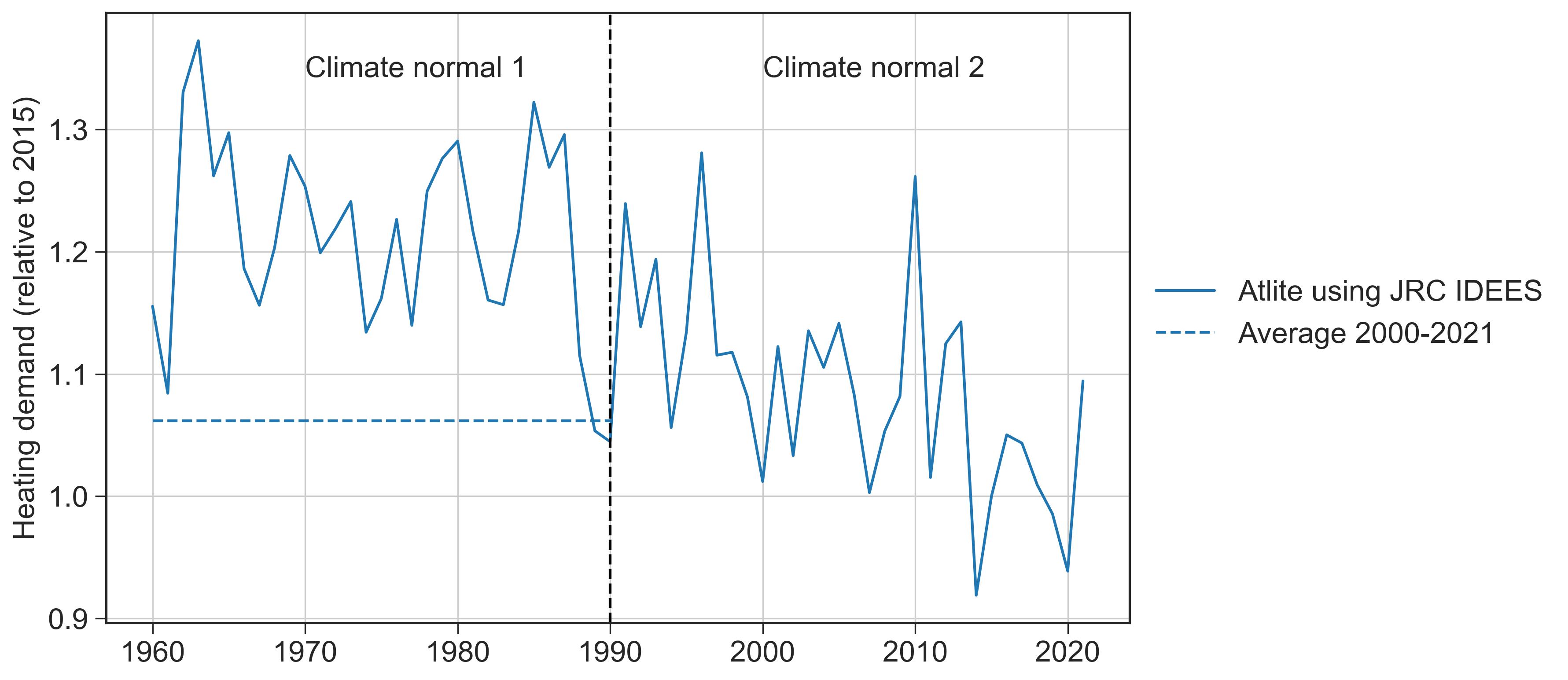}
	\includegraphics[width=0.85\textwidth]{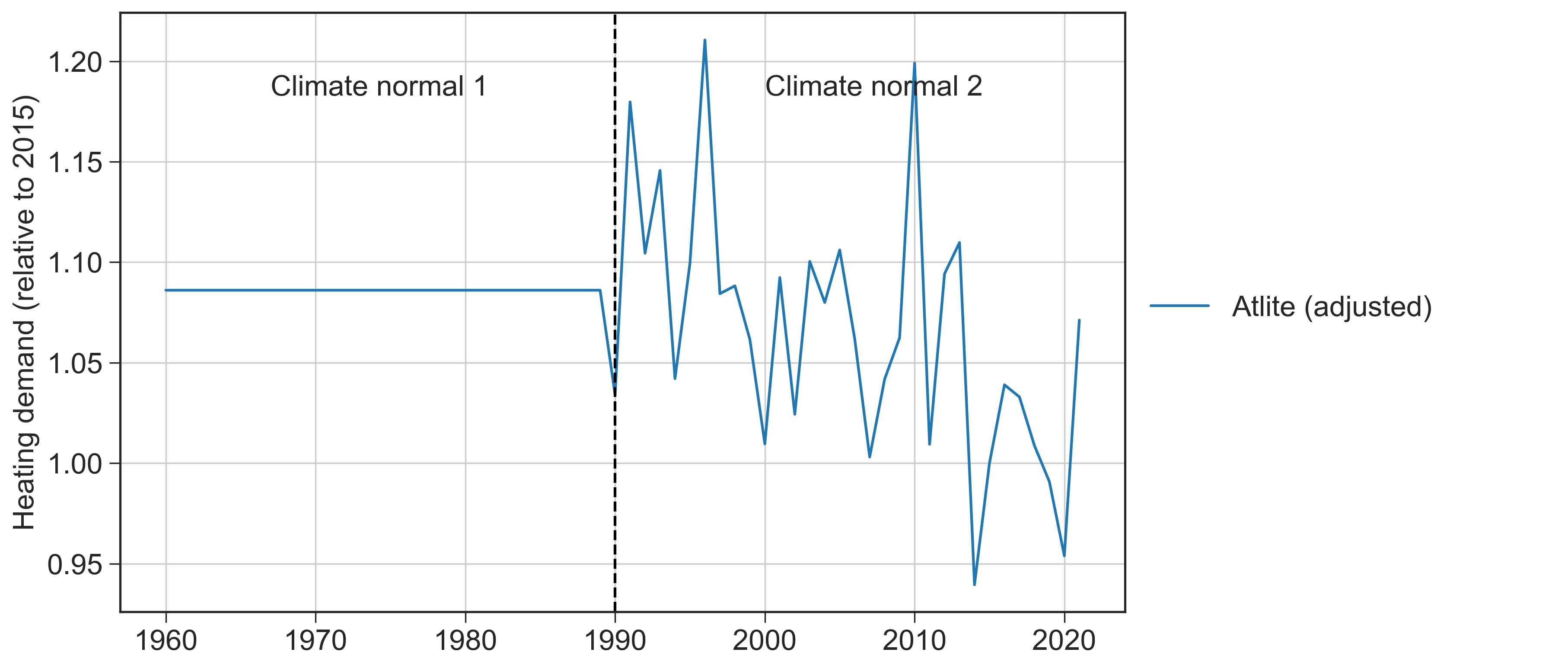}
	\captionsetup{width=14.8cm}
	\caption{Estimated heating demand from heating degree days from Atlite using ERA5 reanalysis data. The top figure shows the default estimate for every year, and the bottom figur shows the detrended heating demand. The latter was obtained by scaling the first 30 years ("Climate normal 1") according to the average of the last 32 years ("Climate normal 2"), to avoid climate change masking the steady-state natural variability of heating demand which was our main focus in this paper. Intraannual variation according to the ERA5 data is retained.}
	\label{sfig:annual_heating_demand_compare}
\end{figure}

\newpage
\subsection{Interannual variability of wind and solar PV}
\begin{figure}[!htbp]
	\centering
	\includegraphics[width=0.85\textwidth]{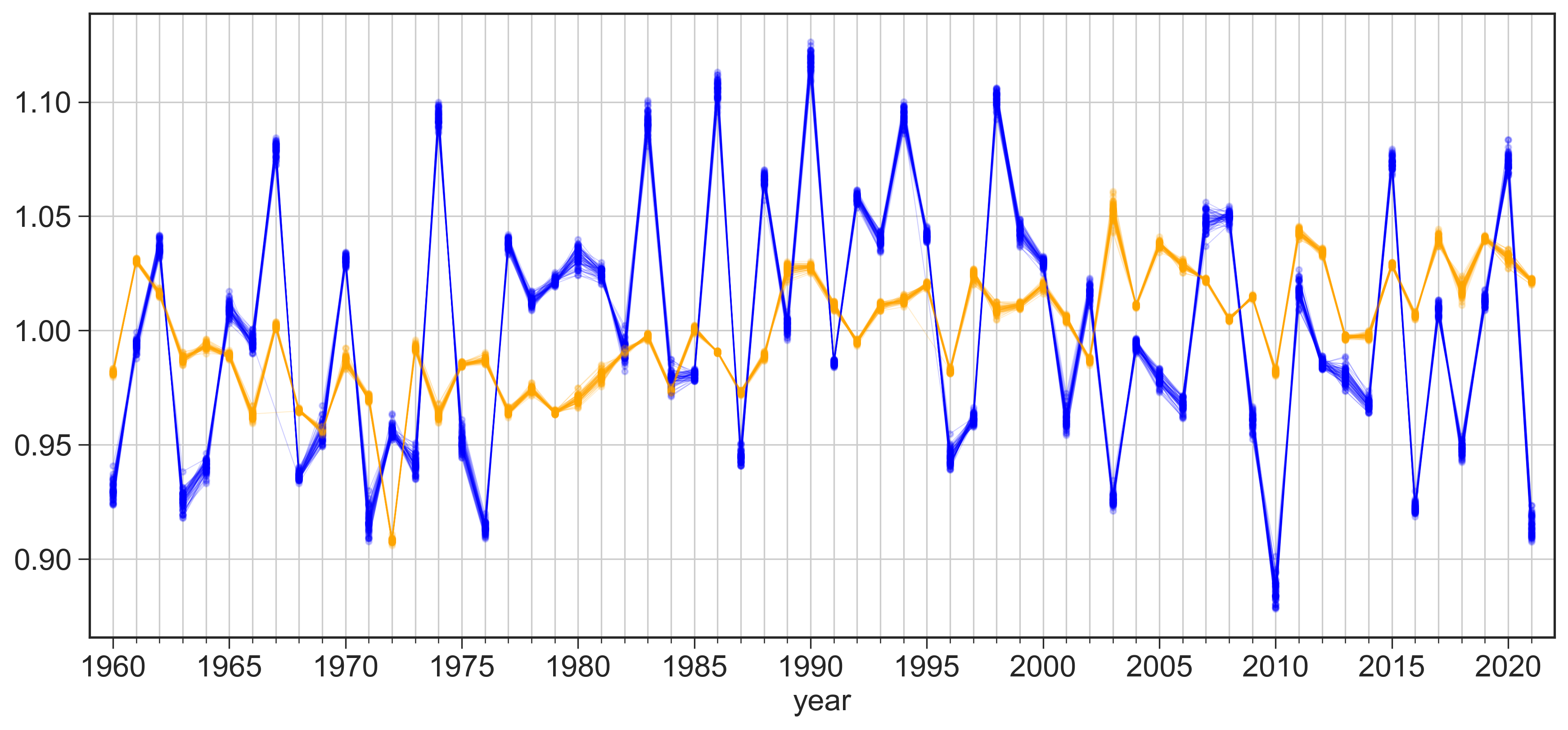}
	\includegraphics[width=0.85\textwidth]{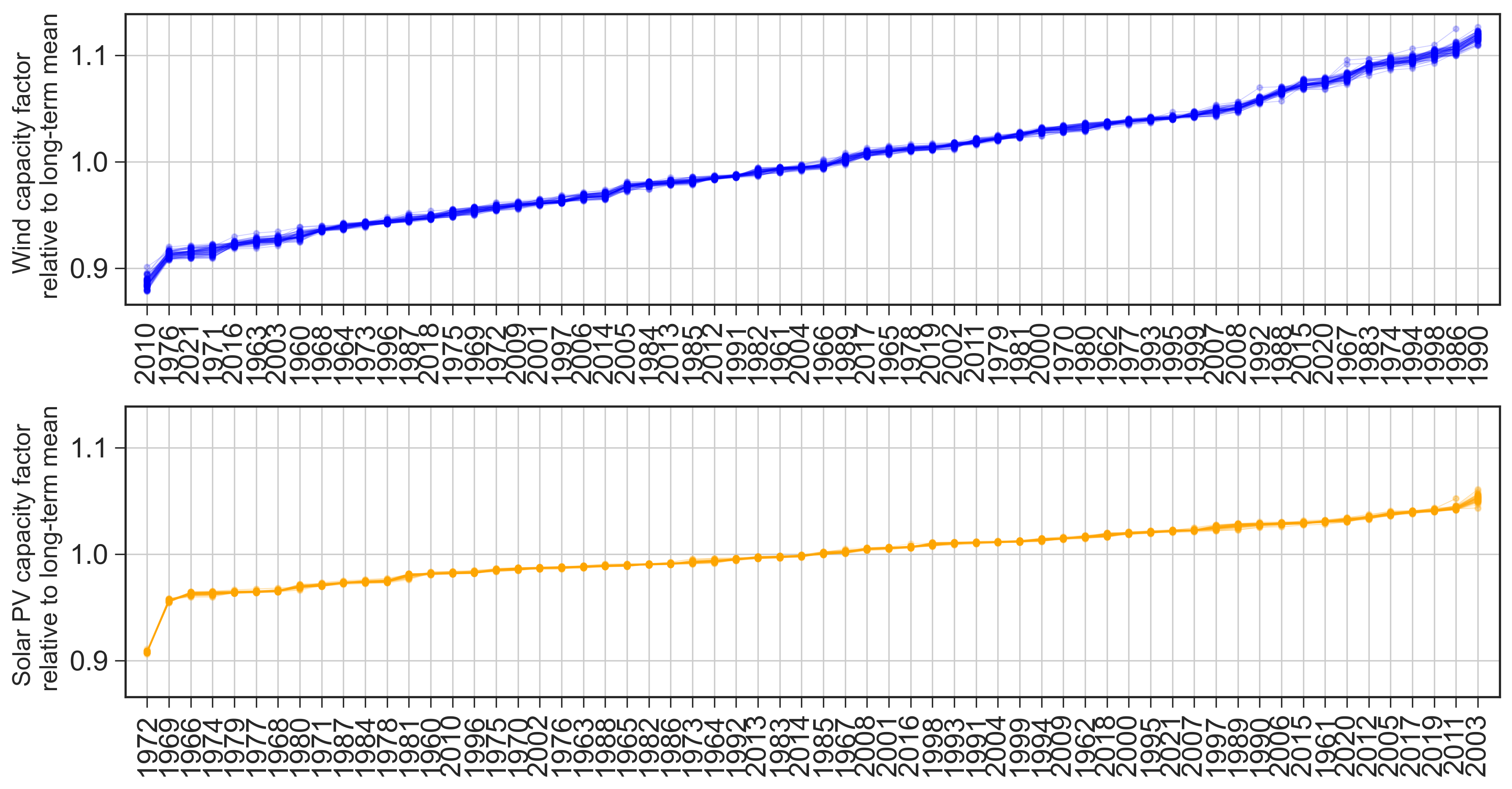}
	\captionsetup{width=14.8cm}
	\caption{Europe-aggregated wind and solar PV resources unsorted (top) and sorted (bottom), normalized by the long-term mean. Each line represent the resources available for a specific optimized capacity layout (shown for every design year).}
	\label{sfig:aggregated_wind_and_solarPV}
\end{figure}

\newpage
\subsection{Interannual variability of Coefficient of Performance}
\begin{figure}[!htbp]
	\centering
	\includegraphics[width=0.85\textwidth]{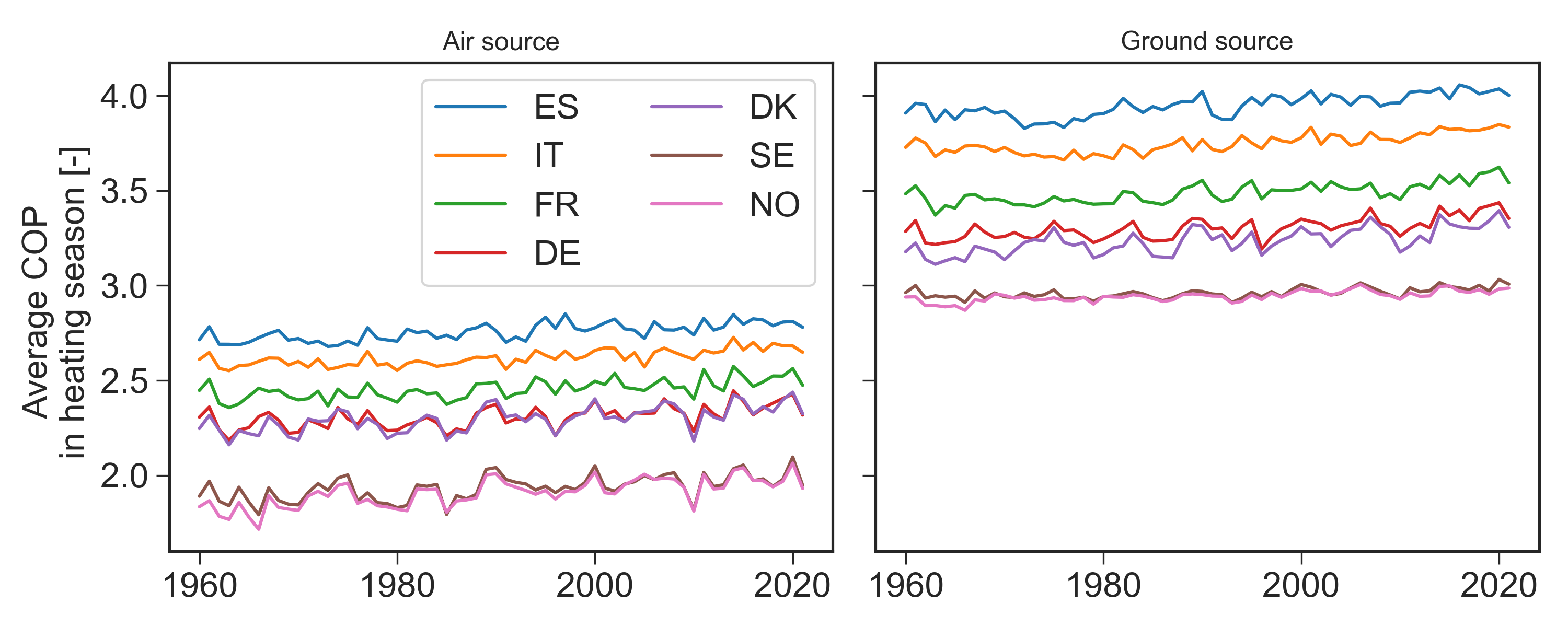}
	\captionsetup{width=14.8cm}
	\caption{Average coefficient of performance (COP) for seven countries estimated from ERA5 temperature data. The average COP is calculated for the heating season (October to April).}
	\label{sfig:average_COP}
\end{figure}

\newpage
\subsection{Electricity generation share in all 62 capacity layouts}
\begin{figure}[!h]
	\centering
	\includegraphics[width=0.85\textwidth]{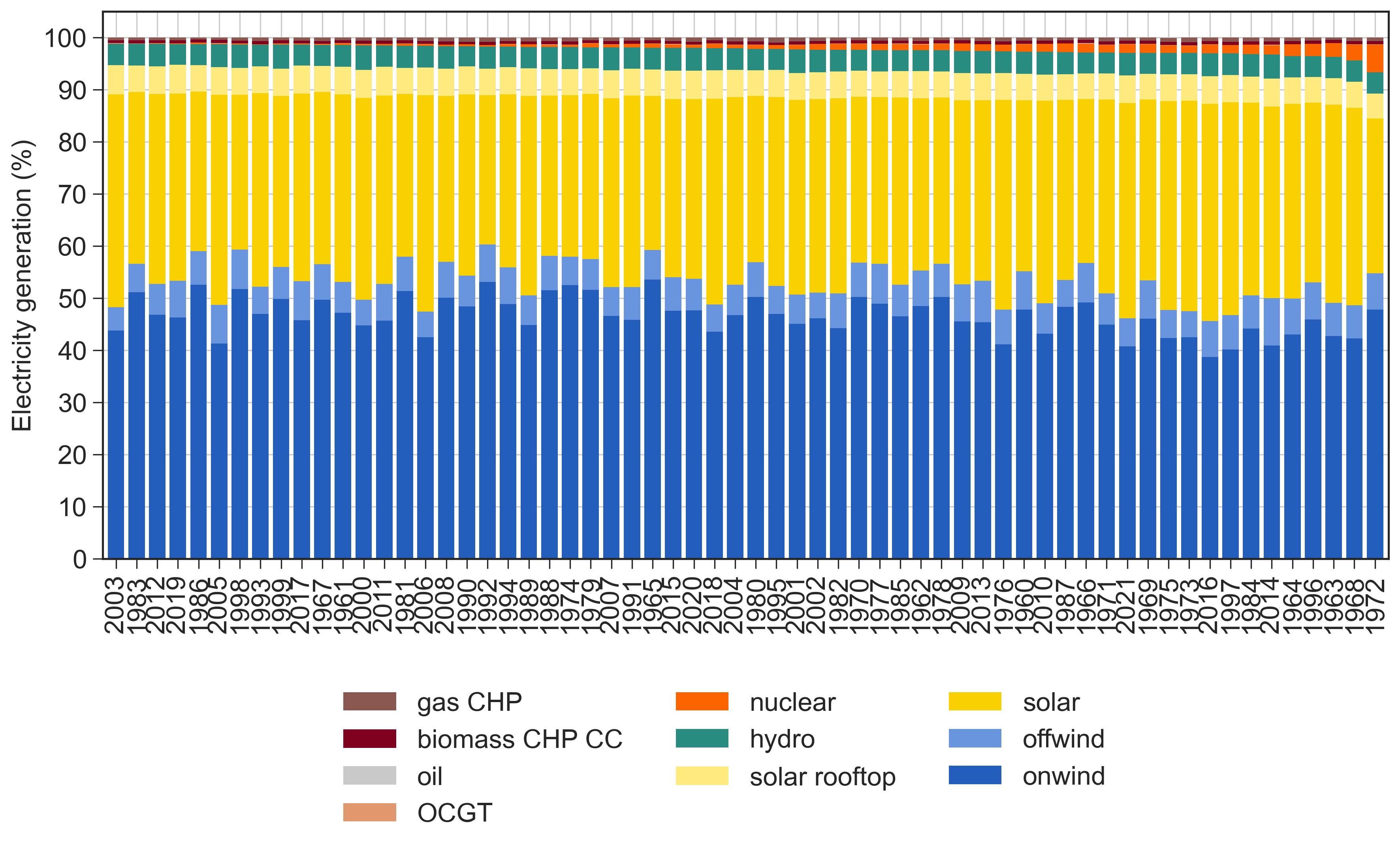}
	\captionsetup{width=14.8cm}
	\caption{Electricity generation mix from the capacity optimization for design years 1960 to 2021.}
	\label{sfig:vre_share}
\end{figure}

\newpage
\subsection{Firm and flexible electricity generation in all 62 capacity layouts}
\begin{figure}[!h]
	\centering
	\includegraphics[width=0.85\textwidth]{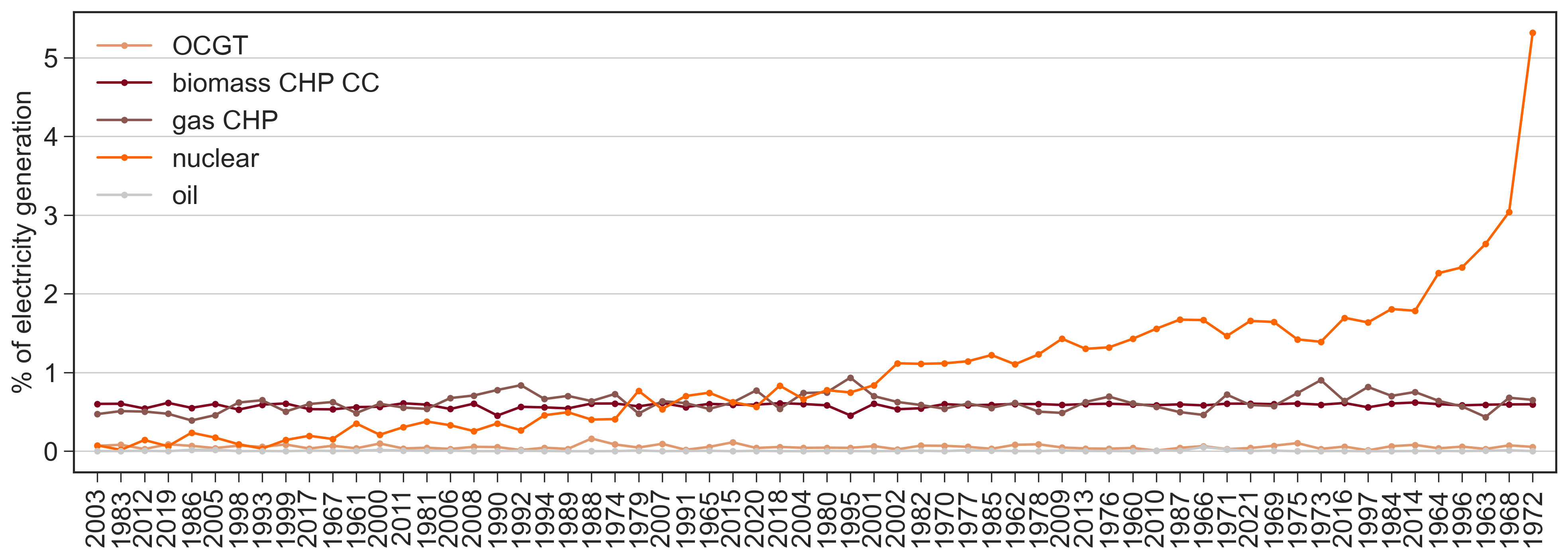}		\includegraphics[width=0.6\textwidth]{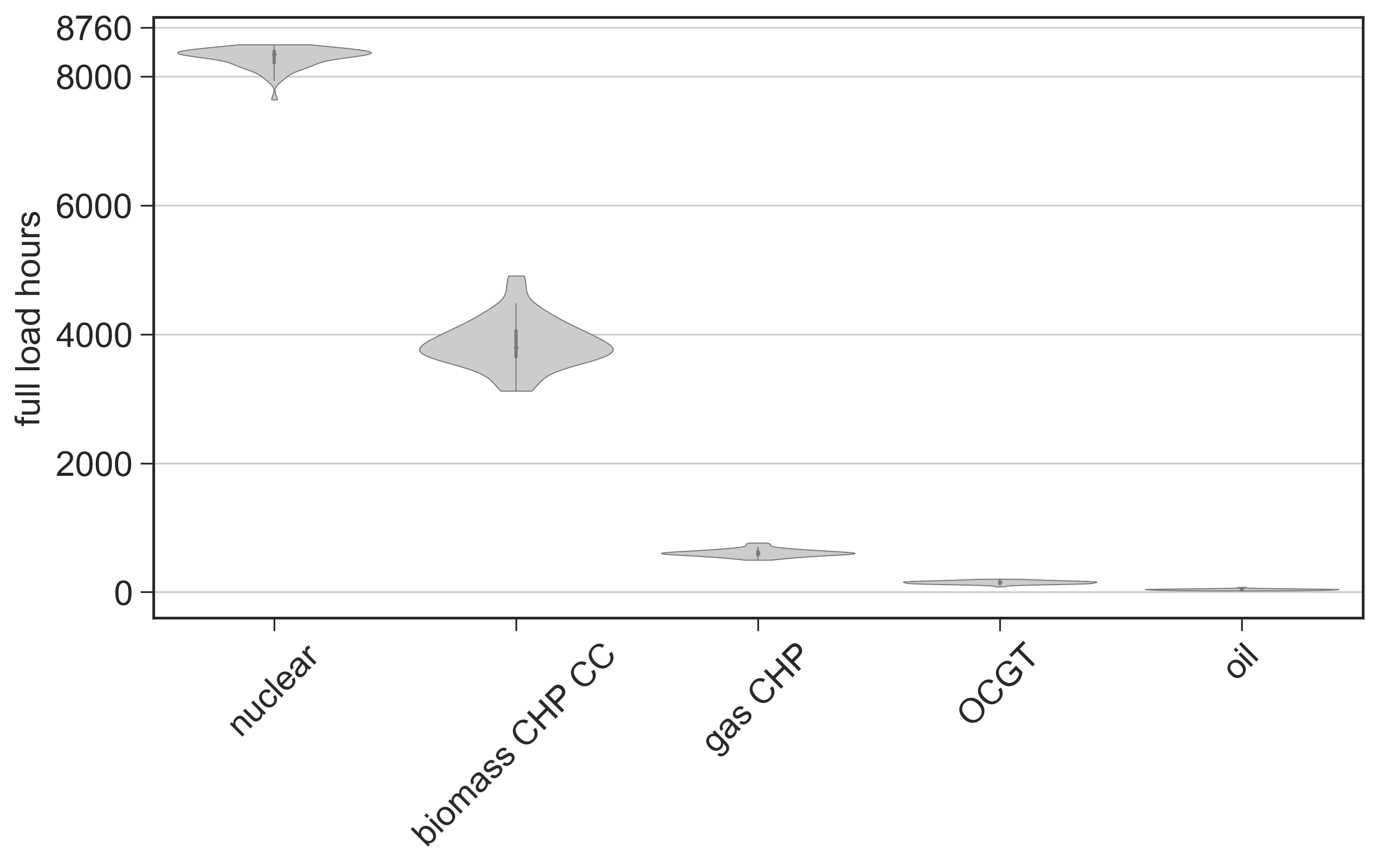}
	\captionsetup{width=14.8cm}
	\caption{Firm and flexible generation share of the electricity generation (top) and full load hours (bottom) from the capacity optimization for design years 1960 to 2021.}
	\label{sfig:firm_and_flexible_generation}
\end{figure}

\newpage
\subsection{CO2 emissions prices derived from Lagrangian multipliers}
\begin{figure}[!h]
	\centering
	\includegraphics[width=0.55\textwidth]{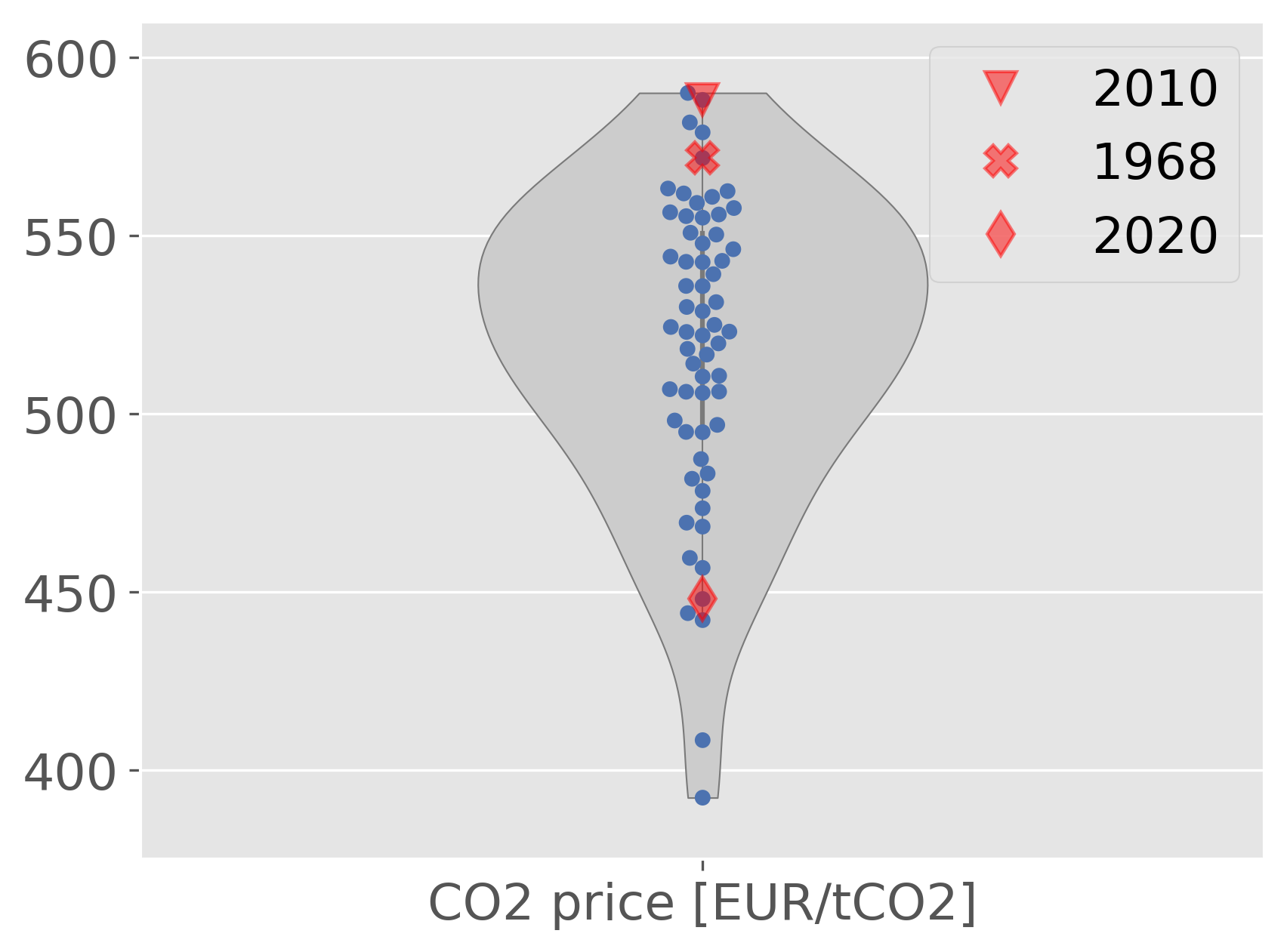}
	\captionsetup{width=14.8cm}
	\caption{CO2 emissions price derived from the Lagrangian multiplier associated with the CO$_2$ emissions cap, obtained for the design years ranging from 1960 to 2021.}
	\label{sfig:co2_emissions_price}
\end{figure}

\newpage
\subsection{Temporal distribution of loss of load}
\begin{figure}[!h]
	\centering
	\includegraphics[width=0.85\textwidth]{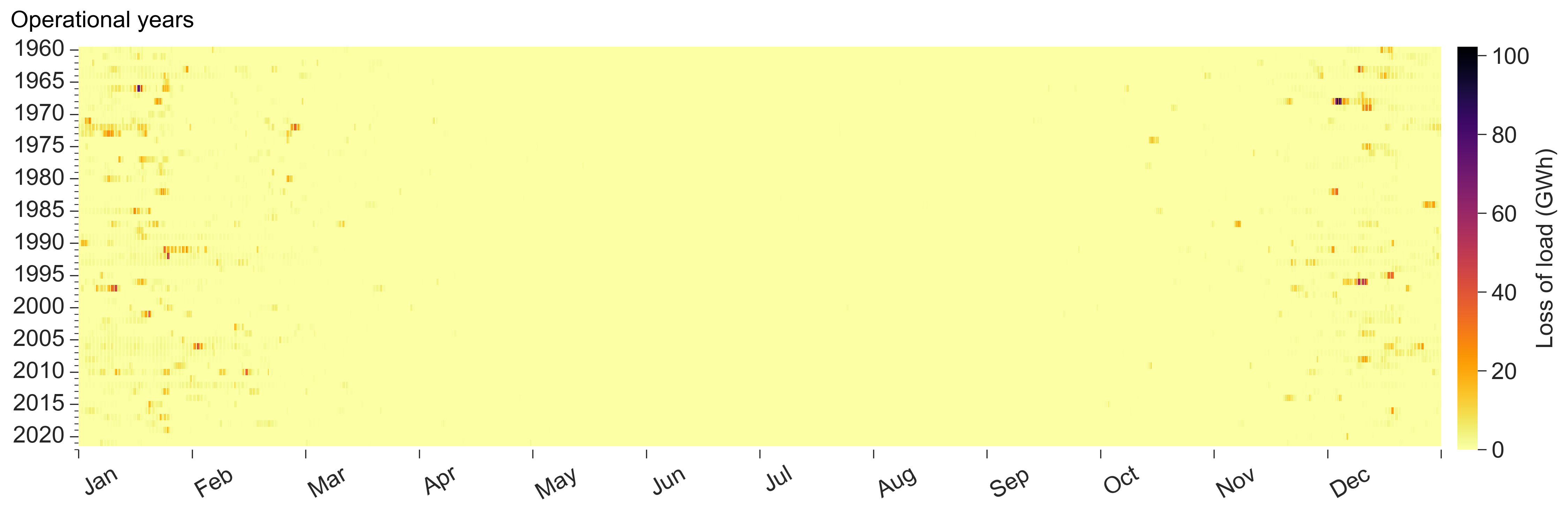}
	\captionsetup{width=14.8cm}
	\caption{Loss of load for every 3 hour timestep in every operational year, averaged for all design years}	
	\label{sfig:unserved_energy_avg}
\end{figure}

\newpage
\subsection{Unserved energy for all design years in every operational year}
\begin{figure}[!h]
	\centering
	\includegraphics[width=0.85\textwidth]{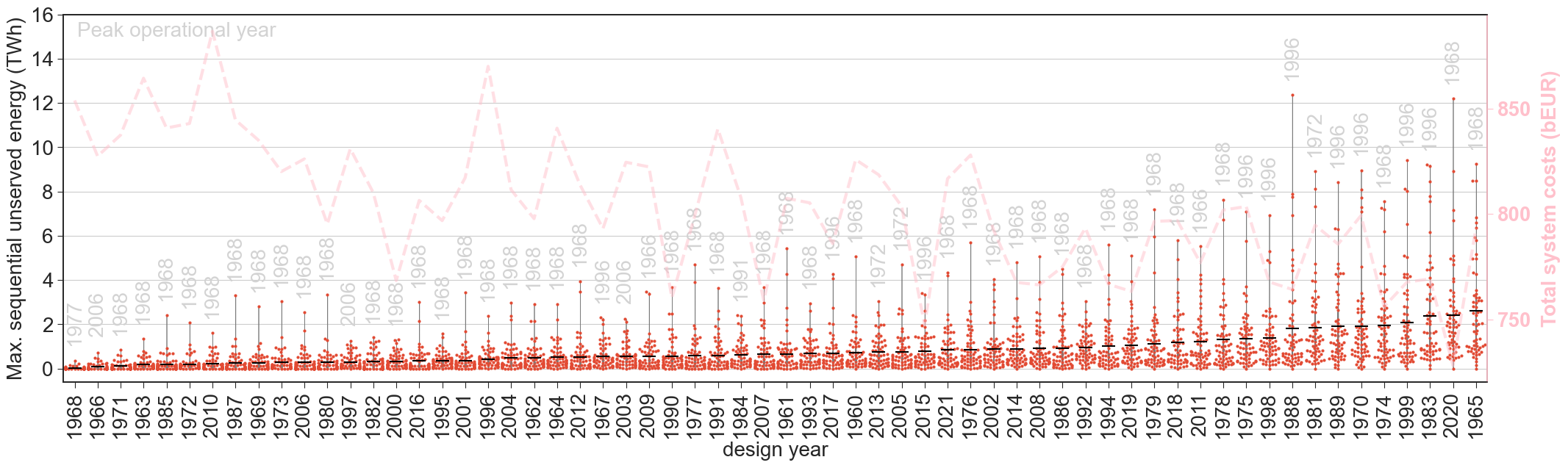}
	\includegraphics[width=0.85\textwidth]{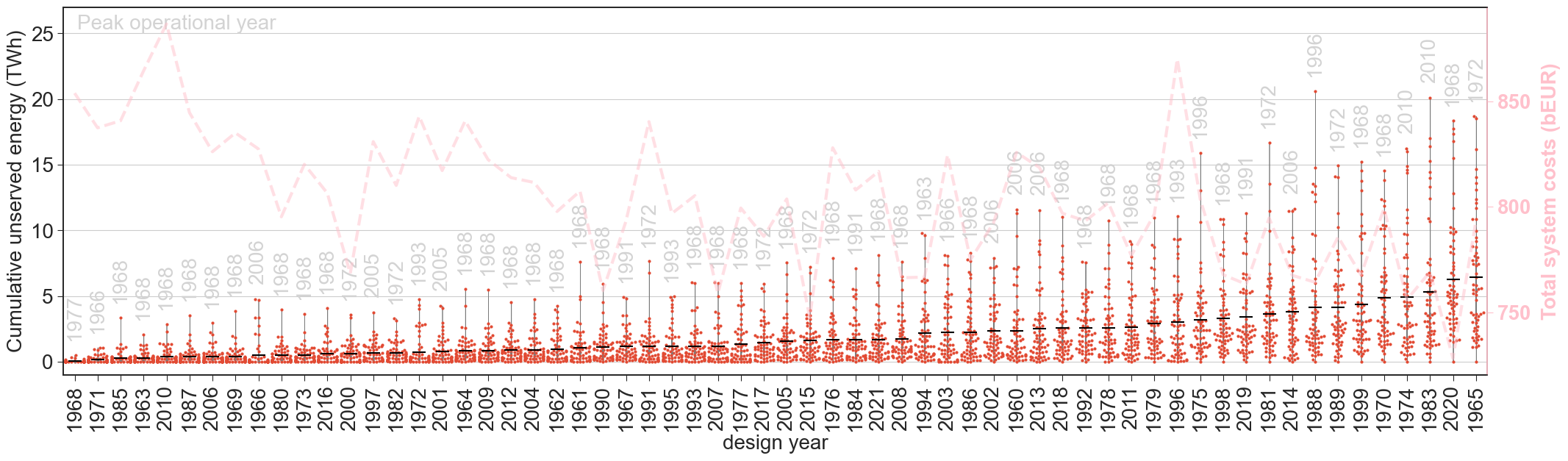}
	\captionsetup{width=14.8cm}
	\caption{Loss of load aggregated for Europe calculated as (top) the maximum sequential unserved energy, and (bottom) the cumulative unserved energy during a full operational year.}	
	\label{sfig:dispatch_optimization_all}
\end{figure}

\newpage
\subsection{Geographical distribution of country-specific cumulative unserved energy for the 2013 design year}

\begin{figure}[!h]
	\centering
	\includegraphics[width=0.85\textwidth]{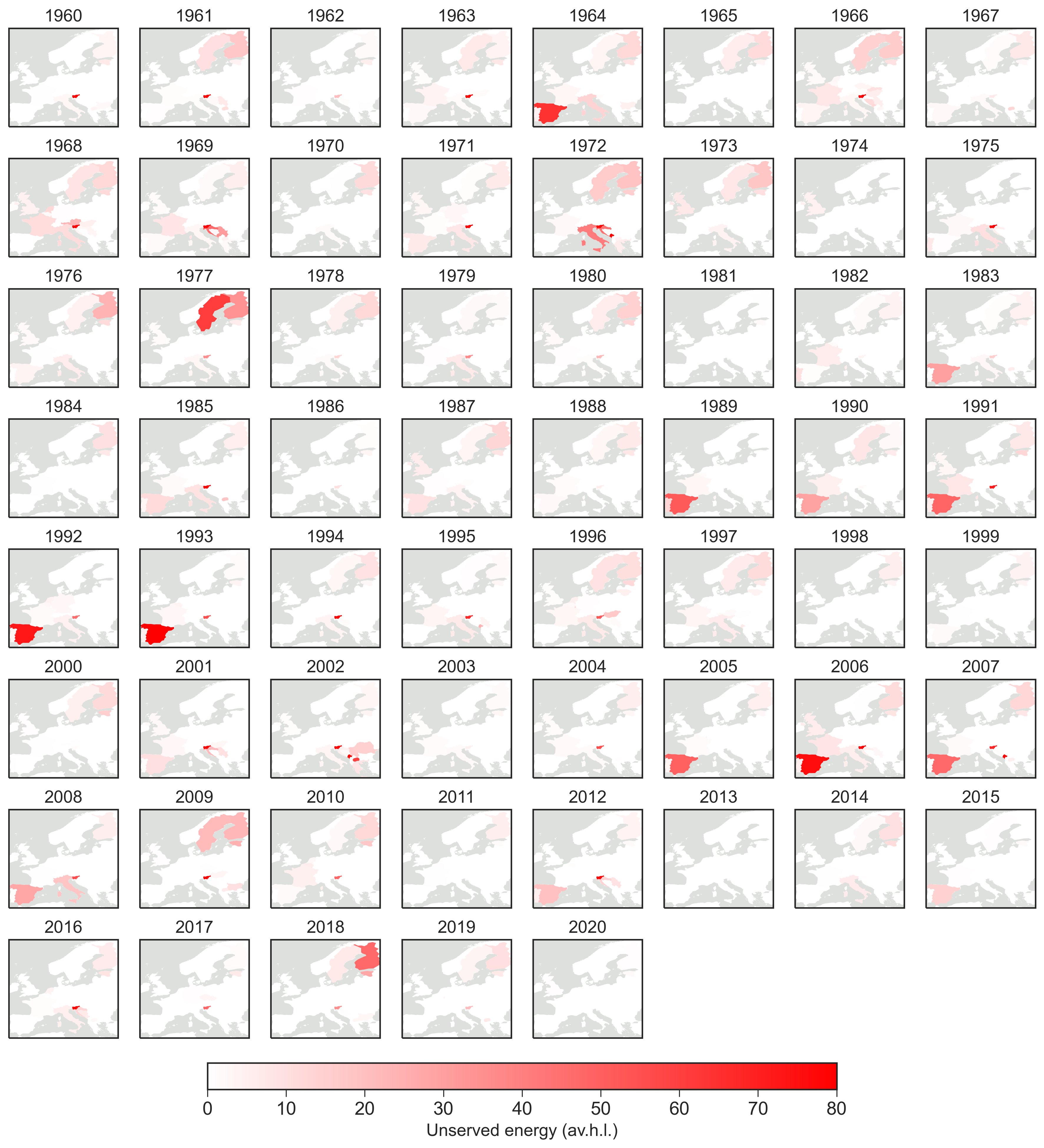}
	\captionsetup{width=14.8cm}
	\caption{Geographical distribution of the annual unserved energy in units of the countries' average hourly electricity load (av.h.l.) shown for the 2013 capacity layout, which corresponds to a total system cost close to the average, simulated for every operational year.}
	\label{sfig:map_unserved_energy_all_weather_years_dy2013}
\end{figure}

\newpage
\subsection{Geographical distribution of peak loss of load for the 2013 design year}

\begin{figure}[!h]
	\centering
	\includegraphics[width=0.85\textwidth]{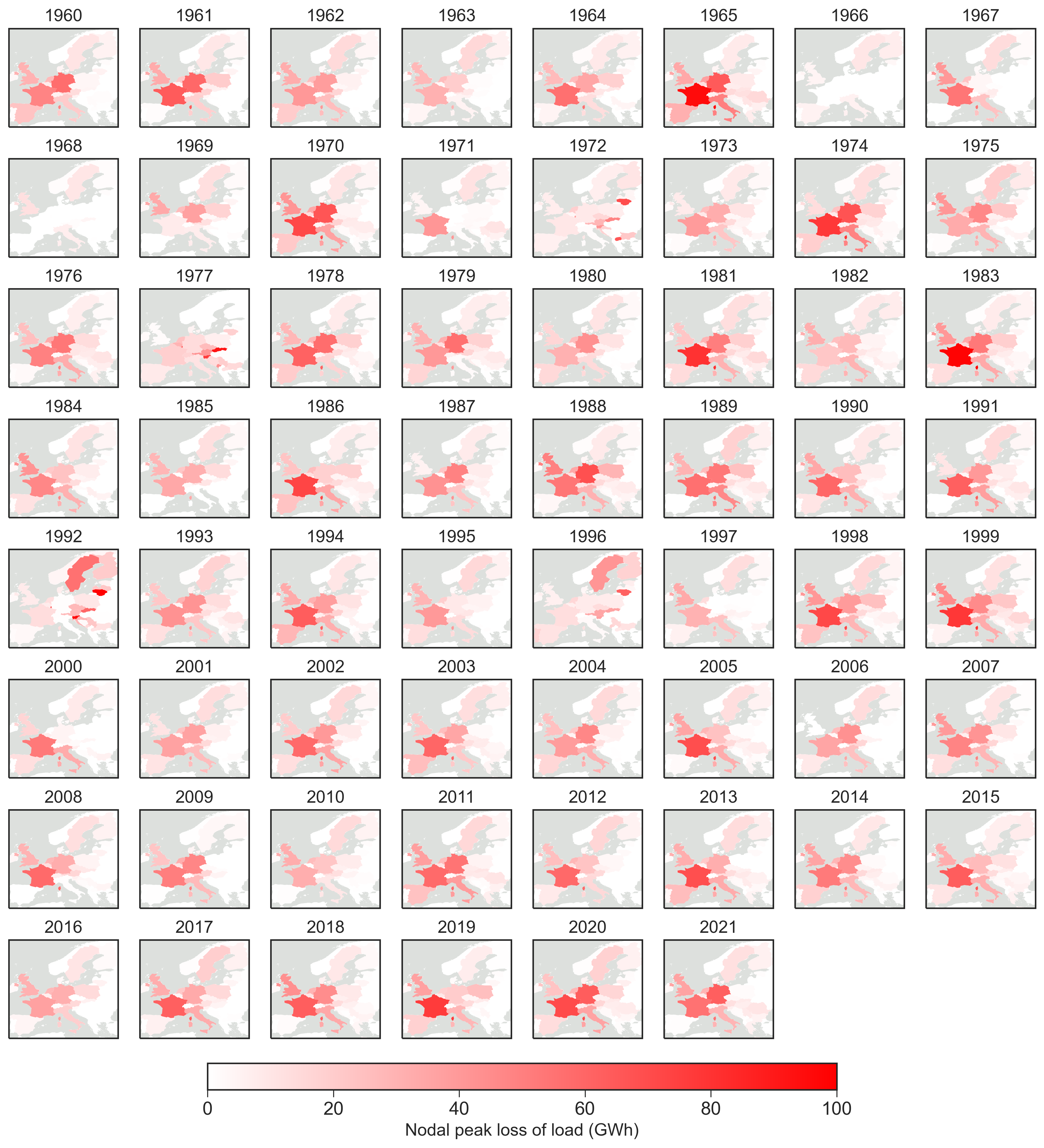}
	\captionsetup{width=14.8cm}
	\caption{Geographical distribution of country-specific peak loss of load shown for different design years simulated in every operational year.}
	\label{sfig:map_nodal_loss_of_load}
\end{figure}

\newpage
\subsection{Geographical distribution of cumulative unserved energy for every operational year}
\begin{figure}[!h]
	\centering
	\includegraphics[width=0.85\textwidth]{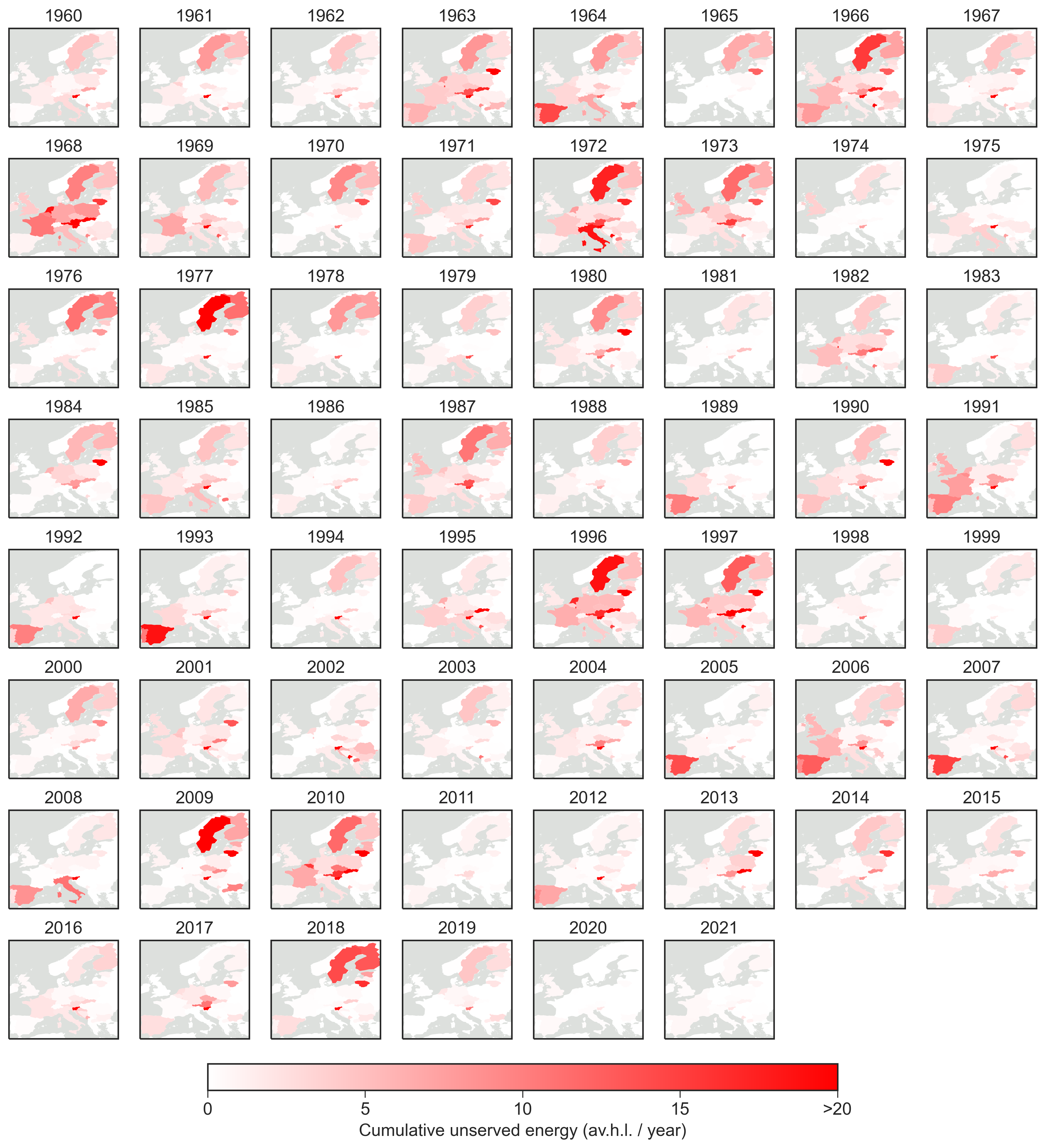}
	\captionsetup{width=14.8cm}
	\caption{Annual unserved energy shown for different operational years averaged across all capacity layouts. The unserved energy is normalized by the country's average hourly electricity load (av.h.l.).}
	\label{sfig:map_unserved_energy_all_operations}
\end{figure}

\newpage
\subsection{Geographical distribution of cumulative unserved energy for every capacity layout}
\begin{figure}[!h]
	\centering
	\includegraphics[width=0.85\textwidth]{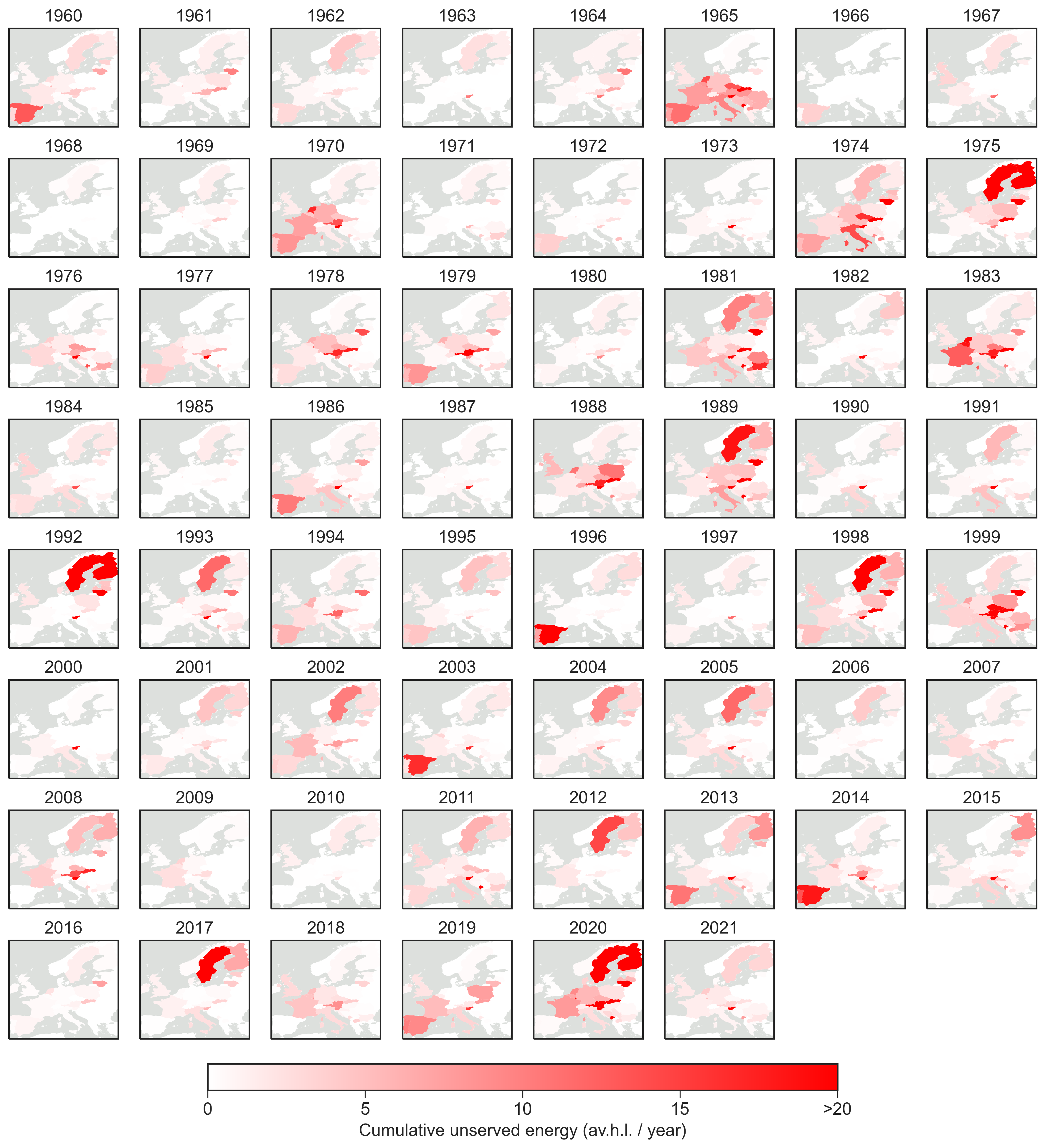}
	\captionsetup{width=14.8cm}
	\caption{Annual unserved energy shown for every design year \textbf{averaged across all operational years}. The unserved energy is normalized by the country's average hourly electricity load (av.h.l.).}
	\label{sfig:map_unserved_energy_all_designs}
\end{figure}

\newpage
\subsection{Unserved energy and renewable droughts}
\textbf{Operational year: 1966}
\begin{figure}[!h]
	\centering
	\includegraphics[width=0.85\textwidth]{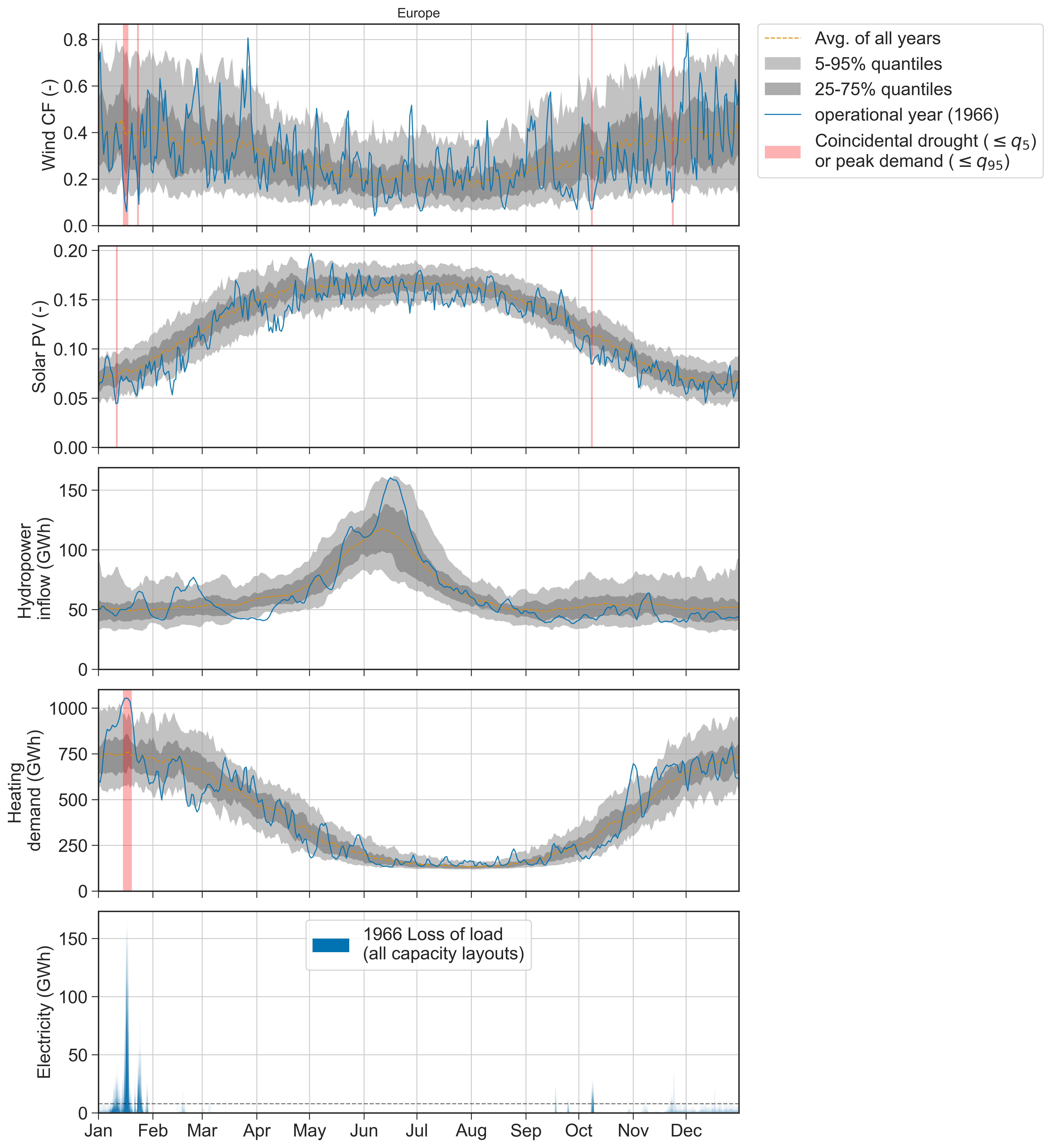}
	\captionsetup{width=14.8cm}
	\caption{Daily average renewable resources, heating demand, electricity load, and loss of load for Europe simulated for all capacity layouts in the 1966 operational year. The red bars indicate where the resources/demand differ from 90\% of the data coinciding with a loss of load event. This is only shown for loss of load events exceeding 10~GWh (represented by the dashed line in the bottom subfigure).}
	\label{sfig:unserved_energy_and_renwable_droughts_EU_1966}
\end{figure}

\newpage
\textbf{Operational year: 1968}
\begin{figure}[!h]
	\centering
	\includegraphics[width=0.85\textwidth]{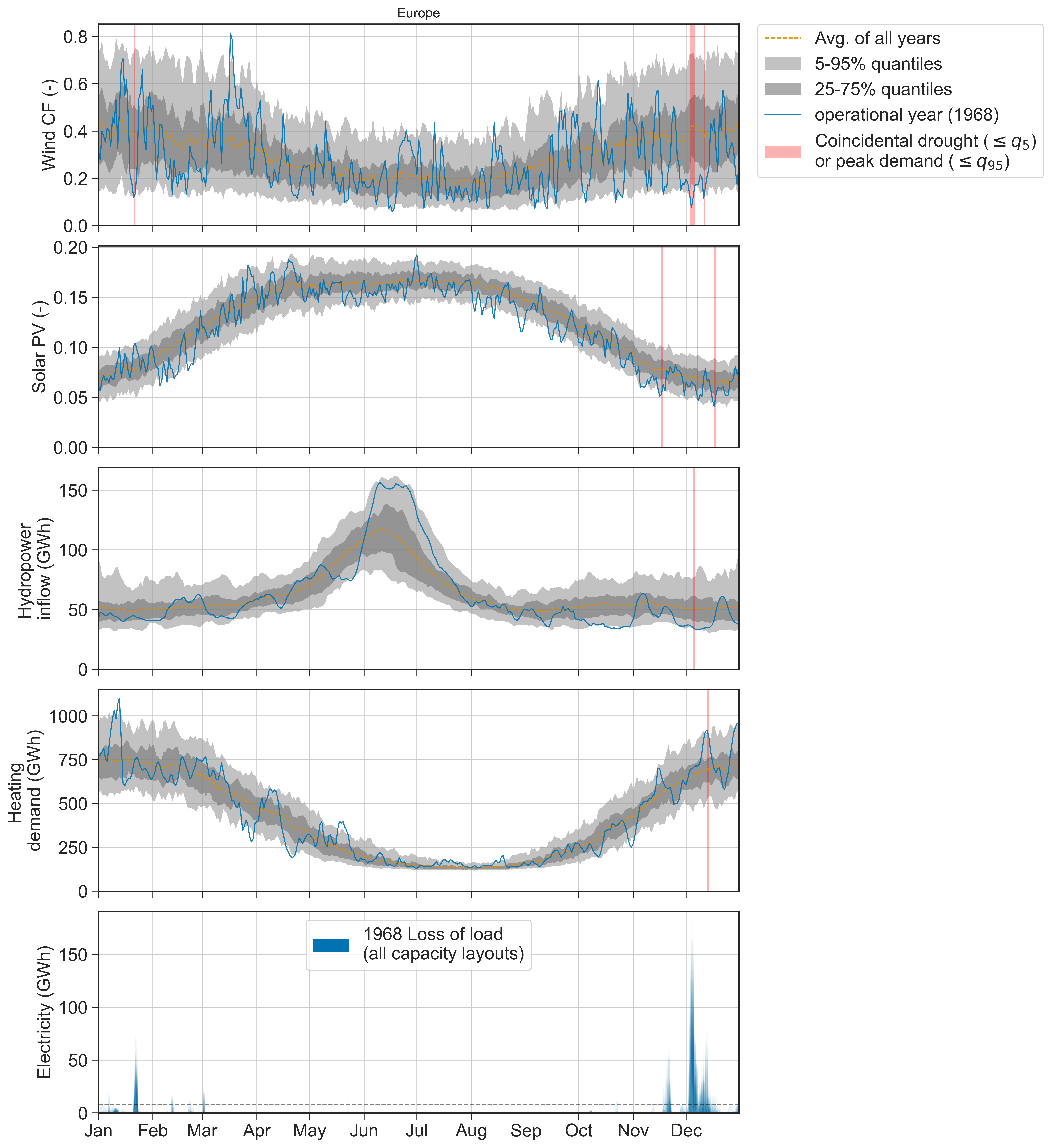}
	\captionsetup{width=14.8cm}
	\caption{Daily average renewable resources, heating demand, electricity load, and loss of load for Europe simulated for all capacity layouts in the 1968 operational year. The red bars indicate where the resources/demand differ from 90\% of the data coinciding with a loss of load event. This is only shown for loss of load events exceeding 10~GWh (represented by the dashed line in the bottom subfigure).}
	\label{sfig:unserved_energy_and_renwable_droughts_EU_1968}
\end{figure}

\newpage
\textbf{Operational year: 1972}
\begin{figure}[!h]
	\centering
	\includegraphics[width=0.85\textwidth]{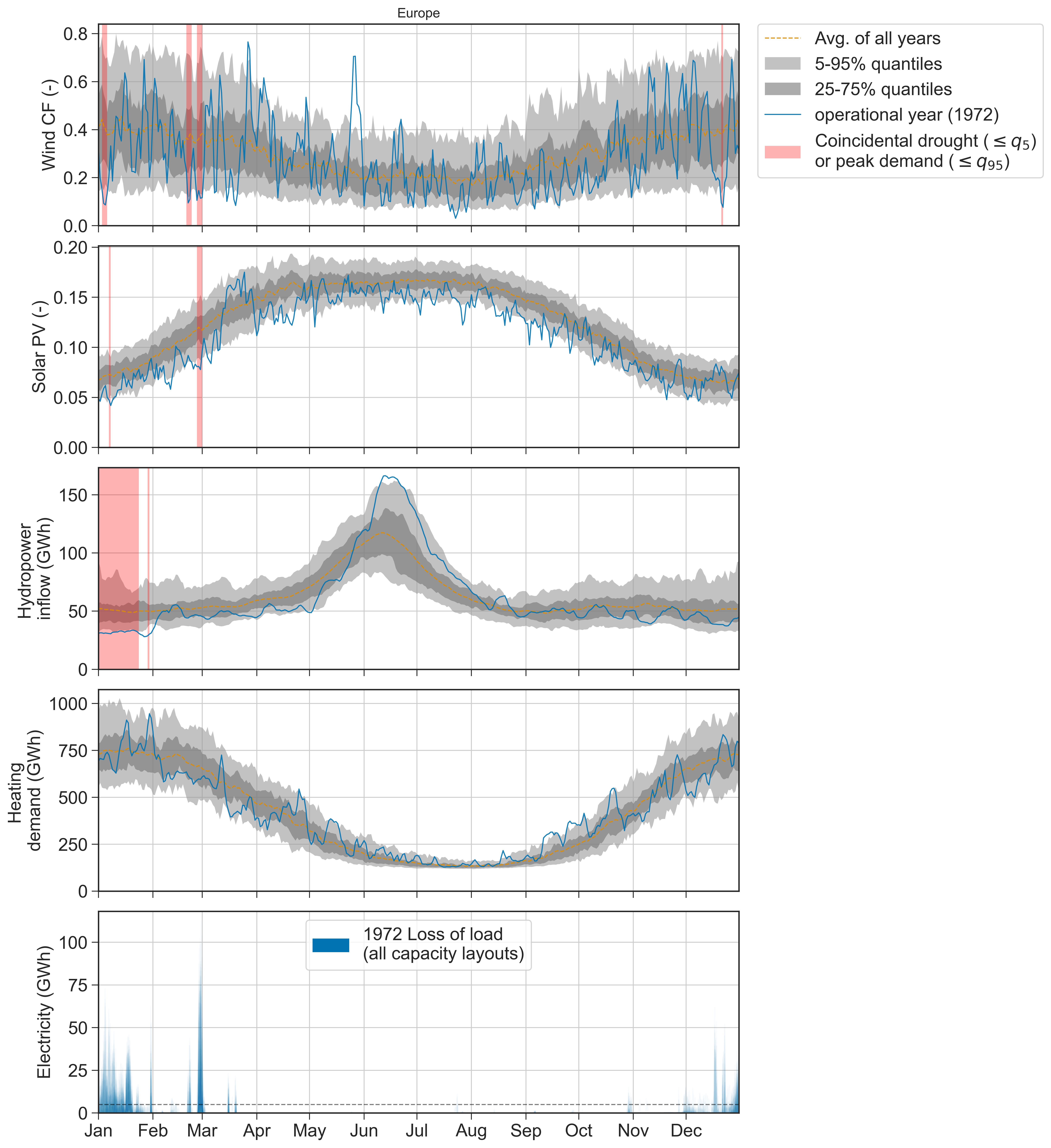}
	\captionsetup{width=14.8cm}
	\caption{Daily average renewable resources, heating demand, electricity load, and loss of load for Europe simulated for all capacity layouts in the 1972 operational year. The red bars indicate where the resources/demand differ from 90\% of the data coinciding with a loss of load event. This is only shown for loss of load events exceeding 10~GWh (represented by the dashed line in the bottom subfigure).}
	\label{sfig:unserved_energy_and_renwable_droughts_EU_1972}
\end{figure}

\newpage
\textbf{Operational year: 1996}
\begin{figure}[!h]
	\centering
	\includegraphics[width=0.85\textwidth]{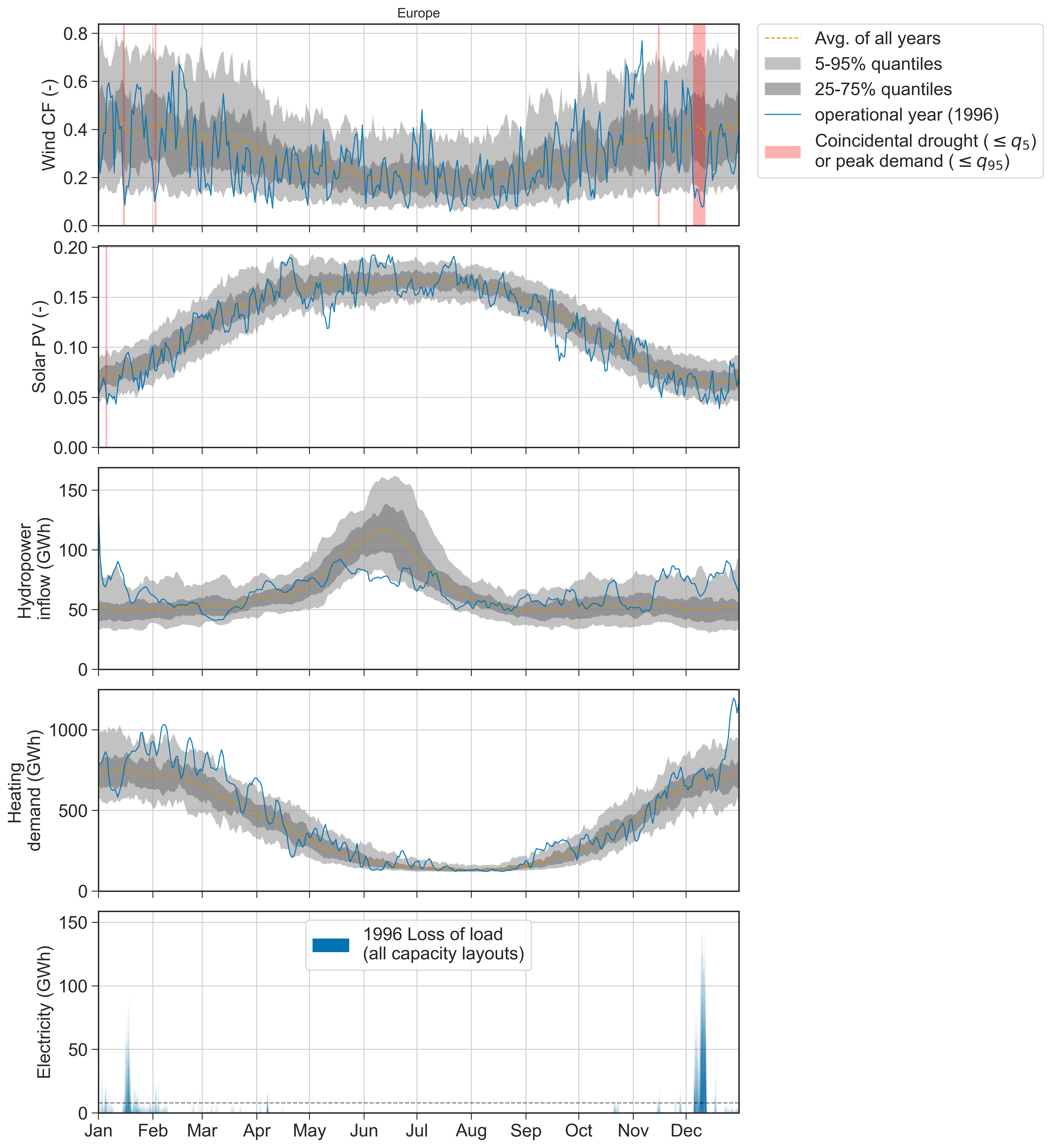}
	\captionsetup{width=14.8cm}
	\caption{Daily average renewable resources, heating demand, electricity load, and loss of load for Europe simulated for all capacity layouts in the 1996 operational year. The red bars indicate where the resources/demand differ from 90\% of the data coinciding with a loss of load event. This is only shown for loss of load events exceeding 10~GWh (represented by the dashed line in the bottom subfigure).}
	\label{sfig:unserved_energy_and_renwable_droughts_EU_1996}
\end{figure}

\newpage
\textbf{Operational year: 2010}
\begin{figure}[!h]
	\centering
	\includegraphics[width=0.85\textwidth]{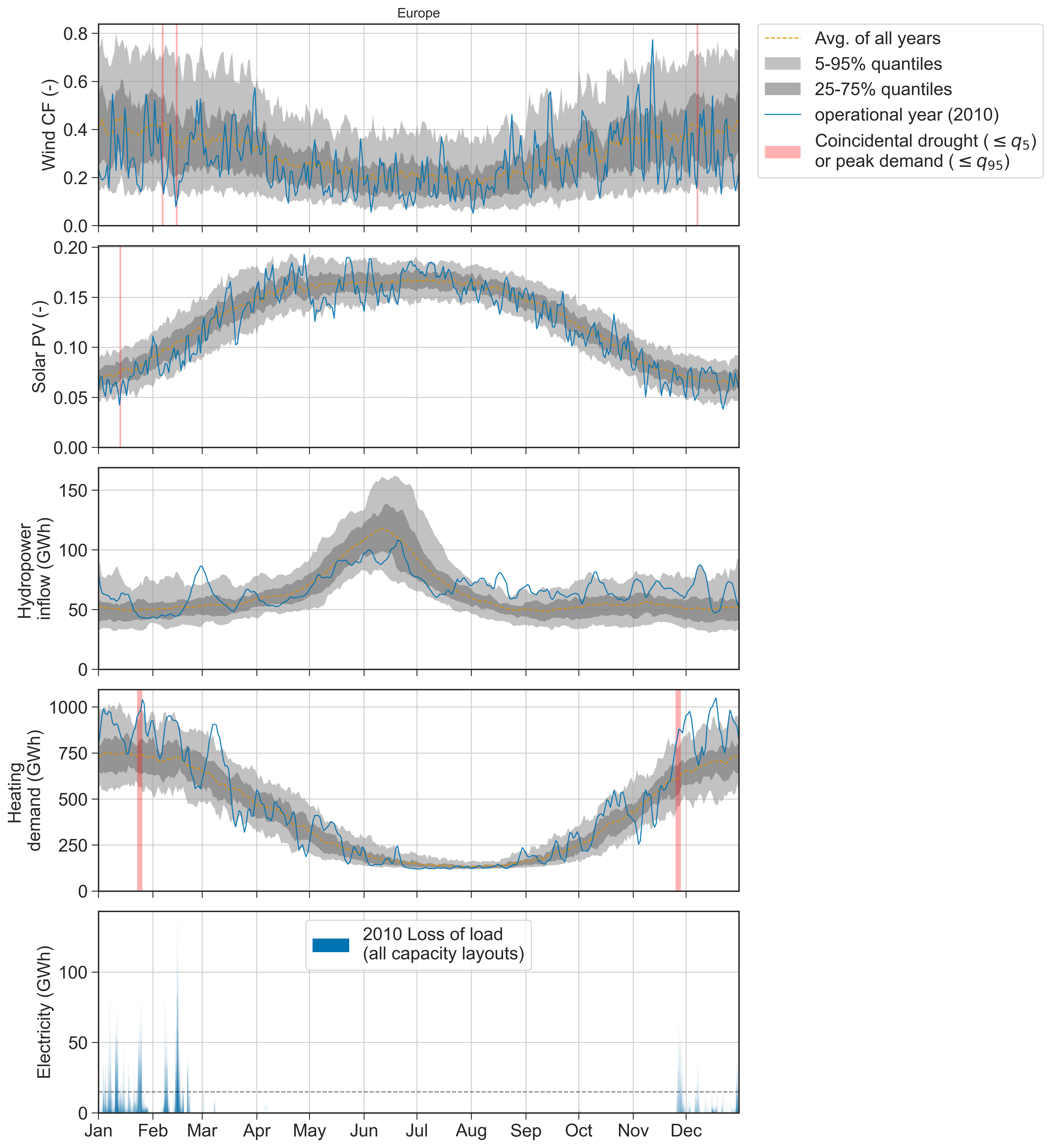}
	\captionsetup{width=14.8cm}
	\caption{Daily average renewable resources, heating demand, electricity load, and loss of load for Europe simulated for all capacity layouts in the 2010 operational year. The red bars indicate where the resources/demand differ from 90\% of the data coinciding with a loss of load event. This is only shown for loss of load events exceeding 10~GWh (represented by the dashed line in the bottom subfigure).}
	\label{sfig:unserved_energy_and_renwable_droughts_EU_2010}
\end{figure}

\newpage
\subsection{Wind resources}
\textbf{First two weeks of January}
\begin{figure}[!h]
	\centering
	\includegraphics[width=0.75\textwidth]{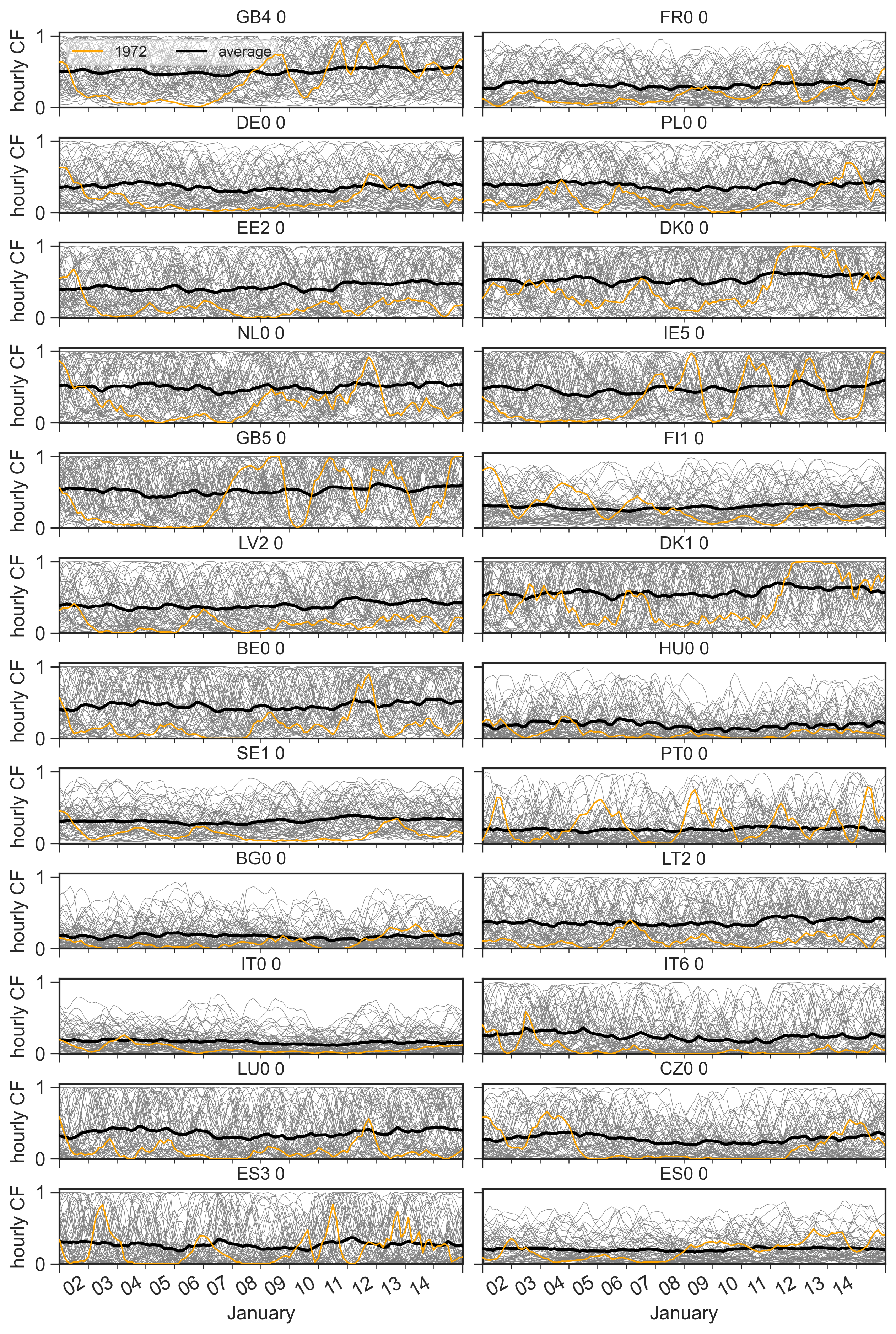}
	\captionsetup{width=14.8cm}
	\caption{Daily average wind resources during the first two weeks of January. The red bars indicate where the resources/demand differ from 90\% of the data coinciding with a loss of load event. This is only shown for loss of load events exceeding 10~GWh (represented by the dashed line in the bottom subfigure).}
	\label{sfig:wind_first_two_weeks}
\end{figure}

\newpage
\textbf{Middle of January}
\begin{figure}[!h]
	\centering
	\includegraphics[width=0.75\textwidth]{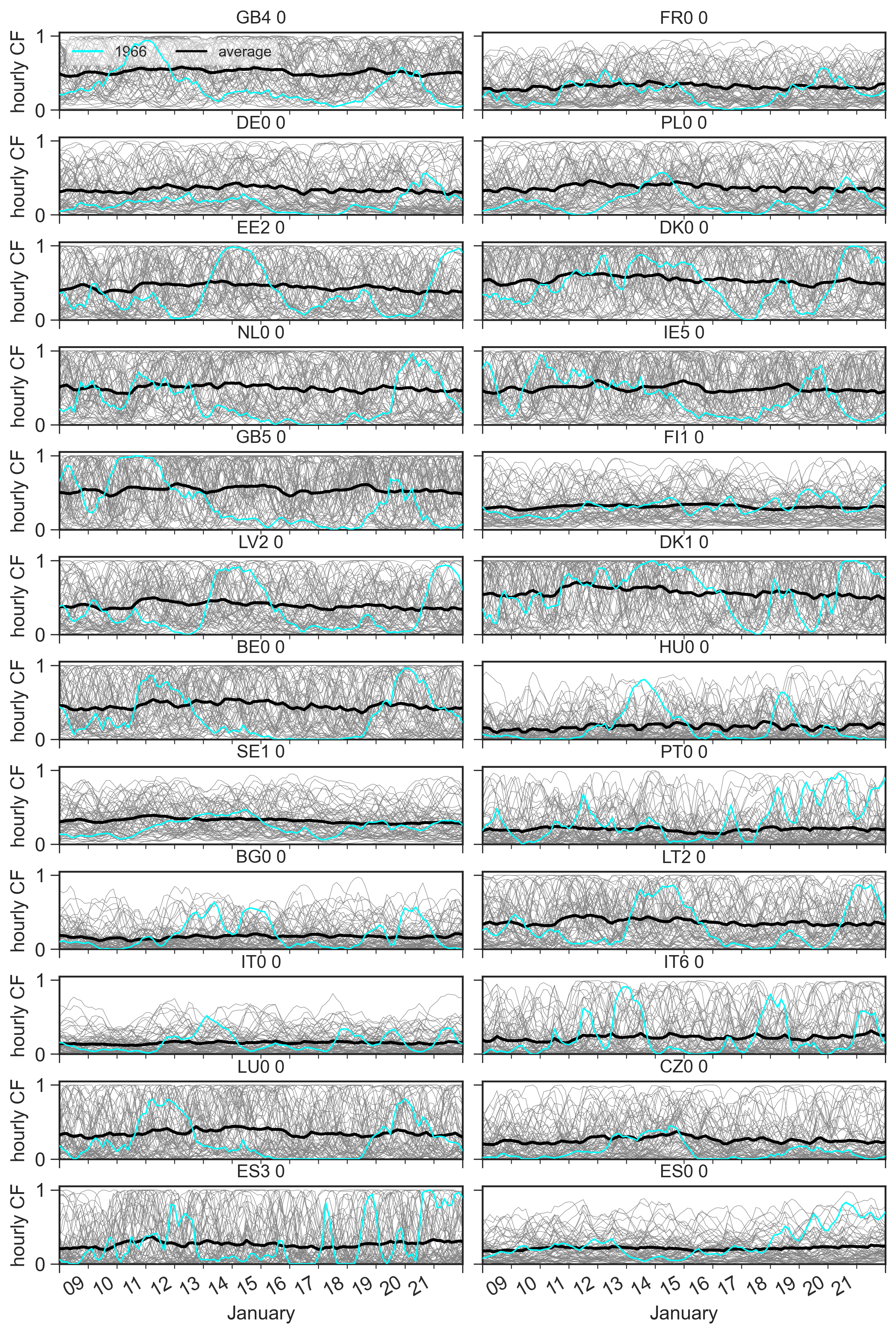}
	\captionsetup{width=14.8cm}
	\caption{Daily average wind resources during two weeks of January.}
	\label{sfig:wind_middle_January}
\end{figure}

\newpage
\textbf{From late November to early December}
\begin{figure}[!h]
	\centering
	\includegraphics[width=0.75\textwidth]{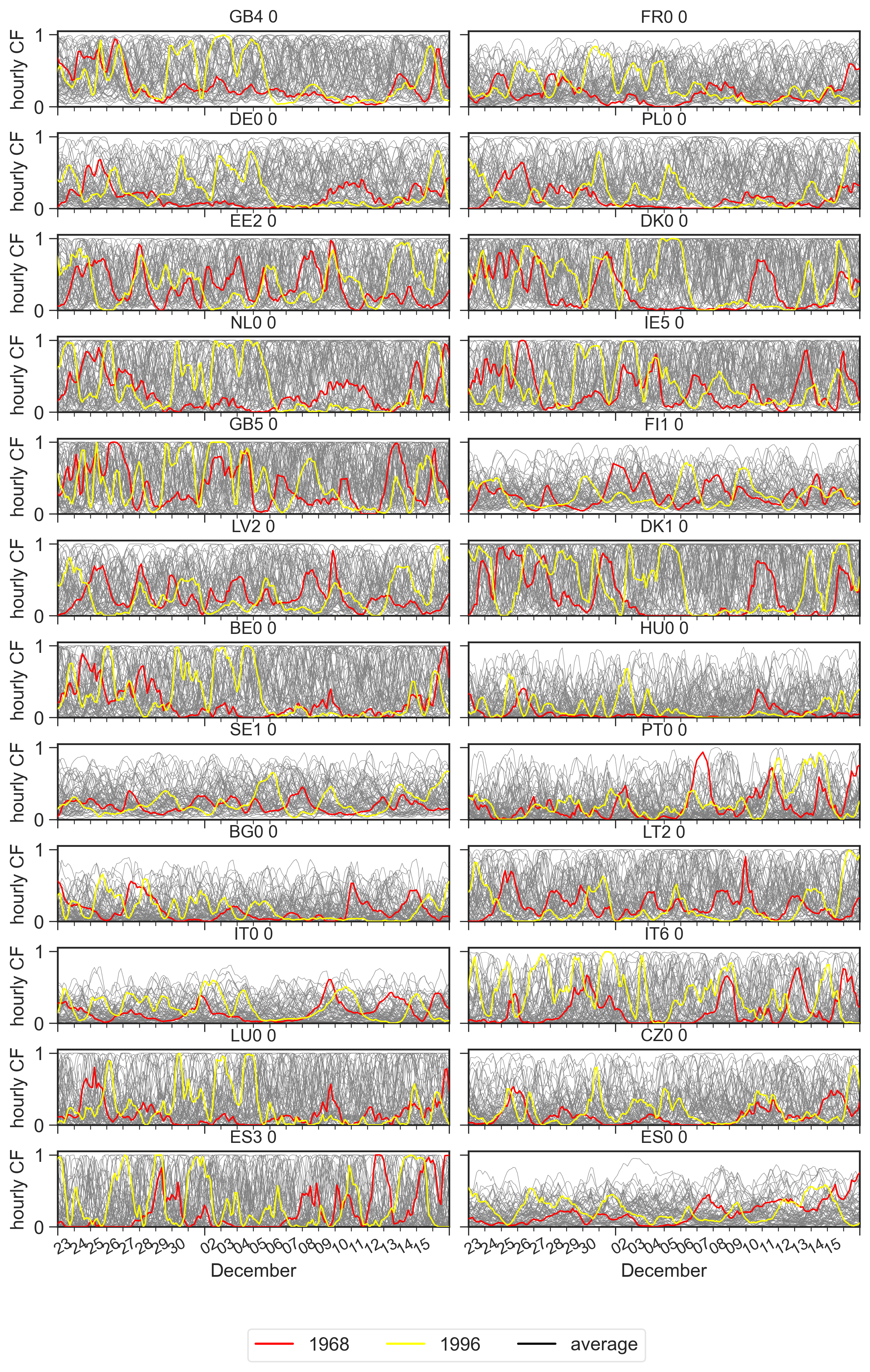}
	\captionsetup{width=14.8cm}
	\caption{Daily average wind resources from late November to early December.}
	\label{sfig:wind_December}
\end{figure}

\newpage
\subsection{Hydro resources}
\begin{figure}[!h]
	\centering
	\includegraphics[width=0.85\textwidth]{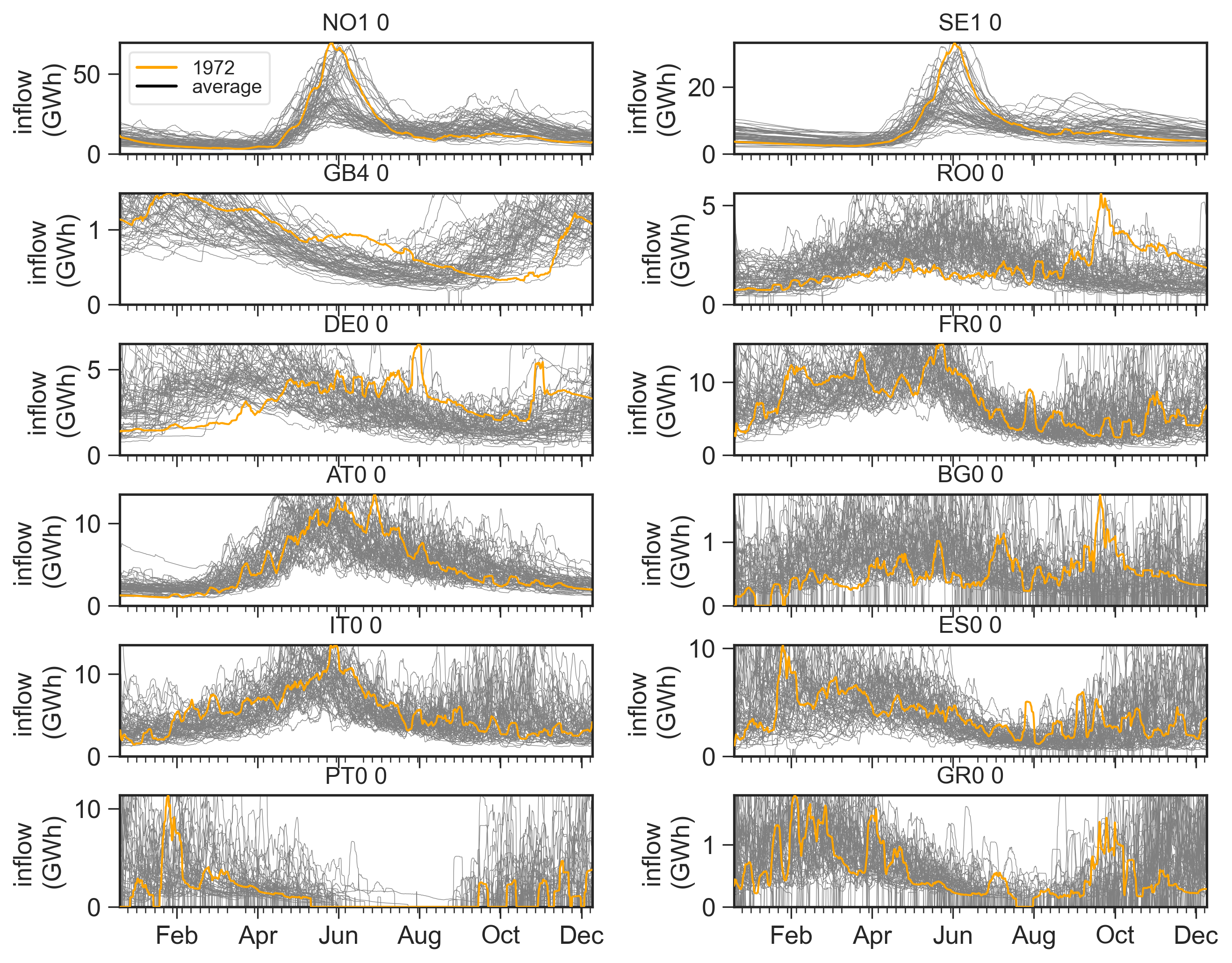}
	\captionsetup{width=14.8cm}
	\caption{Daily hydro inflow during the full calendar year.}
	\label{sfig:hydro}
\end{figure}

\newpage
\subsection{Heating demand}
\begin{figure}[!h]
	\centering
	\includegraphics[width=0.75\textwidth]{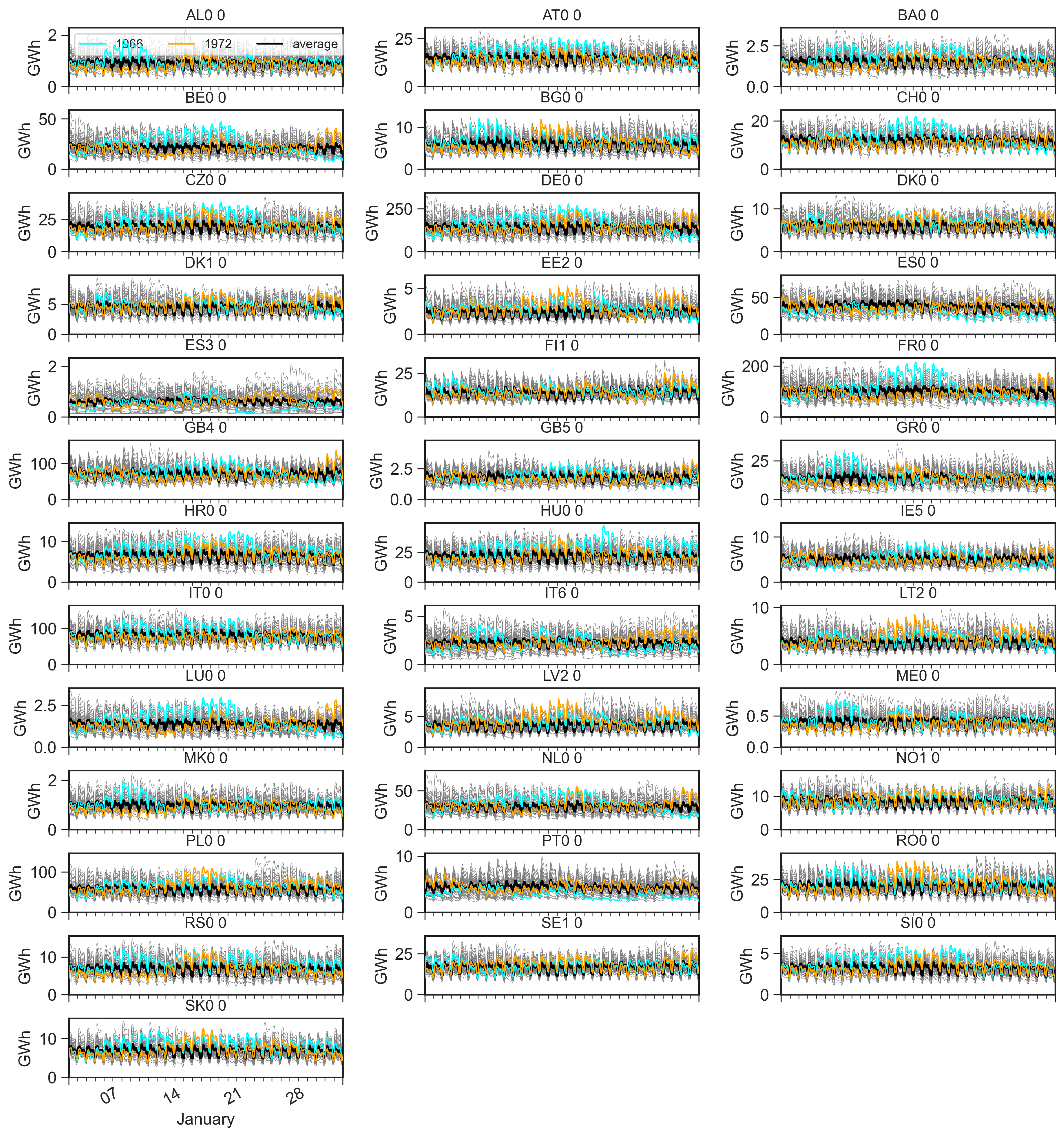}
	\captionsetup{width=14.8cm}
	\caption{Hourly heating demand during January.}
	\label{sfig:heating_demand}
\end{figure}

\newpage
\subsection{Capacities for the aggregated Europe in selected design years}

\textbf{Electricity, heating, and fuel generation technology}
\begin{figure}[!h]
	\centering
	\includegraphics[width=0.85\textwidth]{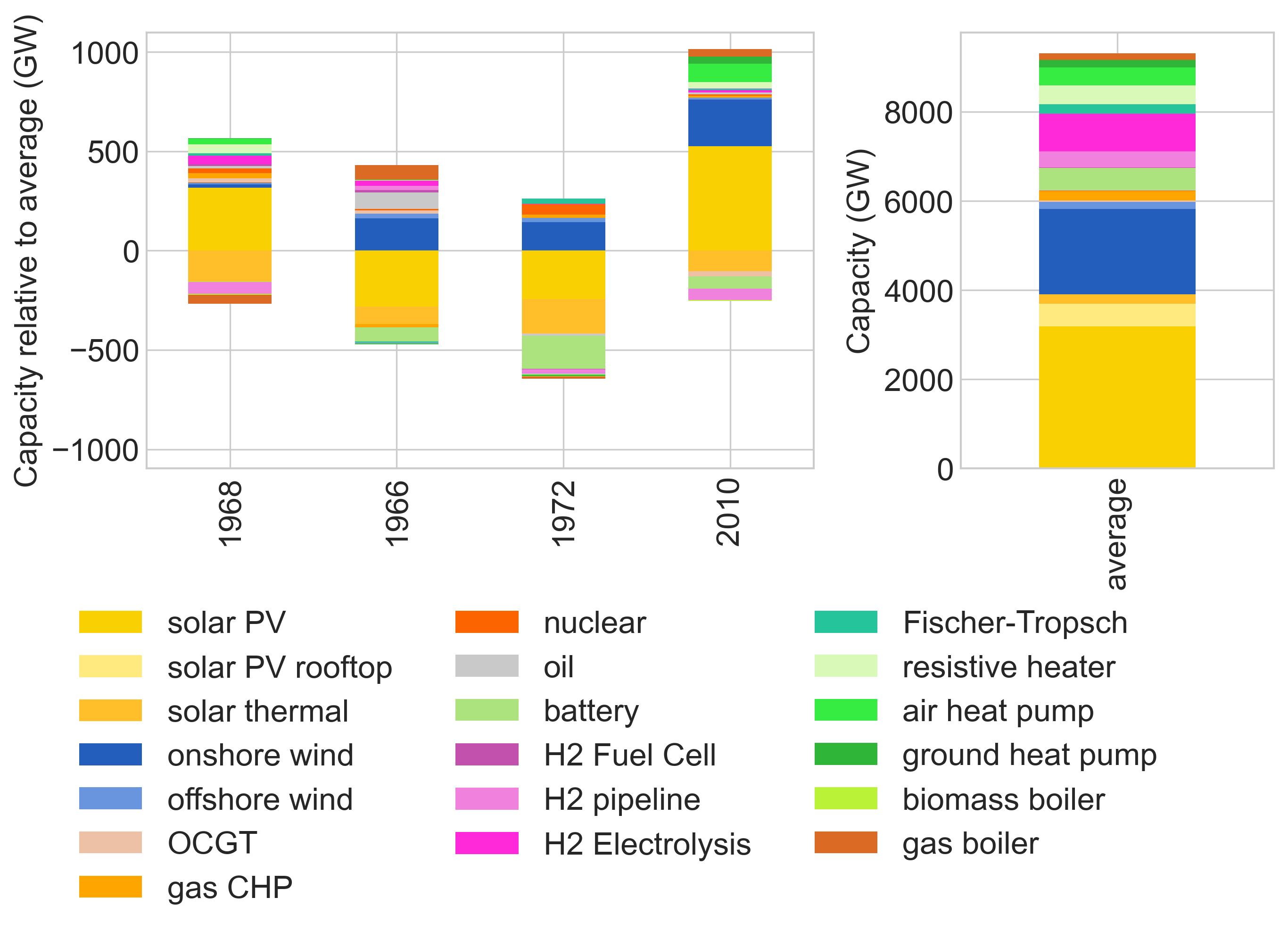}
	\captionsetup{width=14.8cm}
	\caption{European aggregated capacities for electricity, heating, and synthetic fuel production, illustrated for (right) the average of all design years, and (left) three systems designed with extreme operational years (1966, 1968, and 1972) and the most expensive system (2010).}
	\label{sfig:Eurupe_aggregate_capacity}
\end{figure}

\newpage
\textbf{H$_2$ storage capacity}
\begin{figure}[!h]
	\centering
	\includegraphics[width=0.85\textwidth]{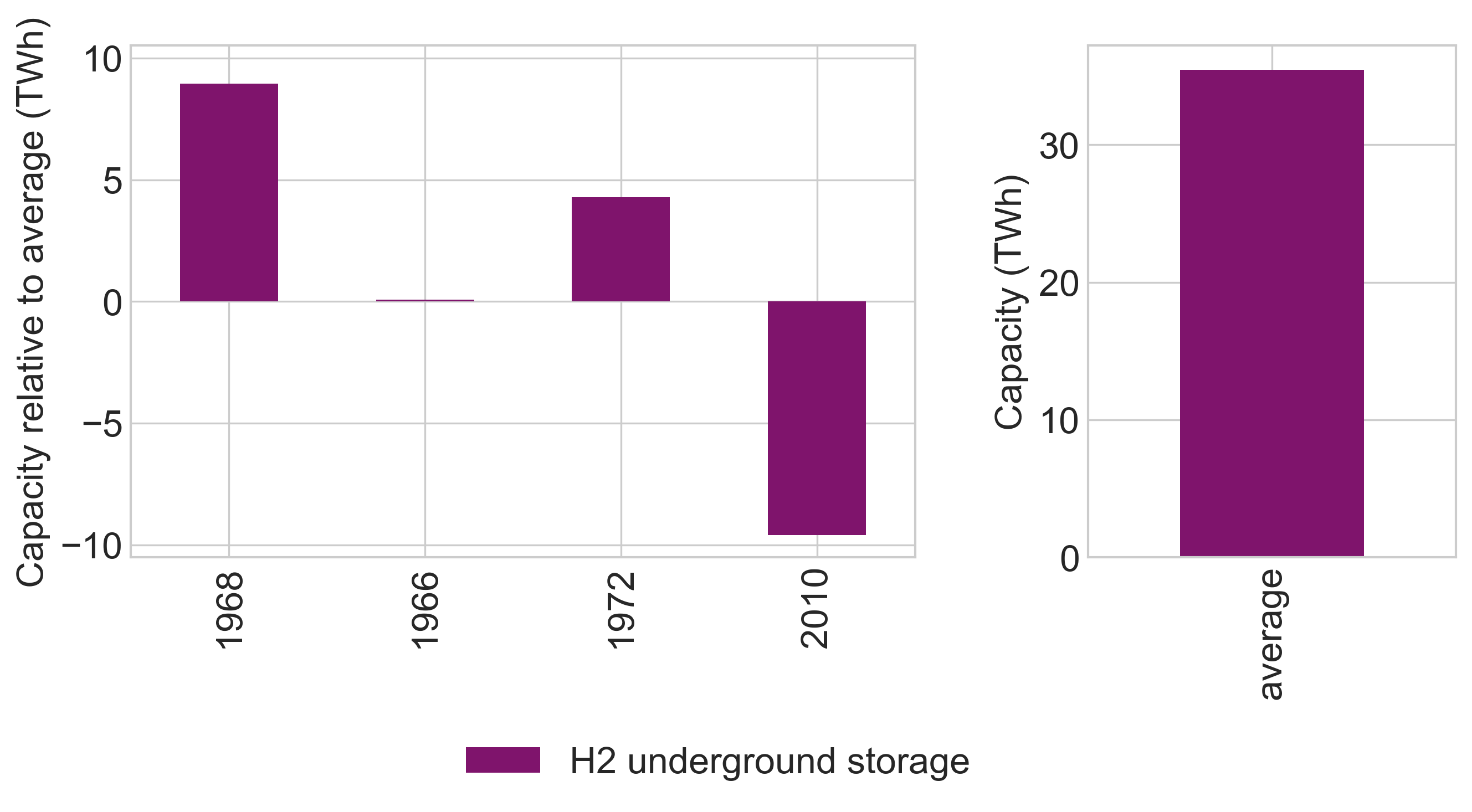}
	\captionsetup{width=14.8cm}
	\caption{European aggregated H$_2$ storage capacity, illustrated for (right) the average of all design years, and (left) three systems designed with extreme operational years (1966, 1968, and 1972) and the most expensive system (2010).}
	\label{sfig:Eurupe_aggregate_capacity_H2_storage}
\end{figure}

\newpage
\textbf{Thermal storage capacity}
\begin{figure}[!h]
	\centering
	\includegraphics[width=0.85\textwidth]{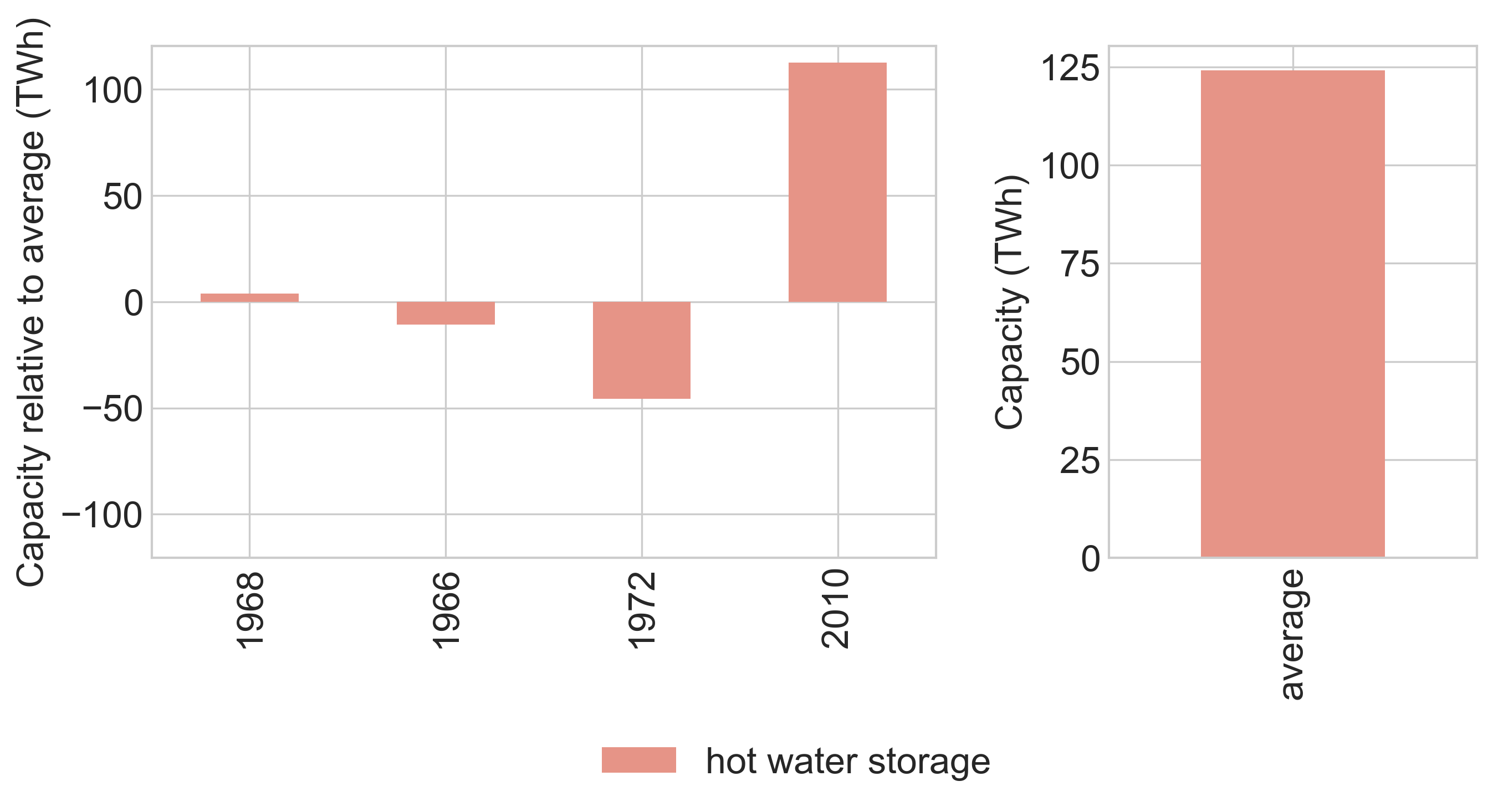}
	\captionsetup{width=14.8cm}
	\caption{European aggregated thermal storage capacity, illustrated for (right) the average of all design years, and (left) three systems designed with extreme operational years (1966, 1968, and 1972) and the most expensive system (2010).}
	\label{sfig:Eurupe_aggregate_capacity_thermal_storage}
\end{figure}

\newpage
\textbf{Direct air capture}
\begin{figure}[!h]
	\centering
	\includegraphics[width=0.85\textwidth]{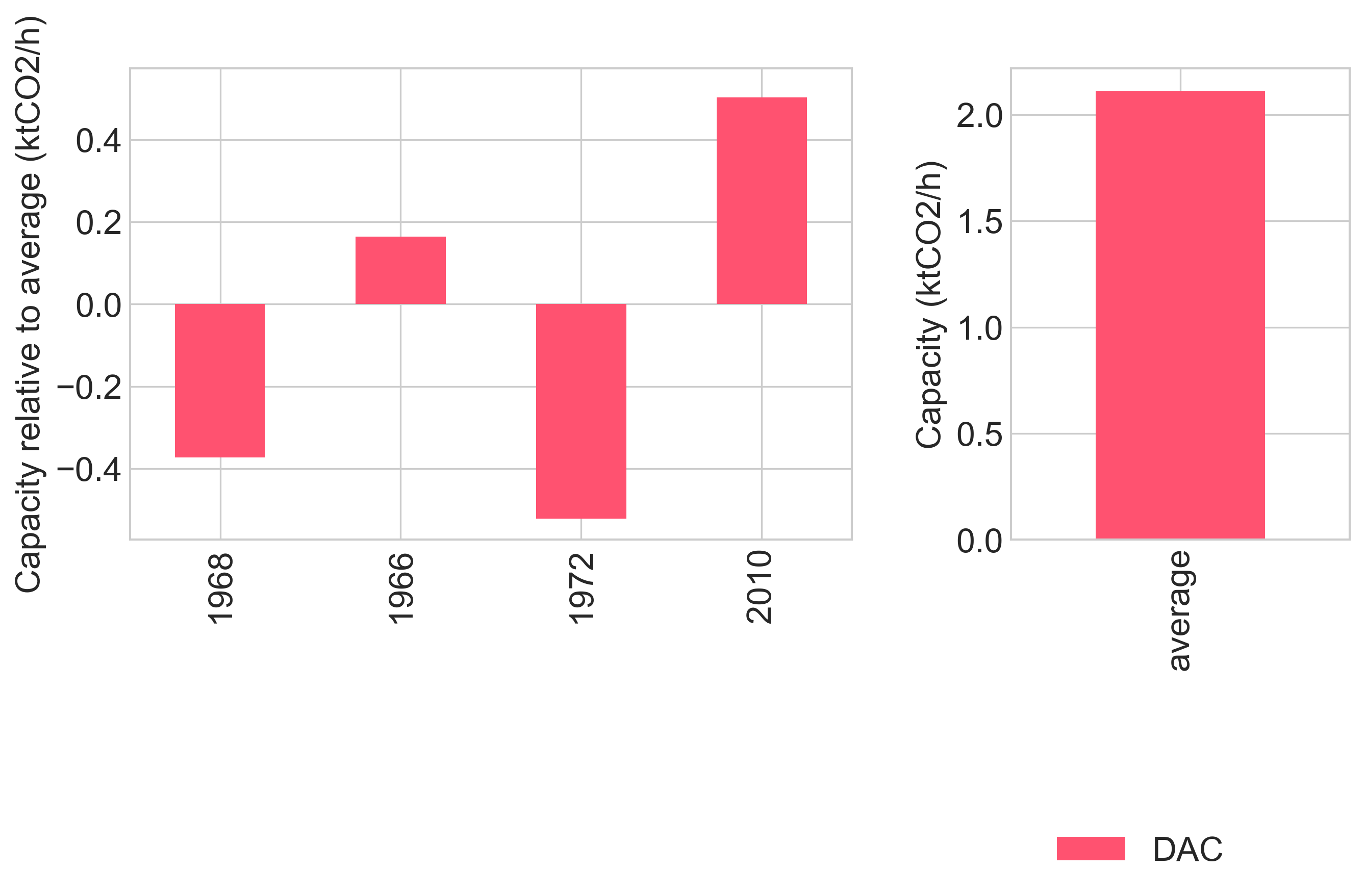}
	\captionsetup{width=14.8cm}
	\caption{European aggregated capacities to capture CO$_2$ from the atmosphere with Direct Air Capture (DAC), illustrated for (right) the average of all design years, and (left) three systems designed with extreme operational years (1966, 1968, and 1972) and the most expensive system (2010).}
	\label{sfig:Eurupe_aggregate_capacity_DAC}
\end{figure}

\newpage
\subsection{Nodal capacities in selected design years}

\textbf{Renewable generation}
\begin{figure}[!h]
	\centering
	\includegraphics[width=0.85\textwidth]{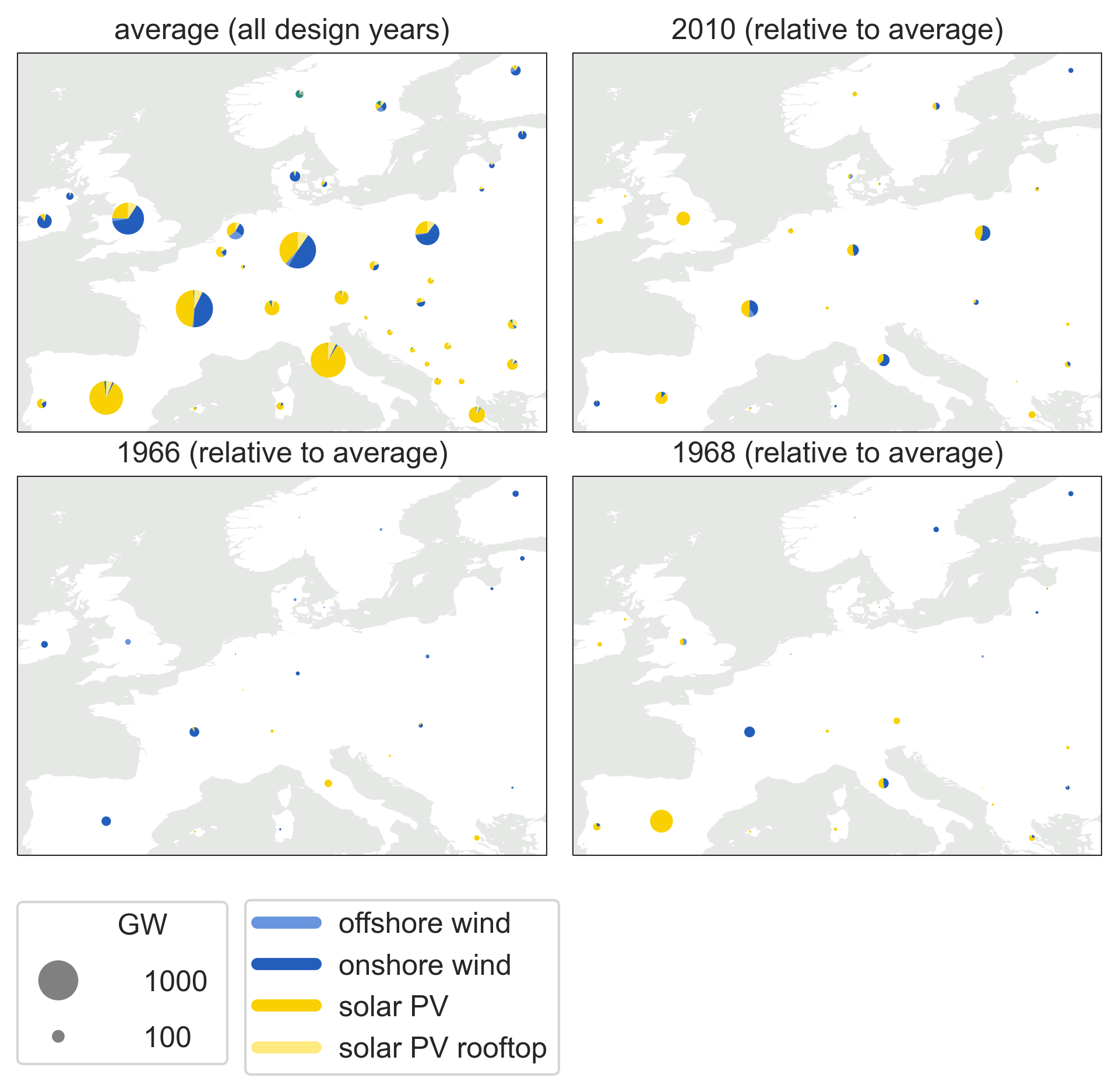}
	\captionsetup{width=14.8cm}
	\caption{Nodal capacities of renewable generation technologies, illustrated for the average of all design years, the two systems with less loss of load (1966 and 1968), and the most expensive system (2010).}
	\label{sfig:renewable_generation_capacity}
\end{figure}

\newpage
\textbf{H$_2$ electrolysis}
\begin{figure}[!h]
	\centering
	\includegraphics[width=0.85\textwidth]{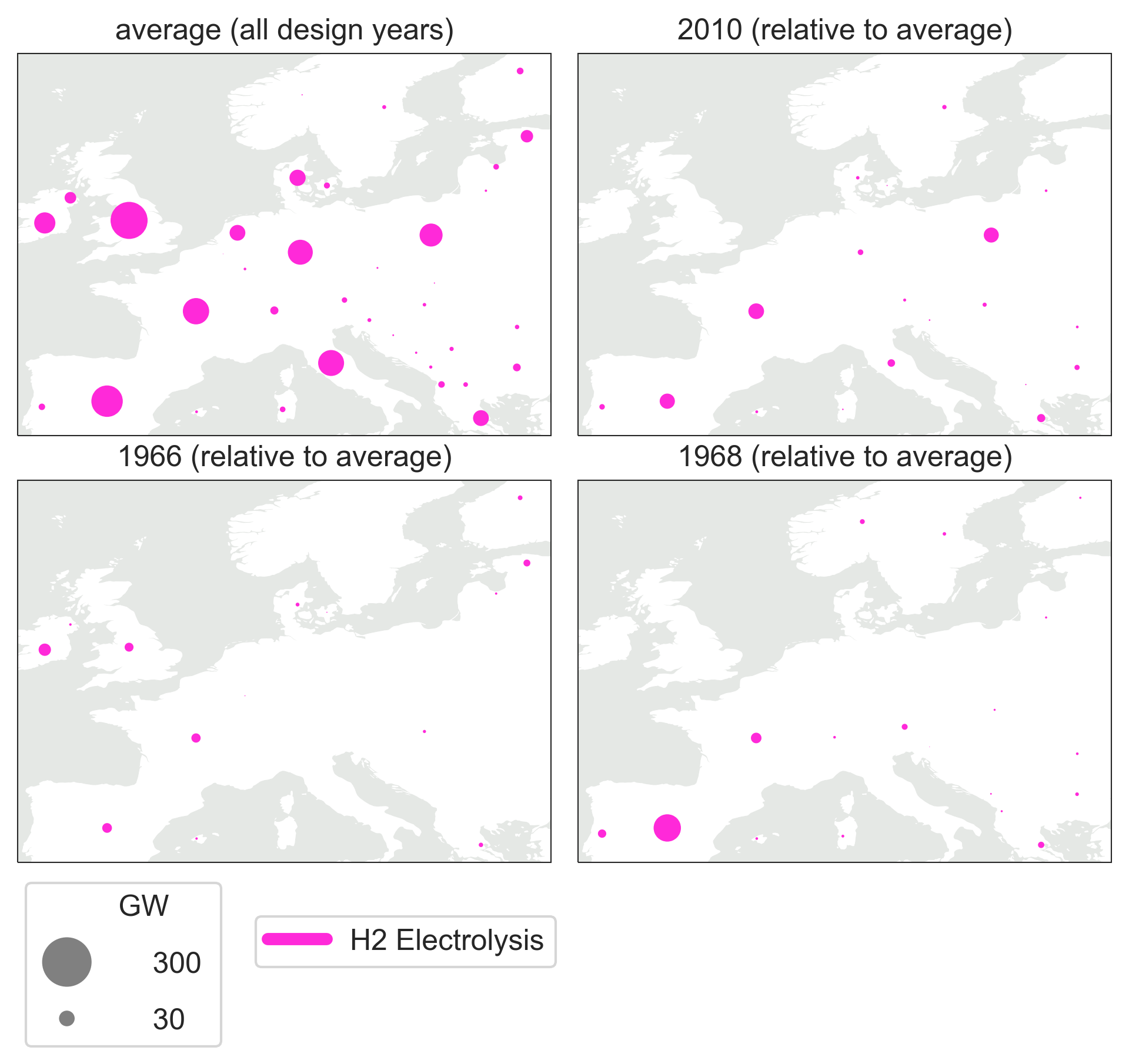}
	\captionsetup{width=14.8cm}
	\caption{Nodal capacities of H$_2$ electrolysis, illustrated for the average of all design years, the two systems with less loss of load (1966 and 1968), and the most expensive system (2010).}
	\label{sfig:H2_electrolysis_maps}
\end{figure}

\newpage
\textbf{Electricity storage (discharge)}
\begin{figure}[!h]
	\centering
	\includegraphics[width=0.85\textwidth]{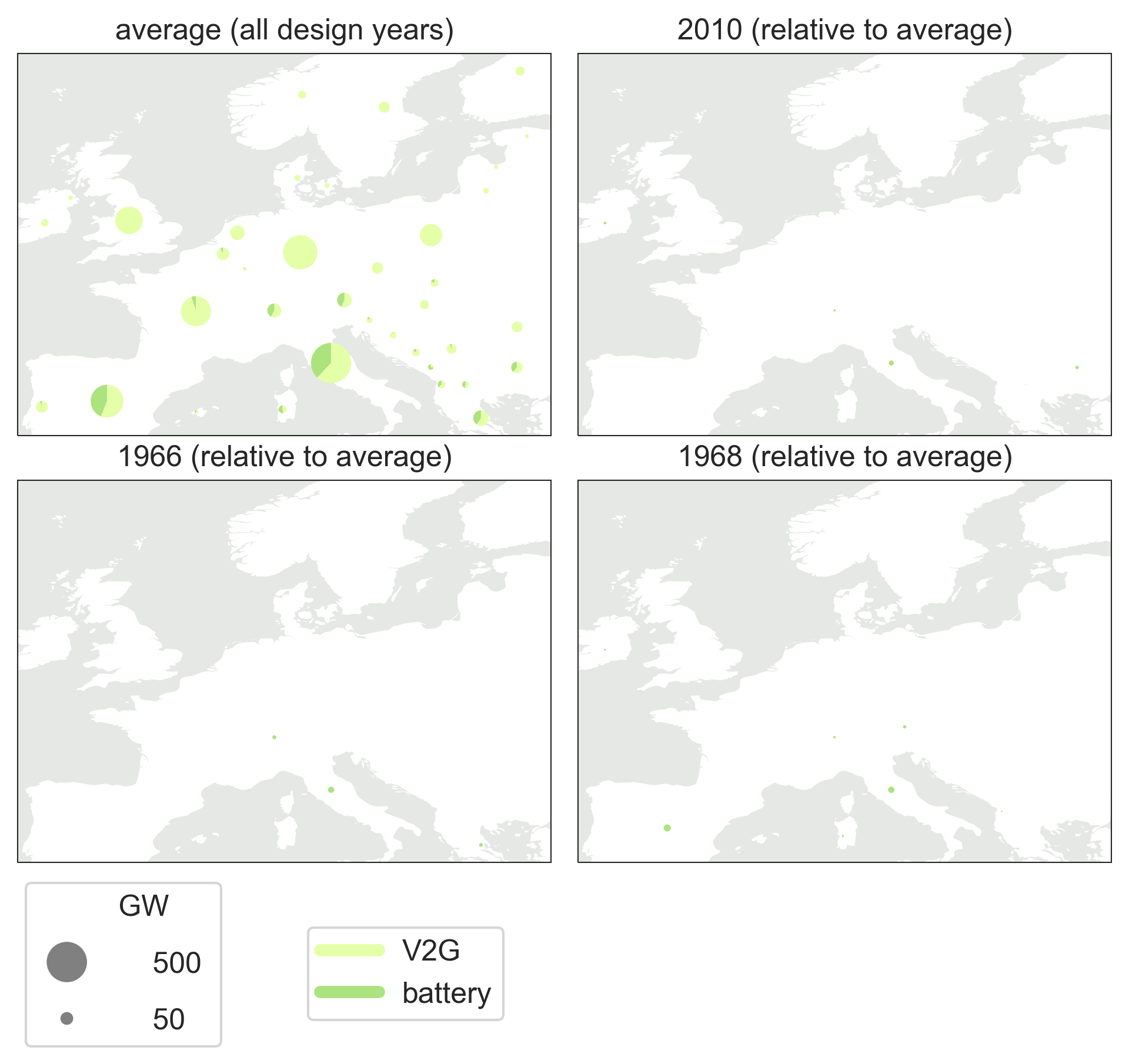}
	\captionsetup{width=14.8cm}
	\caption{Nodal discharge capacities of battery storage, illustrated for the average of all design years, the two systems with less loss of load (1966 and 1968), and the most expensive system (2010).}
	\label{sfig:storage_discharge_maps}
\end{figure}

\newpage
\textbf{Decentral heating technologies}
\begin{figure}[!h]
	\centering
	\includegraphics[width=0.85\textwidth]{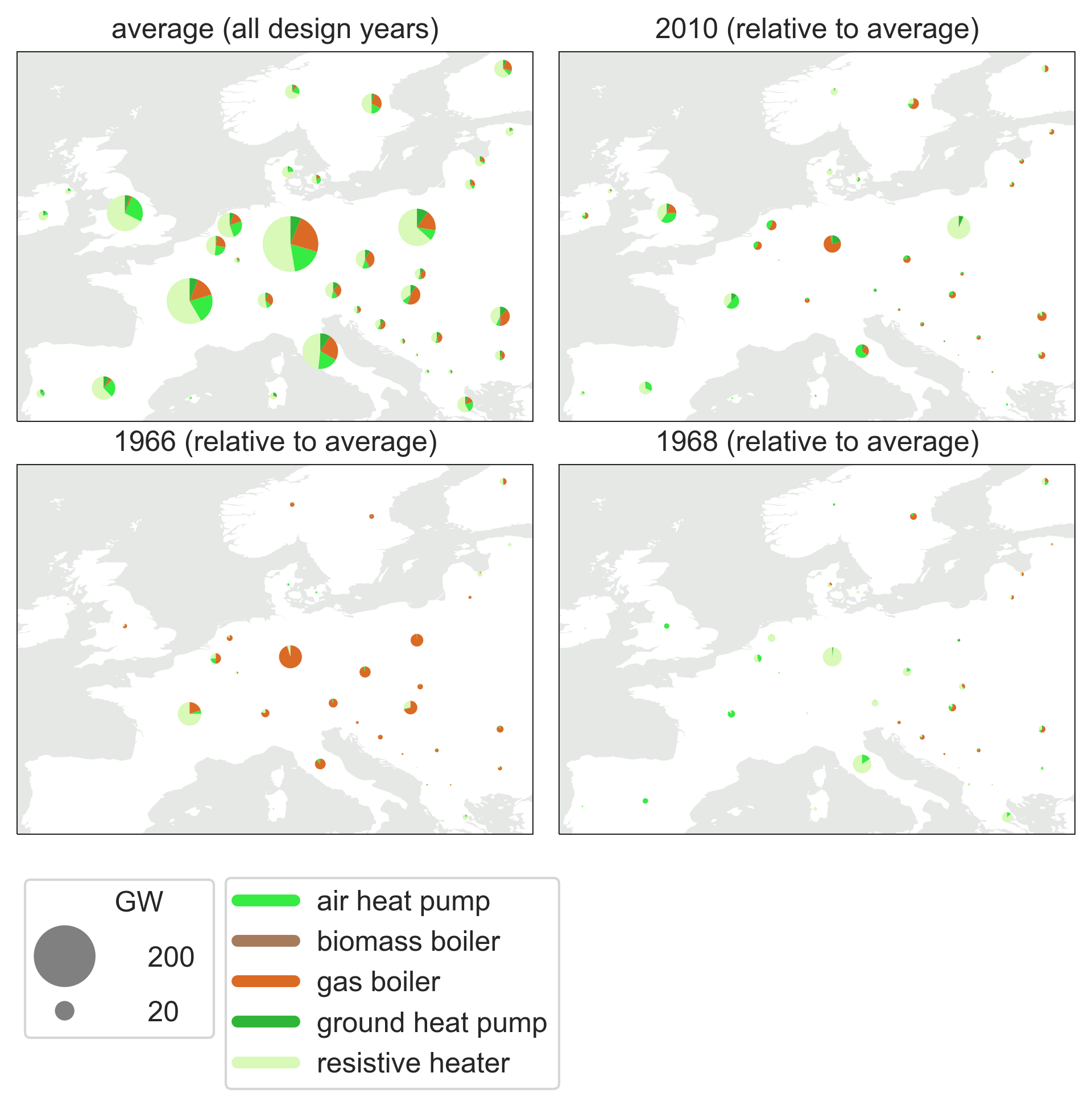}
	\captionsetup{width=14.8cm}
	\caption{Nodal capacities of decentral heating options, illustrated for the average of all design years, the two systems with less loss of load (1966 and 1968), and the most expensive system (2010).}
	\label{sfig:Heating_maps}
\end{figure}

\newpage
\textbf{Hot water storage}
\begin{figure}[!h]
	\centering
	\includegraphics[width=0.85\textwidth]{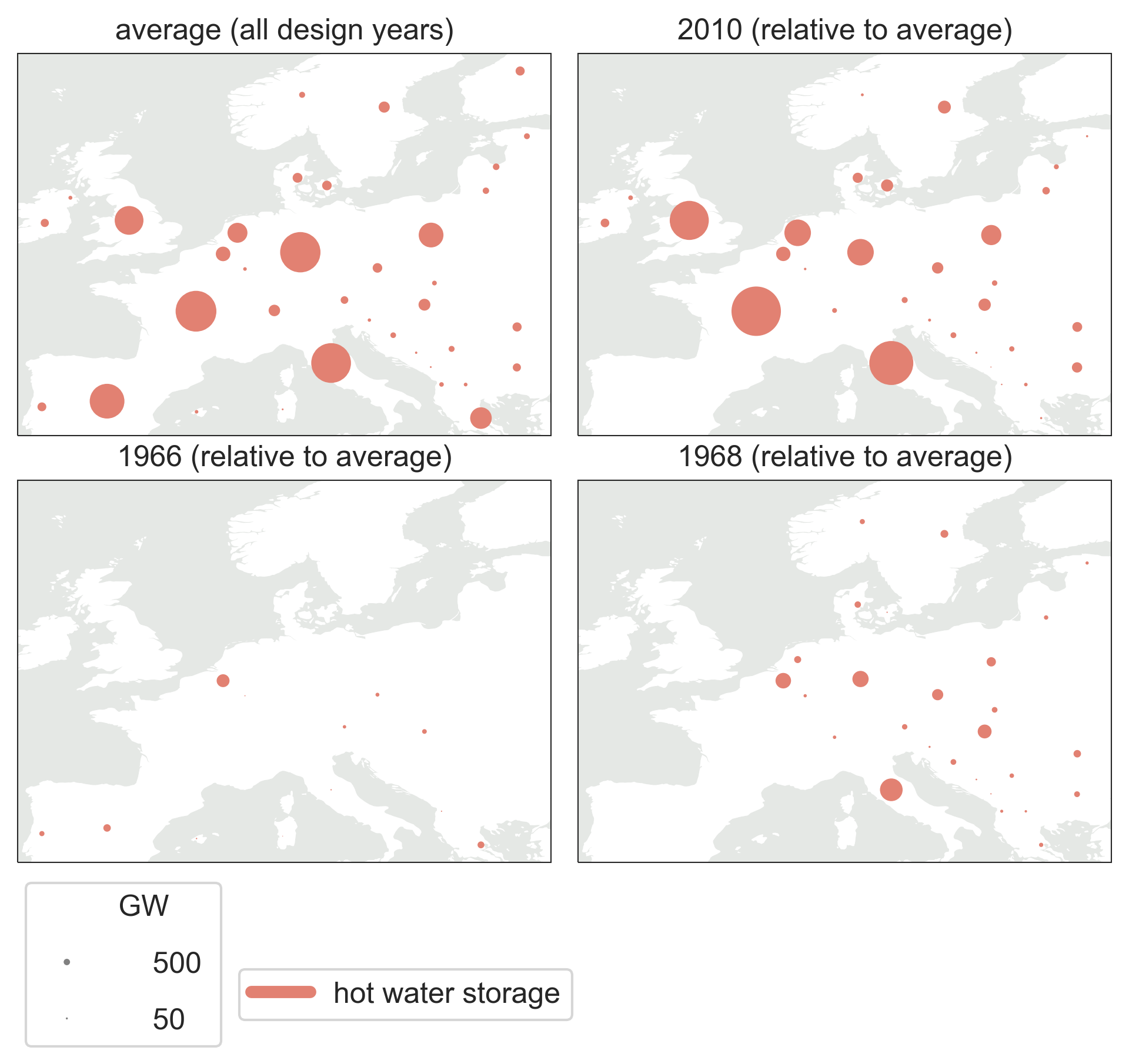}
	\captionsetup{width=14.8cm}
	\caption{Nodal capacities of hot water storage tanks, illustrated for the average of all design years, the two systems with less loss of load (1966 and 1968), and the most expensive system (2010).}
	\label{sfig:hot_water_storage}
\end{figure}

\newpage
\textbf{Direct air capture}
\begin{figure}[!h]
	\centering
	\includegraphics[width=0.85\textwidth]{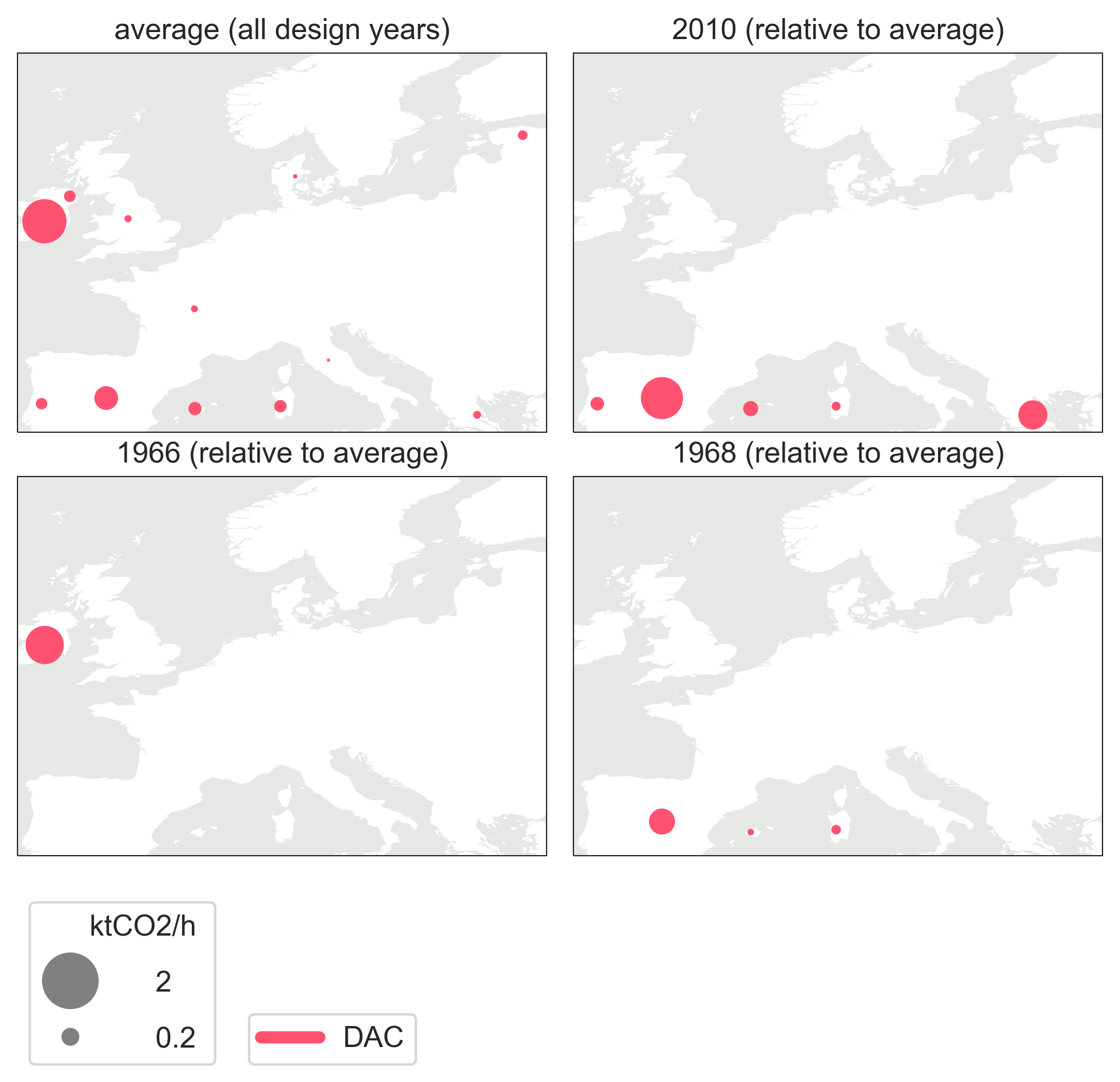}
	\captionsetup{width=14.8cm}
	\caption{Nodal capacities of Direct Air Capture (DAC), illustrated for the average of all design years, the two systems with less loss of load (1966 and 1968), and the most expensive system (2010).}
	\label{sfig:DAC}
\end{figure}

\newpage
\subsection{CO$_2$ emissions by technology}
\begin{figure}[!h]
	\centering
	\includegraphics[width=0.8\textwidth]{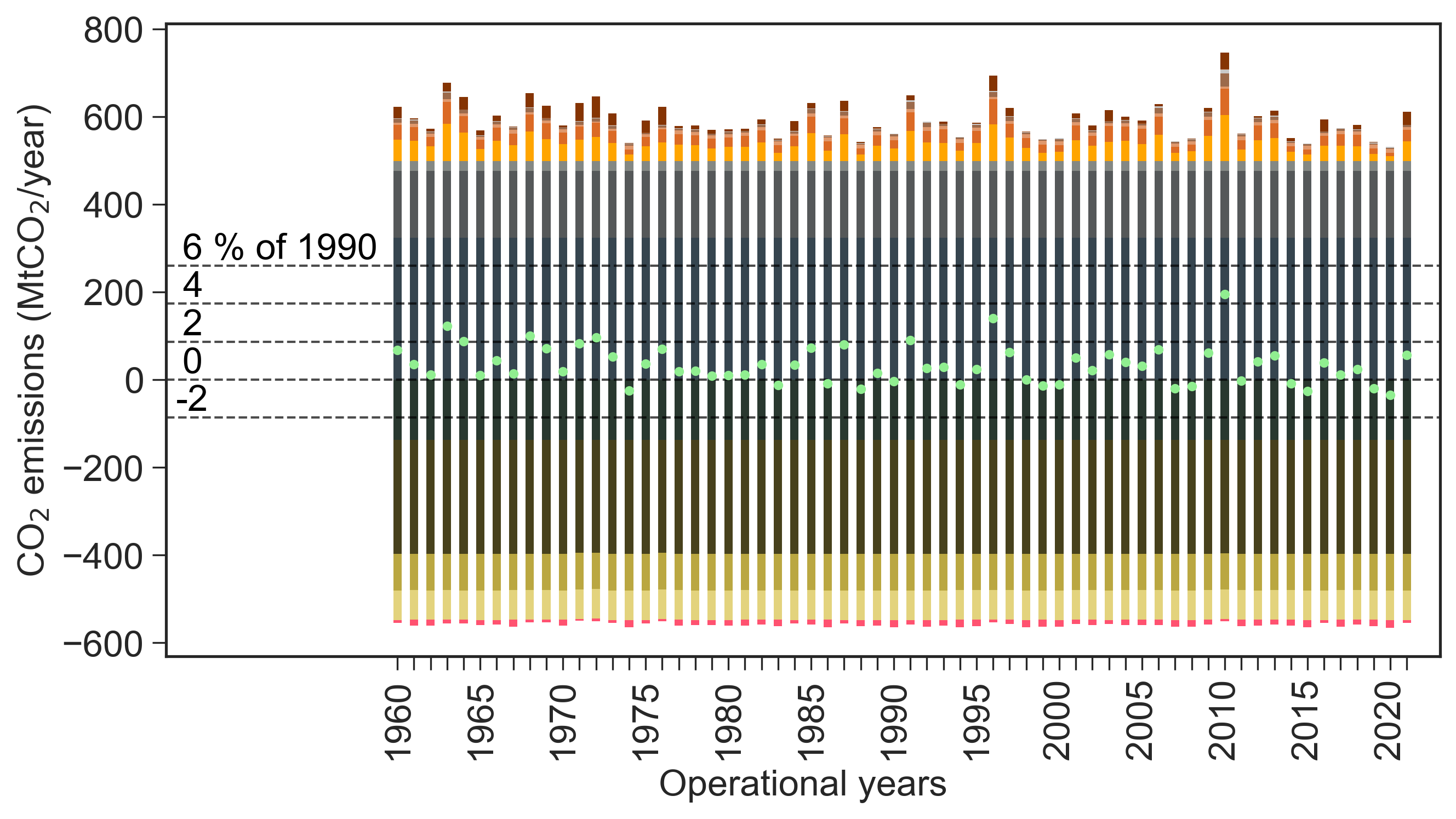}
	\includegraphics[width=0.8\textwidth]{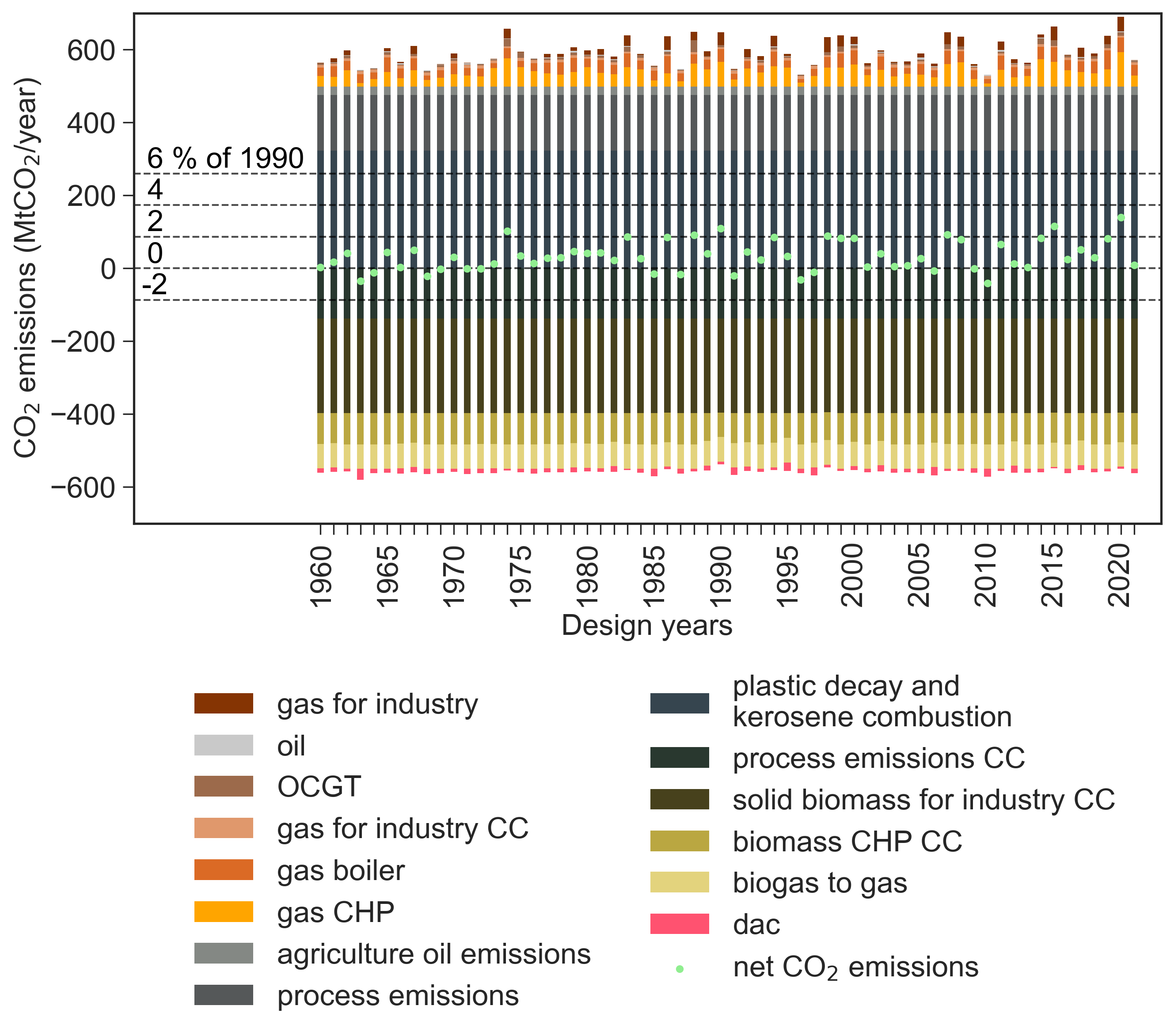}
	\captionsetup{width=14.8cm}
	\caption{CO$_2$ emissions split by technology obtained on average for (top) all operational years and (bottom) all capacity layouts (design years).}
	\label{sfig:CO2_emissions}
\end{figure}

\newpage
\subsection{CO$_2$ emissions temporal distribution}
\begin{figure}[!h]
	\centering
	\includegraphics[width=0.85\textwidth]{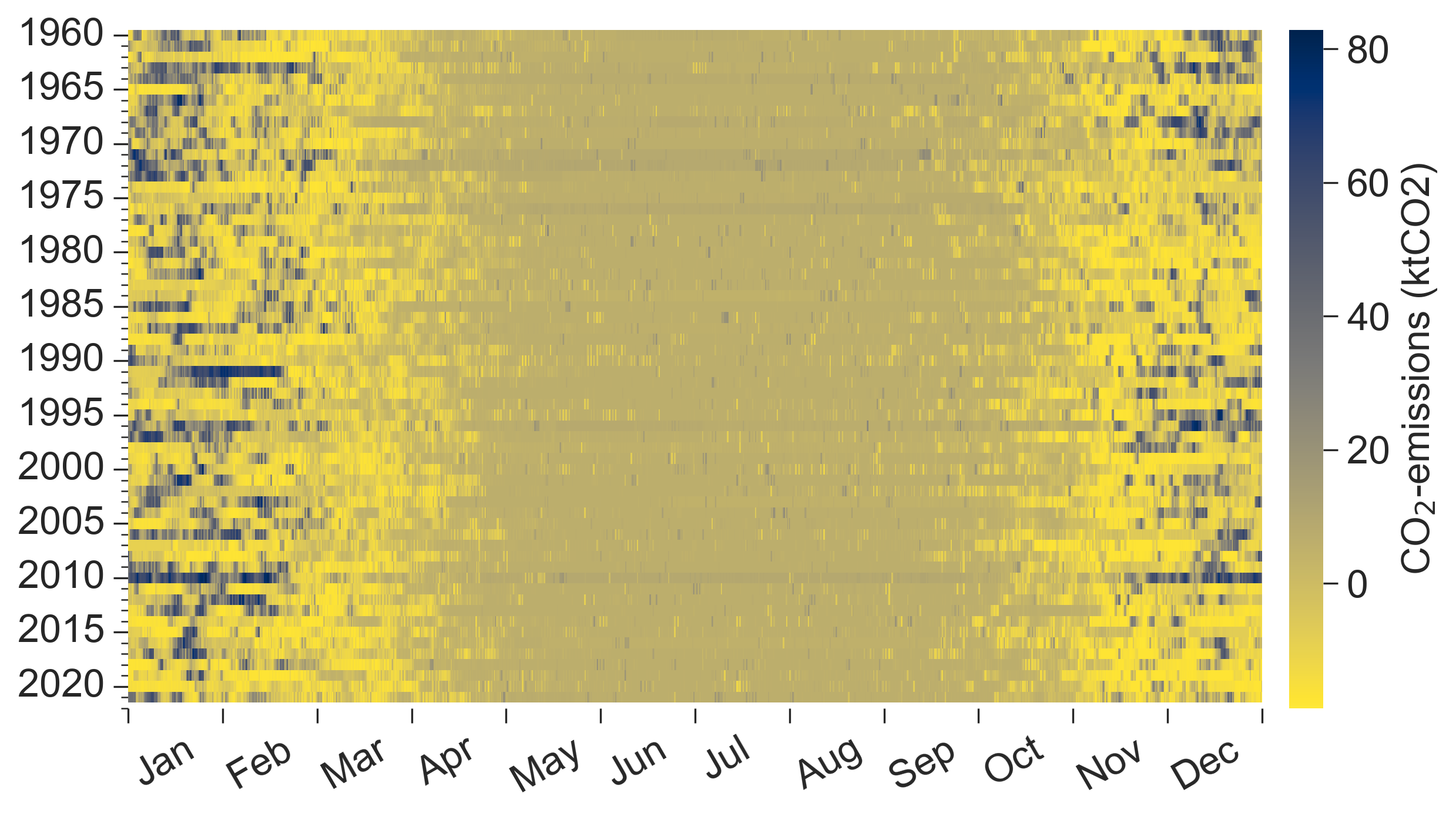}
	\captionsetup{width=14.8cm}
	\caption{Hourly net CO$_2$ emissions for the 2013 design year, which corresponds to a total system cost close to the average. Biomass CHP with carbon capture runs during the heating season, causing more hours with net-negative CO$_2$ emissions to appear during winter.}
	\label{sfig:CO2_emissions_t}
\end{figure}

\newpage
\subsection{Hydropower filling level constraint}
\textbf{Unconstrained operation}
\begin{figure}[!h]
	\centering
	\includegraphics[width=0.85\textwidth]{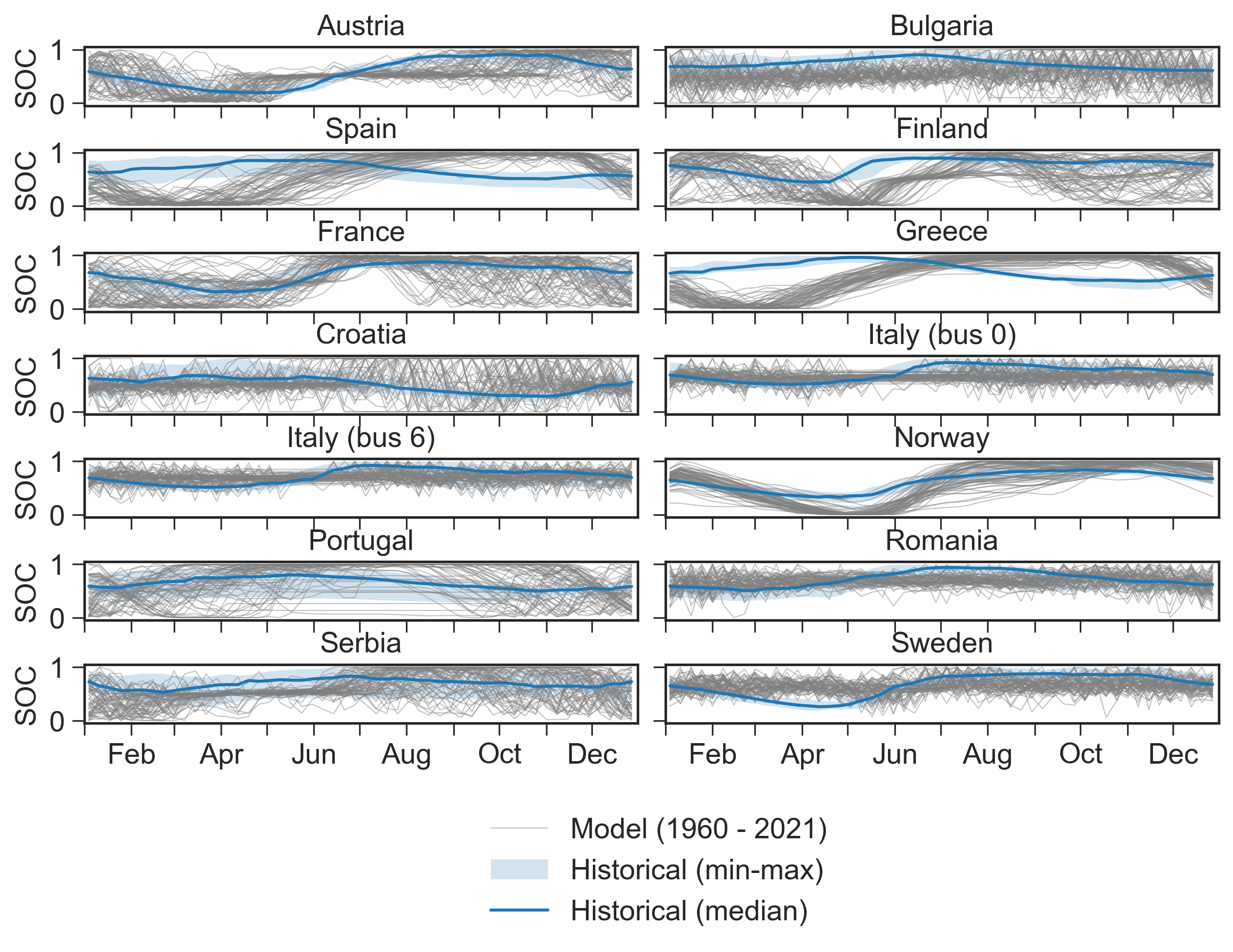}
	\captionsetup{width=14.8cm}
	\caption{State-of-charge (SOC) of country-aggregated hydro reservoirs for which historical data is available from ENTSO-E. The gray lines correspond to unconstrained with weather data from 1960 to 2021.}
	\label{sfig:hydro_constraint}
\end{figure}

\newpage
\textbf{Constrained operation}
\begin{figure}[!h]
	\centering
	\includegraphics[width=0.85\textwidth]{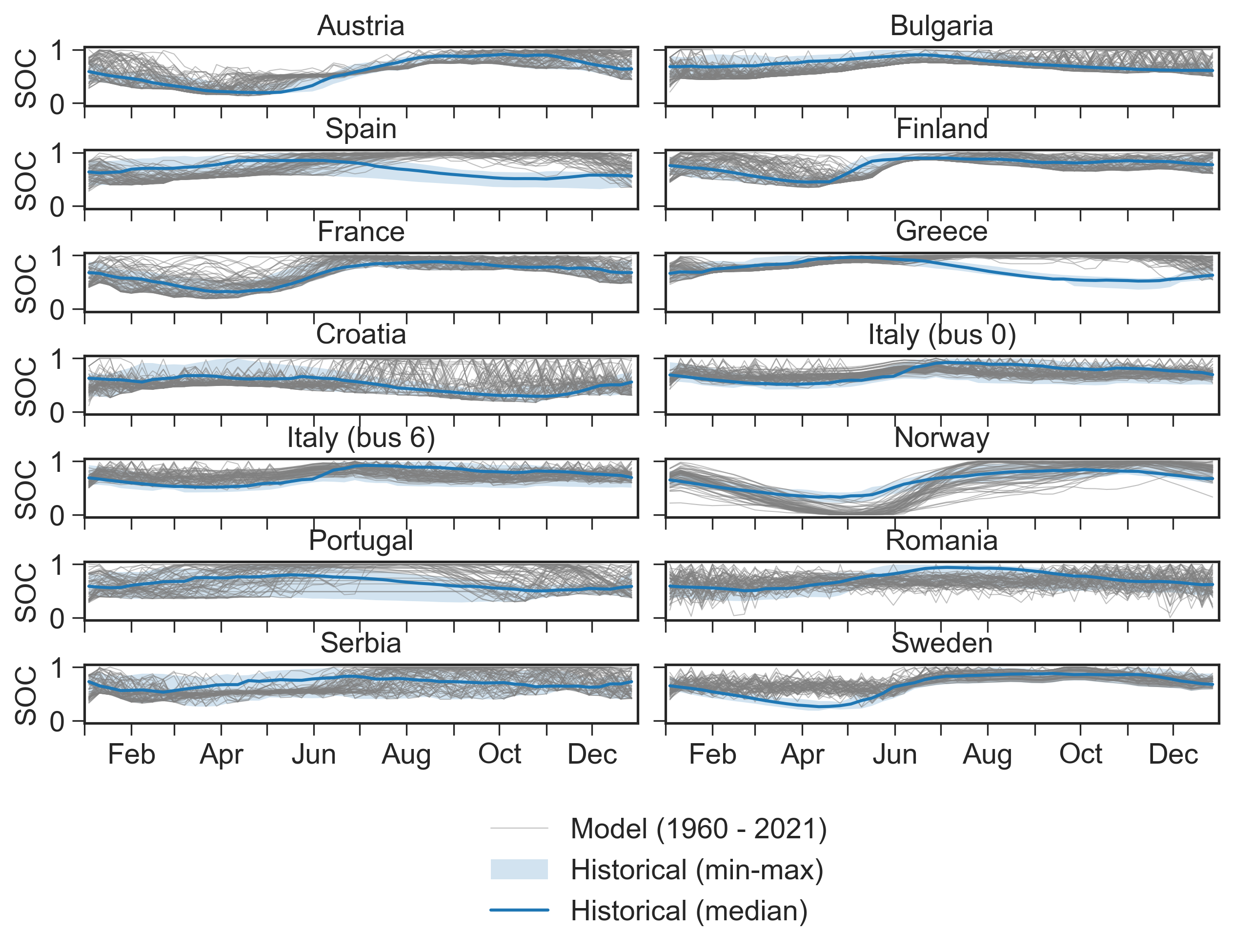}
	\captionsetup{width=14.8cm}
	\caption{\textbf{State-of-charge (SOC) of country-aggregated hydro reservoirs} for which historical data is available from ENTSO-E. The gray lines correspond to the modeled PyPSA-Eur capacity-optimization scenarios with weather data from 1960 to 2021 with the imposed hydropower constraint.}
	\label{sfig:hydro_constraint_result}
\end{figure}

\newpage
\subsection{Capacities}
\textbf{Electricity generation capacity}
\begin{figure}[!h]
	\centering
	\includegraphics[width=0.85\textwidth]{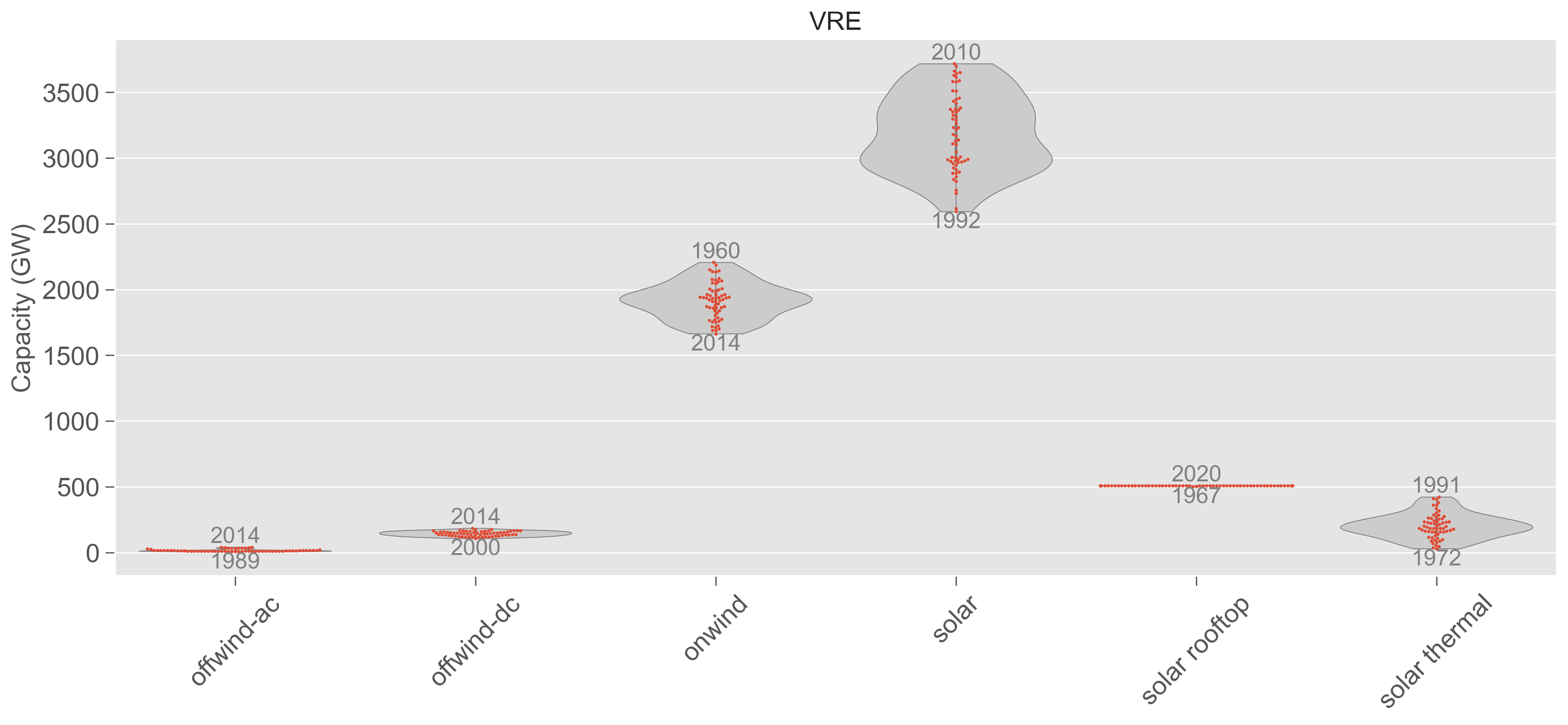}
	\captionsetup{width=14.8cm}
	\caption{Capacity of VRE generation technology. Each red dot represents a design year. Indicated below the first axis is the Pearson's correlation coefficient and p-value between the variation in the capacity and the cumulative unserved energy (a negative value means that it can be associated with less unserved energy).}
	\label{sfig:capacity_VRE}
\end{figure}

\begin{figure}[!h]
	\centering
	\includegraphics[width=0.85\textwidth]{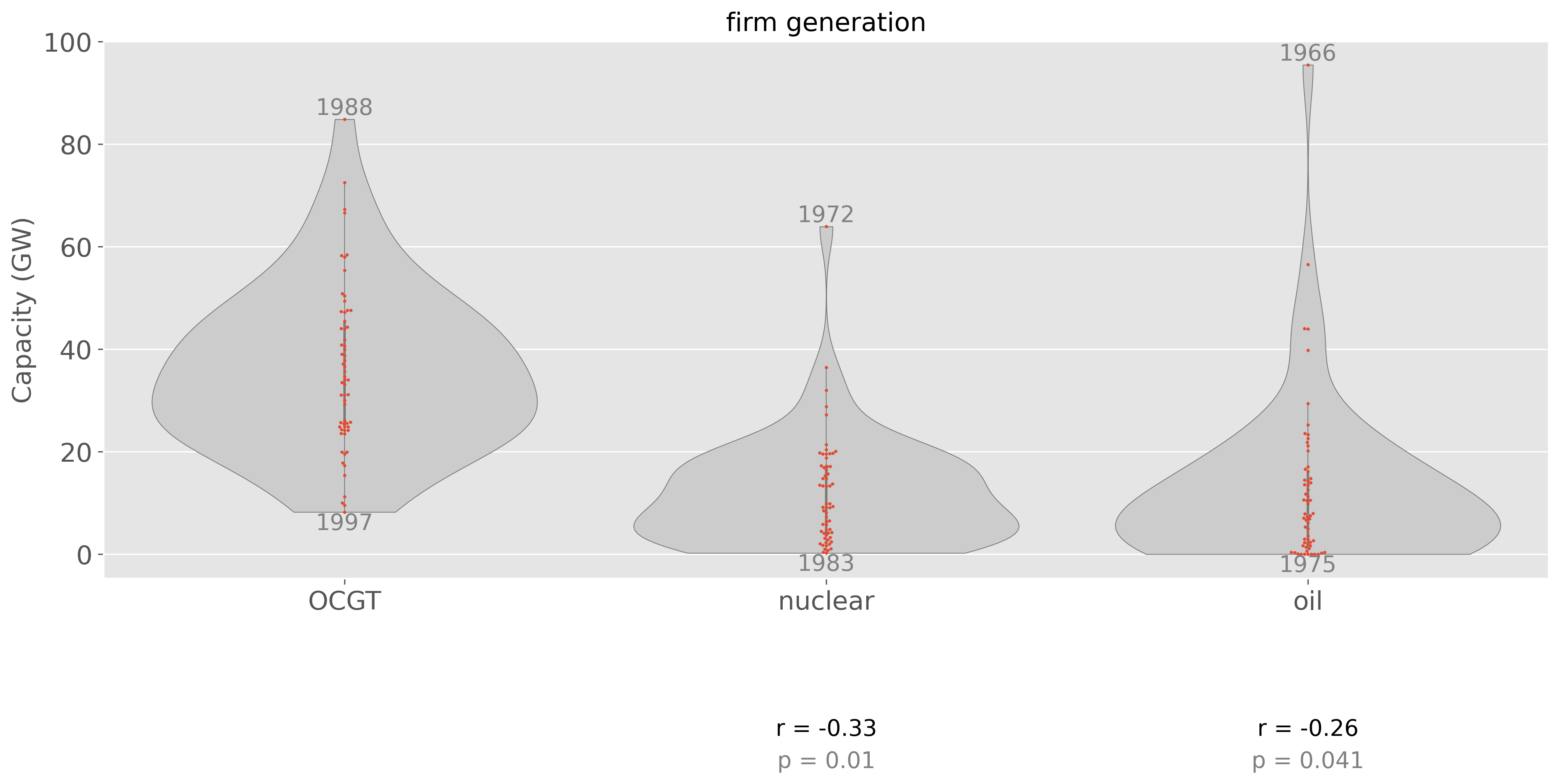}
	\captionsetup{width=14.8cm}
	\caption{Capacity of firm electricity generation. The capacity is reported in units of electricity output. Each red dot represents a design year.}
	\label{sfig:capacity_firm_generation}
\end{figure}

%  Indicated below the first axis is the significant Pearson's correlation coefficients (with a 5\% significance level) between the variation in the capacity and the cumulative unserved energy (a negative value means that it can be associated with less unserved energy).

\begin{figure}[!h]
	\centering
	\includegraphics[width=0.85\textwidth]{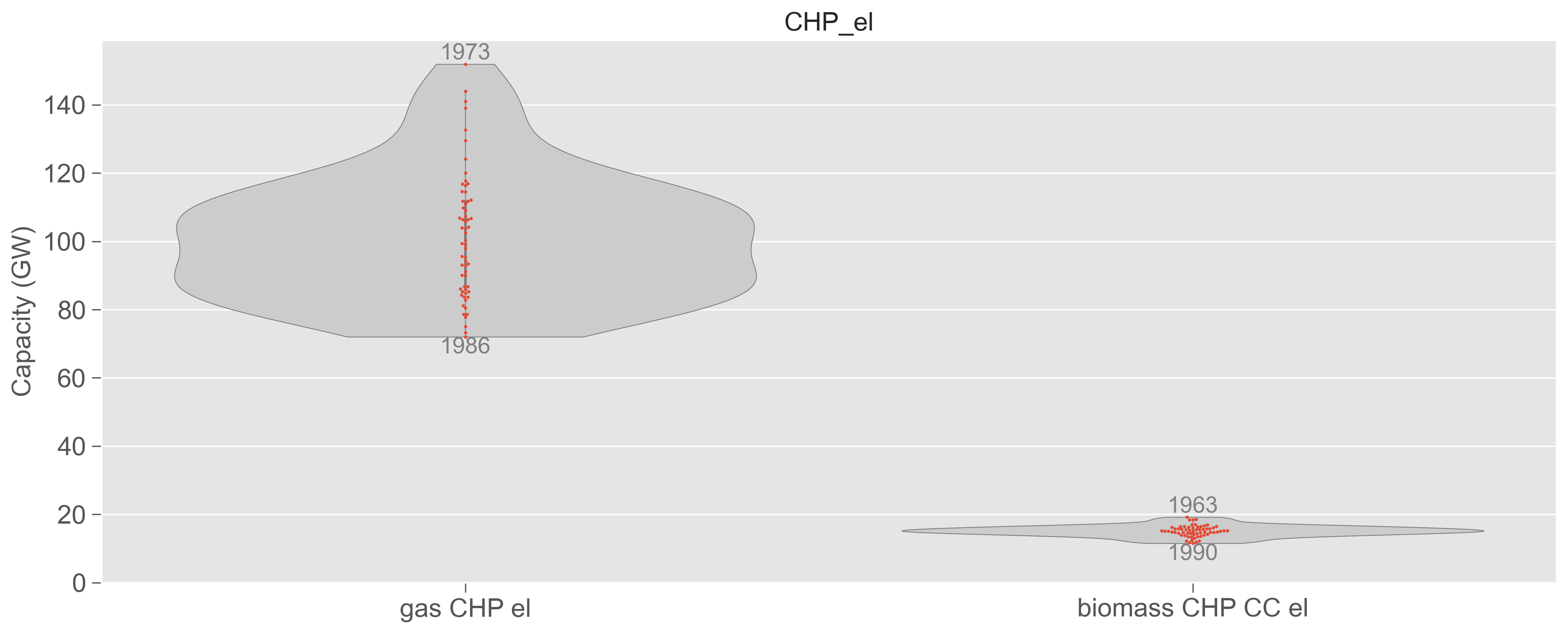}
	\captionsetup{width=14.8cm}
	\caption{Capacity of CHP plants. The capacity is reported in units of electricity output. Each red dot represents a design year.}
	\label{sfig:capacity_CHP_el}
\end{figure}

\newpage
\begin{figure}[!h]
	\centering
	\includegraphics[width=0.85\textwidth]{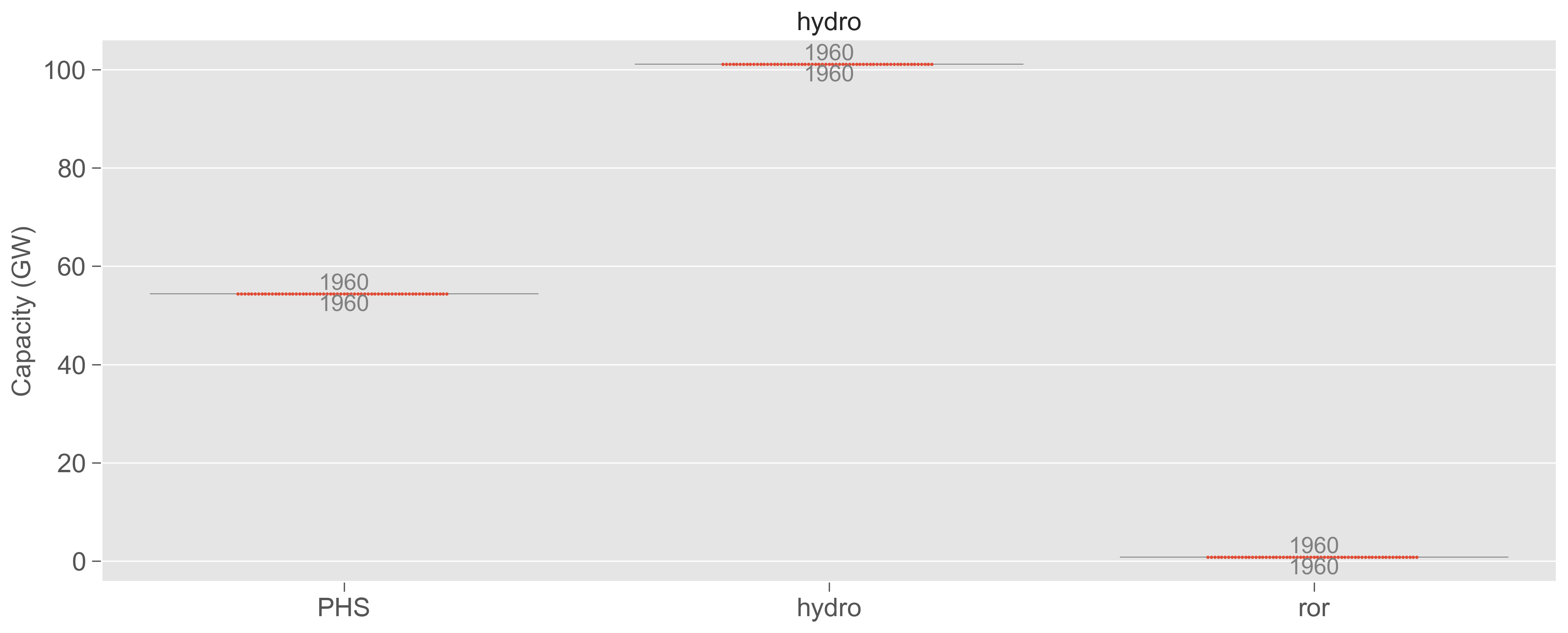}
	\captionsetup{width=14.8cm}
	\caption{Capacity of hydropower (constant for all design years).}
	\label{sfig:capacity_hydro}
\end{figure}

\newpage
\textbf{Energy storage capacity}
\begin{figure}[!h]
	\centering
	\includegraphics[width=0.85\textwidth]{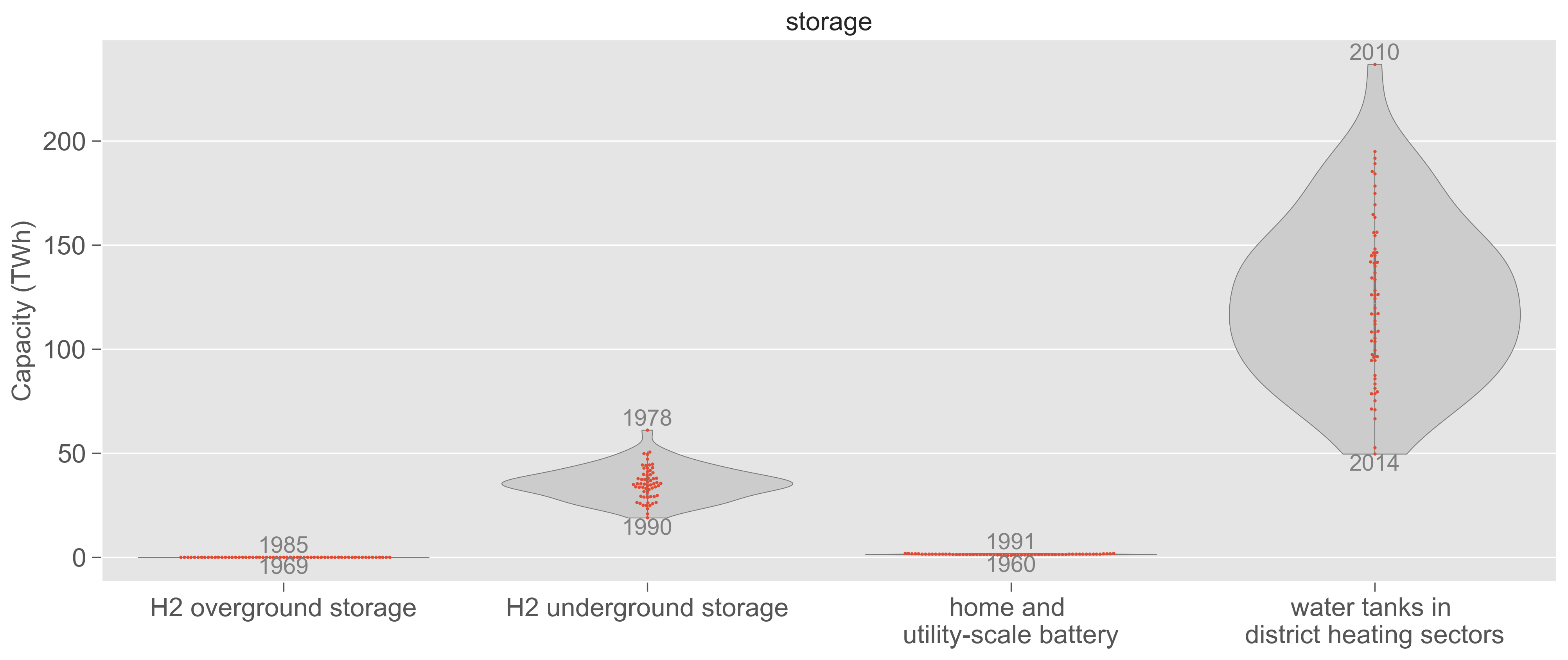}
	\captionsetup{width=14.8cm}
	\caption{Energy capacity of storage technologies. Each red dot represents a design year.}
	\label{sfig:capacity_storage}
\end{figure}

\begin{figure}[!h]
	\centering
	\includegraphics[width=0.85\textwidth]{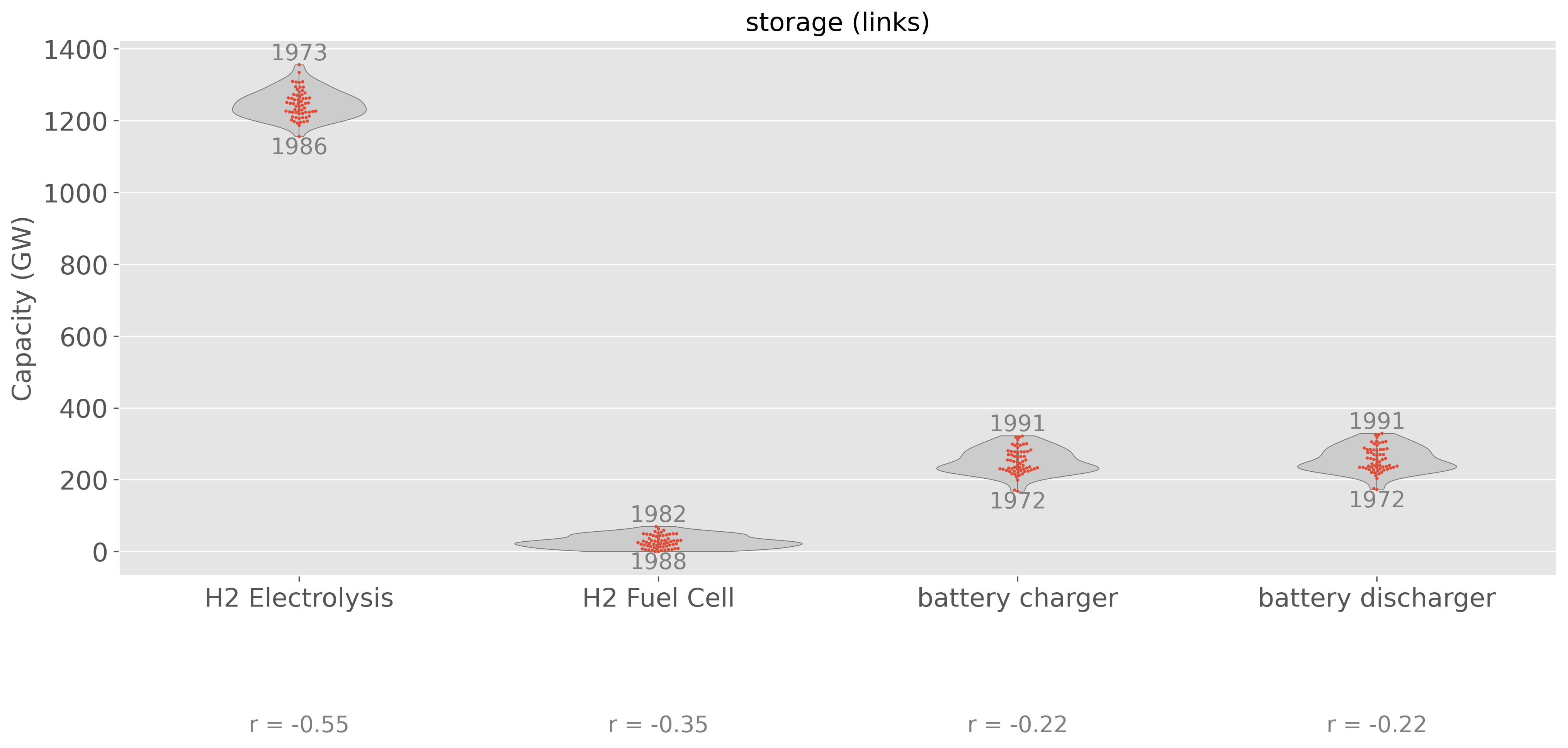}
	\captionsetup{width=14.8cm}
	\caption{Capacity of energy storage charging and discharging links. The capacity is reported in units of the output. Each red dot represents a design year.}
	\label{sfig:capacity_storage_links}
\end{figure}

\newpage
\textbf{Heating technology capacity}
\begin{figure}[!h]
	\centering
	\includegraphics[width=0.85\textwidth]{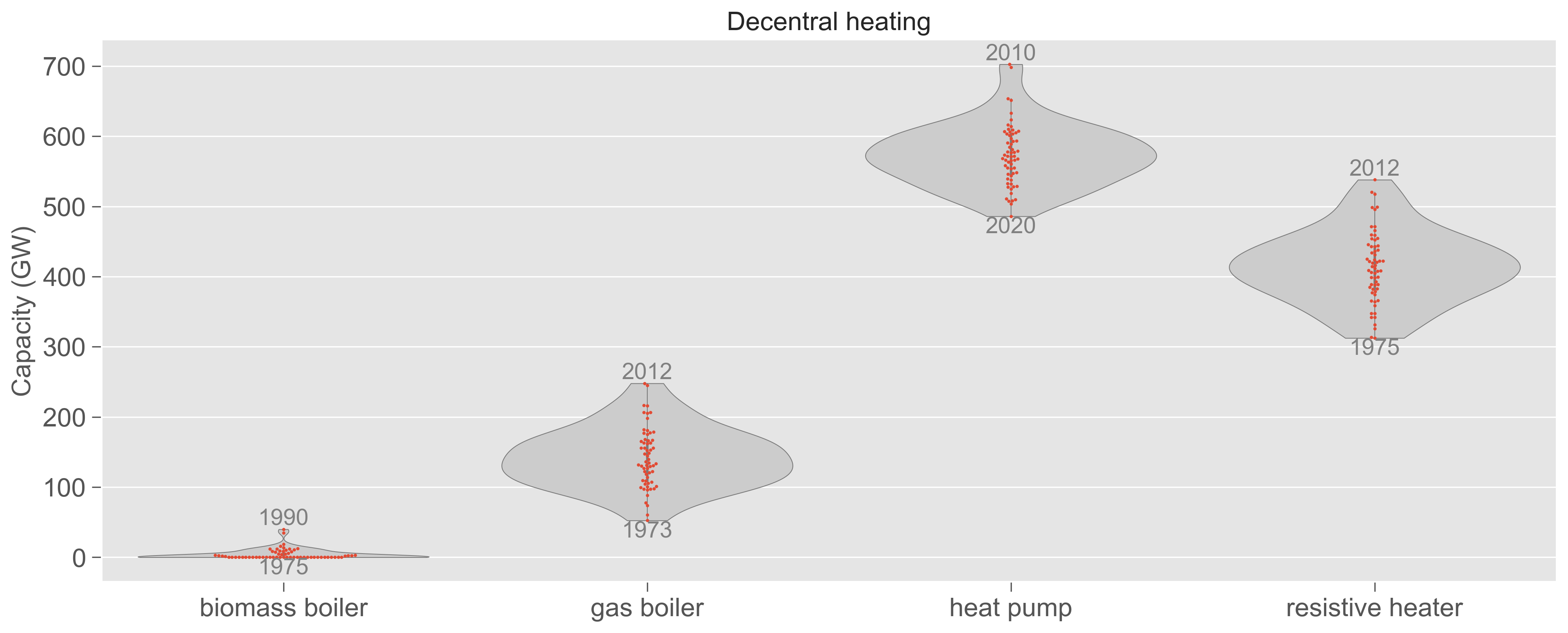}
	\captionsetup{width=14.8cm}
	\caption{Capacity of decentral heating options. The capacity is reported in units of heating output. Each red dot represents a design year.}
	\label{sfig:capacity_decentral_heating}
\end{figure}

\begin{figure}[!h]
	\centering
	\includegraphics[width=0.85\textwidth]{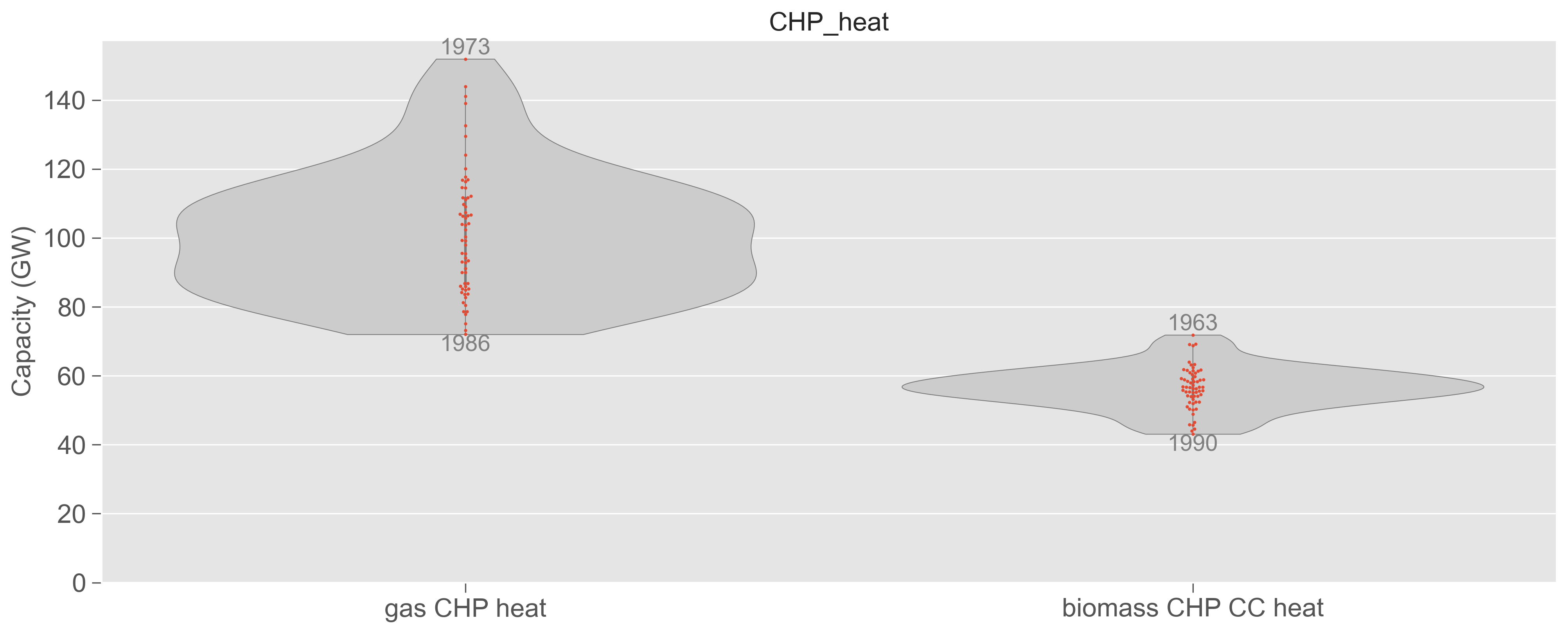}
	\captionsetup{width=14.8cm}
	\caption{Capacity of CHP plants. The capacity is reported in units of heating output. Each red dot represents a design year.}
	\label{sfig:capacity_CHP_heat}
\end{figure}

\newpage
\textbf{Synthetic fuel production capacity}
\begin{figure}[!h]
	\centering
	\includegraphics[width=0.85\textwidth]{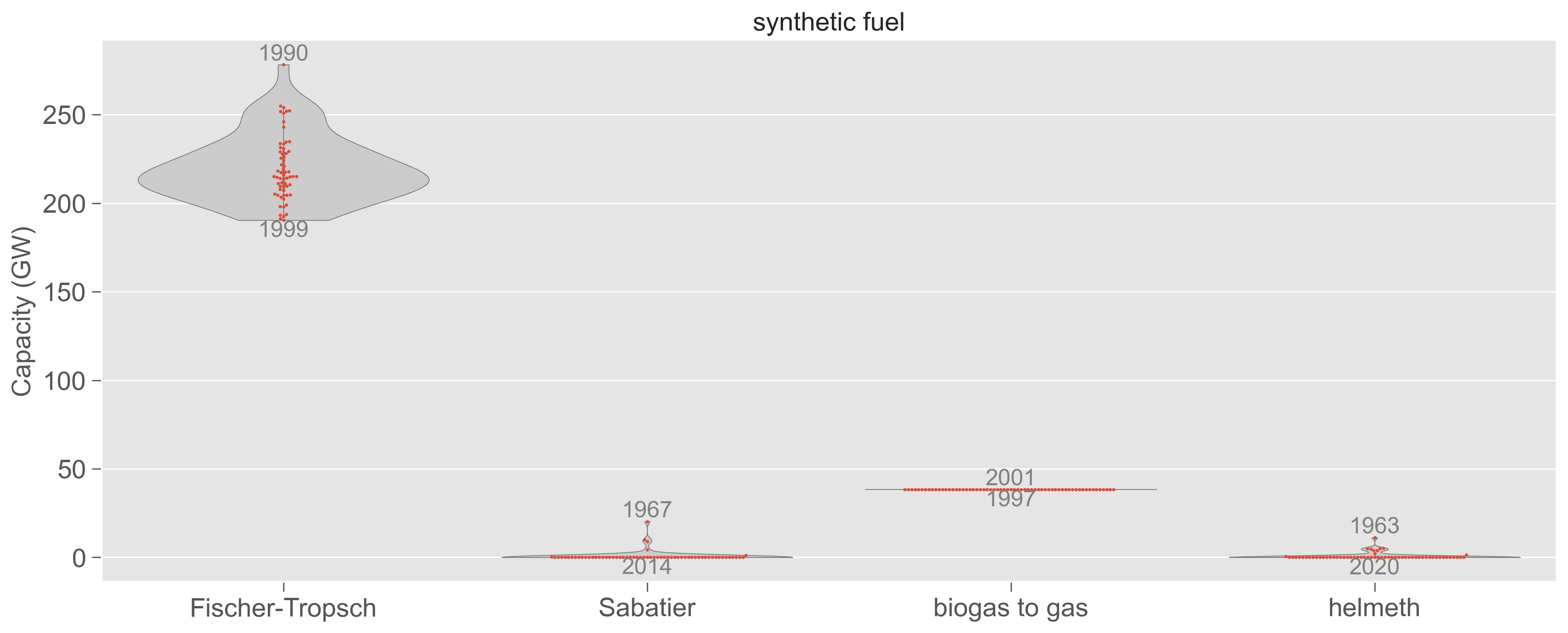}
	\captionsetup{width=14.8cm}
	\caption{Capacity of technologies for synthetic fuel production. The capacity is reported in units of the output. Each red dot represents a design year.}
	\label{sfig:capacity_synfuel}
\end{figure}

\newpage
\textbf{Carbon capture}
\begin{figure}[!h]
	\centering
	\includegraphics[width=0.85\textwidth]{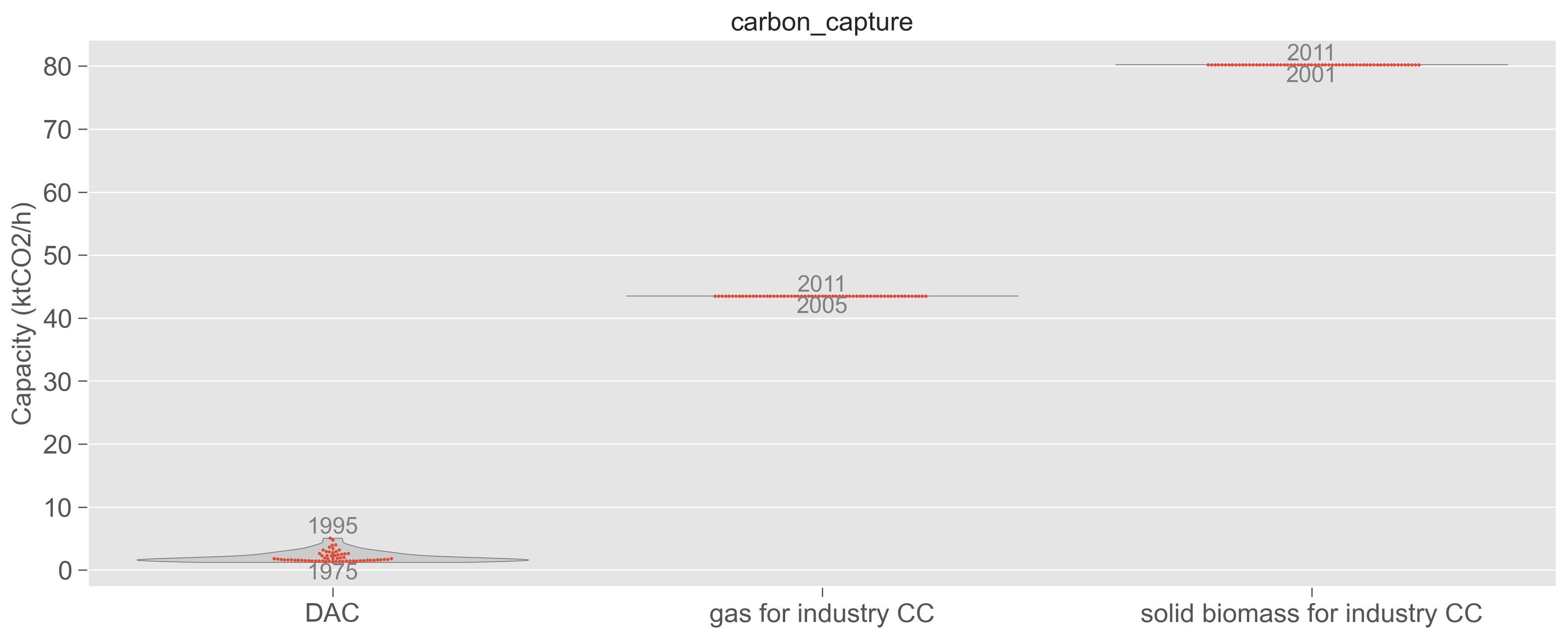}
	\captionsetup{width=14.8cm}
	\caption{Capacity of carbon capture technologies. Each red dot represents a design year.}
	\label{sfig:capacity_carbon_capture}
\end{figure}

\newpage
\textbf{Infrastructure}
\begin{figure}[!h]
	\centering
	\includegraphics[width=0.85\textwidth]{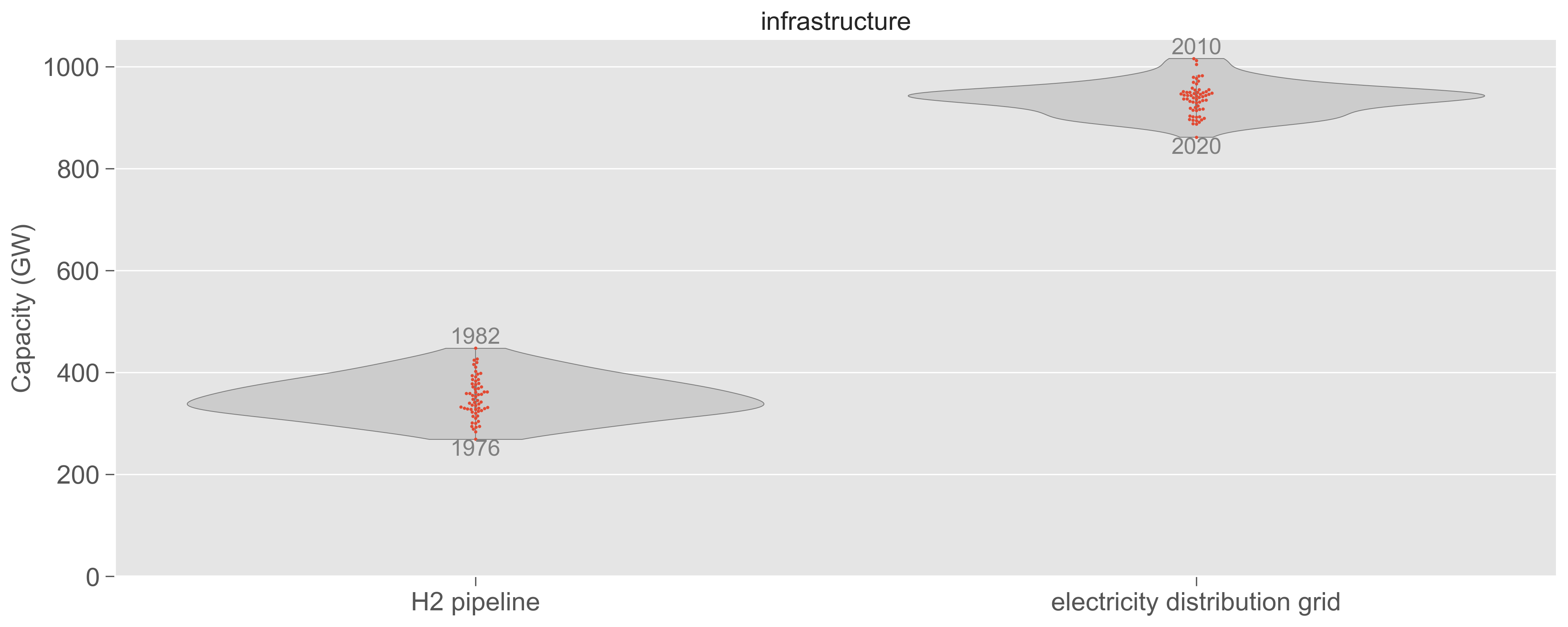}
	\captionsetup{width=14.8cm}
	\caption{Capacity of H$_2$ transmission network and electricity distribution grid. Each red dot represents a design year.}
	\label{sfig:capacity_infrastructure}
\end{figure}

\newpage
\textbf{Electric vehicle batteries}
\begin{figure}[!h]
	\centering
	\includegraphics[width=0.85\textwidth]{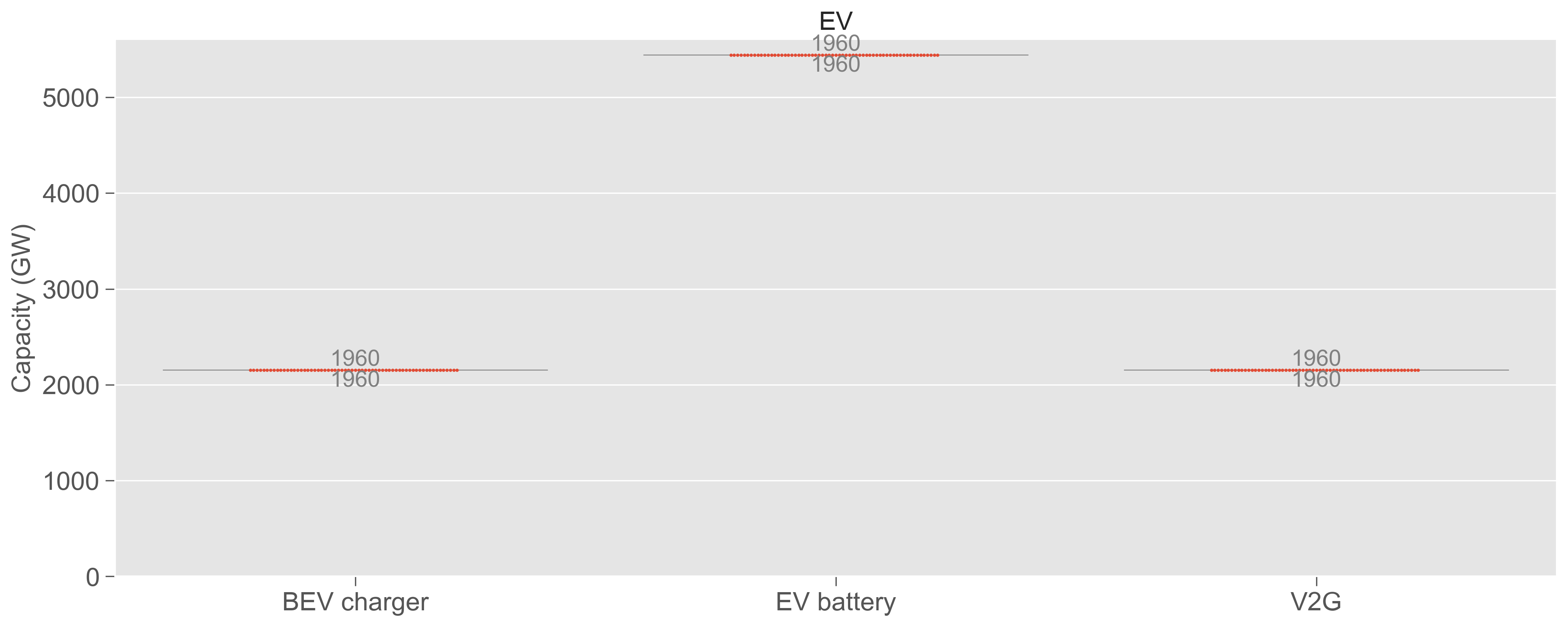}
	\captionsetup{width=14.8cm}
	\caption{Capacity of electric vehicle batteries (constant for all design years).}
	\label{sfig:capacity_EV}
\end{figure}

\newpage
\textbf{Mining or import of fossil fuels}
\begin{figure}[!h]
	\centering
	\includegraphics[width=0.85\textwidth]{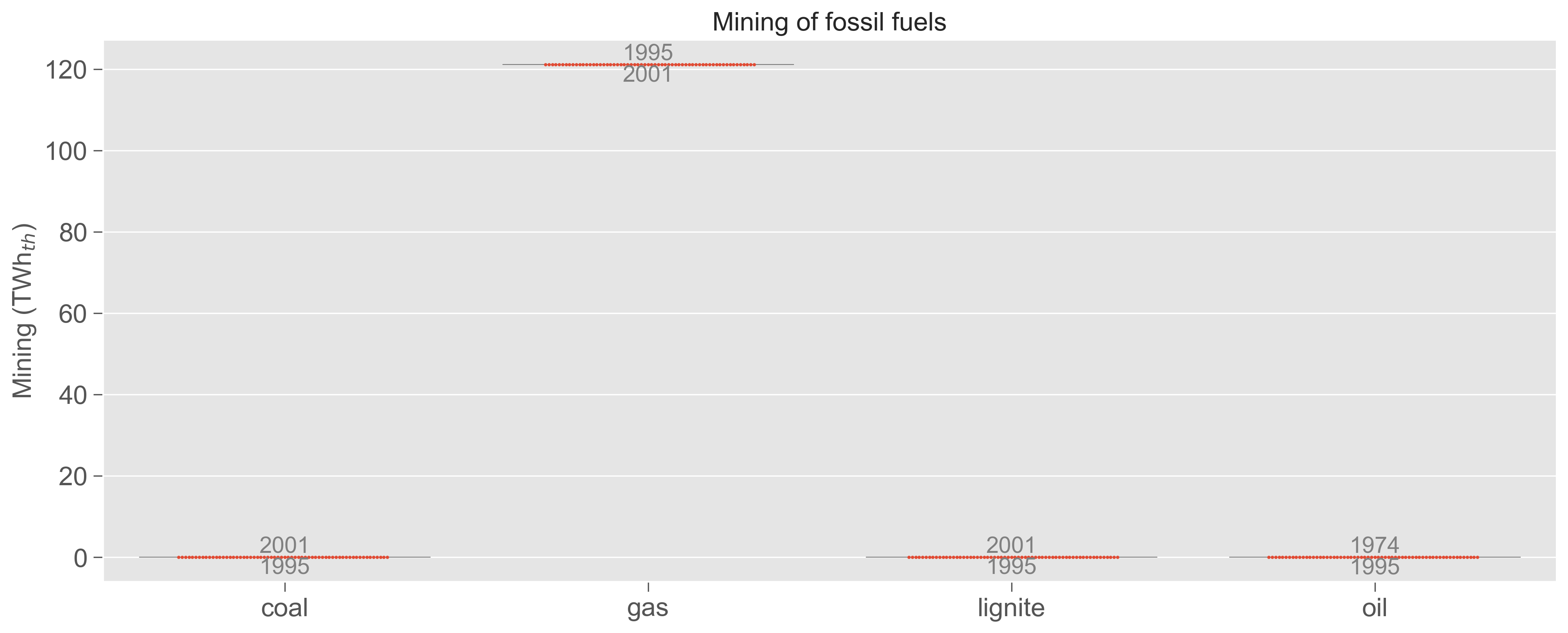}
	\captionsetup{width=14.8cm}
	\caption{Annual extraction or import of fossil fuels in Europe.}
	\label{sfig:capacity_mining}
\end{figure}

\newpage
\textbf{Biogenic and synthetic methane production}
\begin{figure}[!h]
	\centering
	\includegraphics[width=0.85\textwidth]{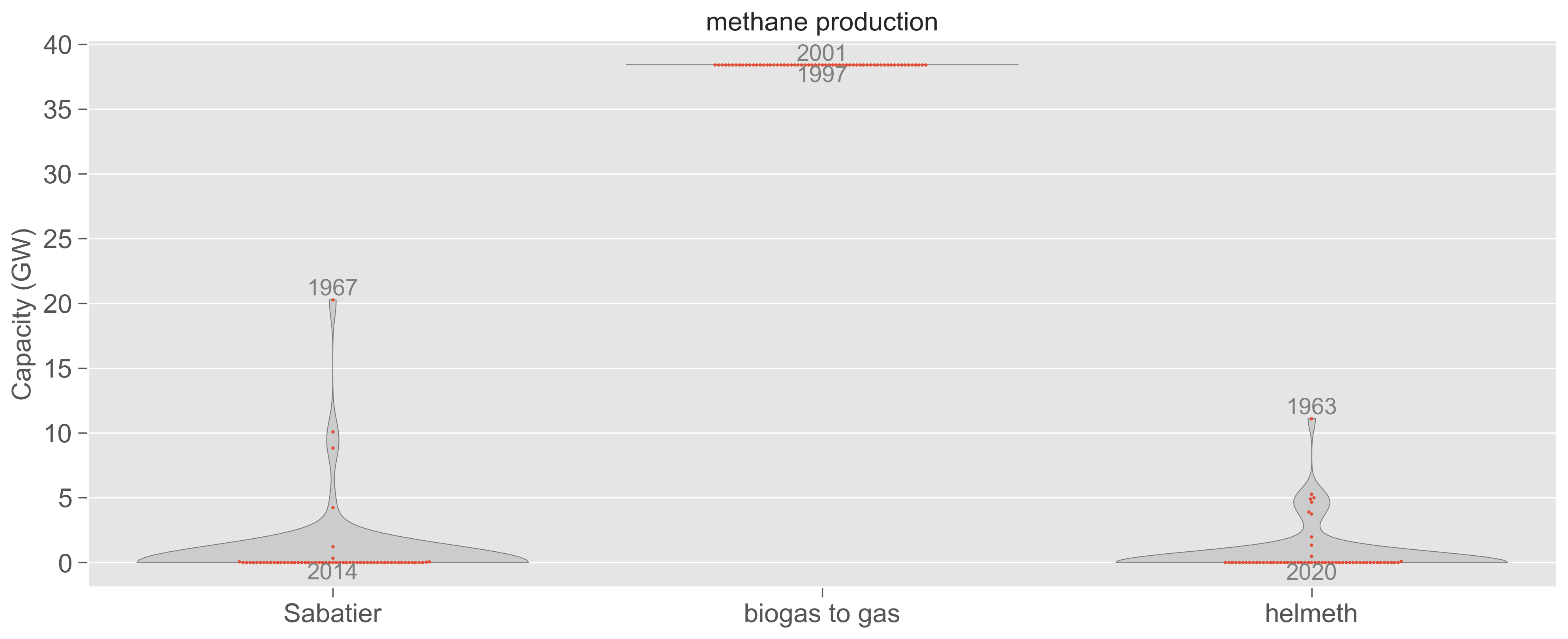}
	\captionsetup{width=14.8cm}
	\caption{Annual production of methane from Sabatier, biogas, and helmeth.}
	\label{sfig:methane_production}
\end{figure}

\newpage
\textbf{Other}
\begin{figure}[!h]
	\centering
	\includegraphics[width=0.85\textwidth]{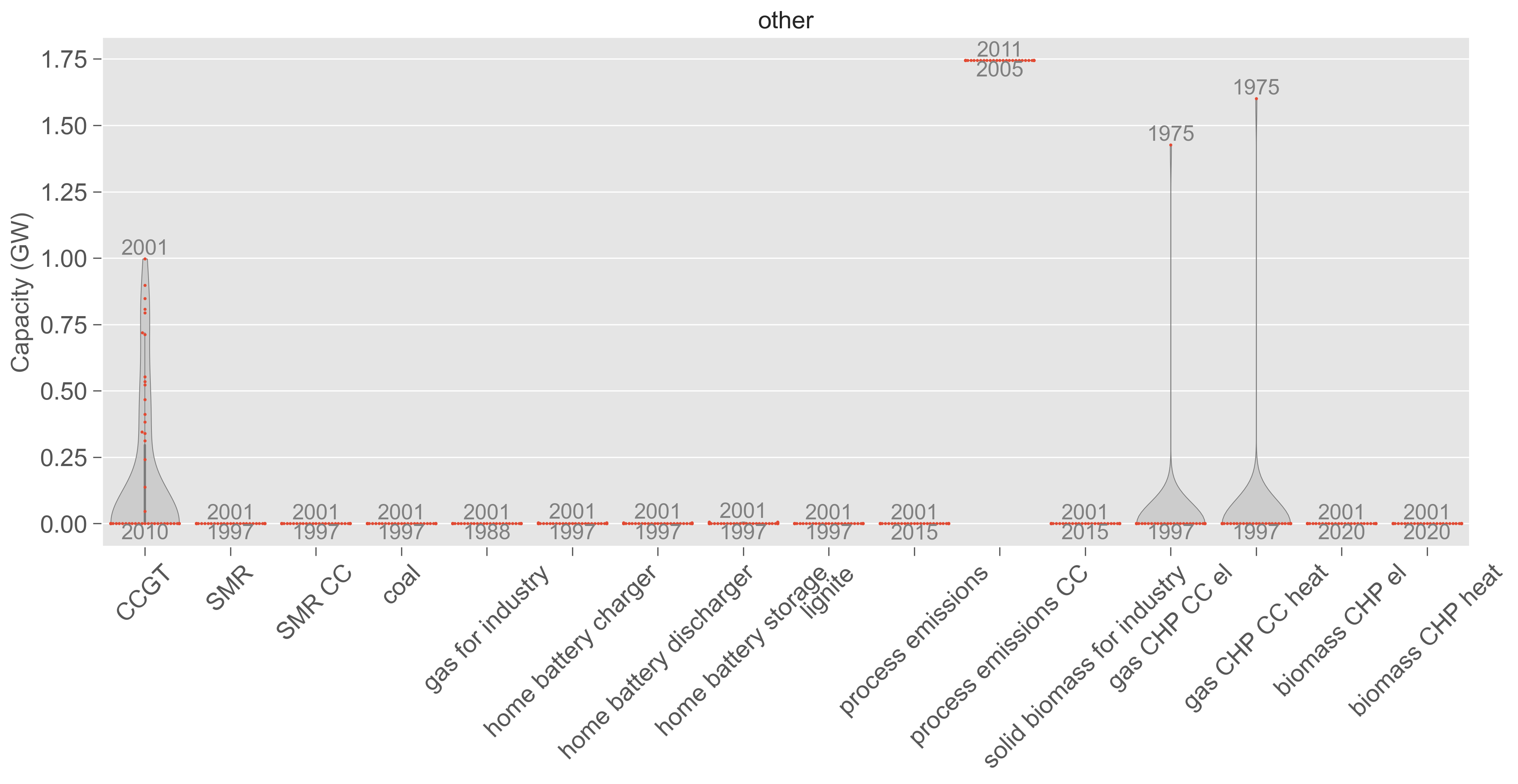}
	\captionsetup{width=14.8cm}
	\caption{Capacity of remaining technologies. Each red dot represents a design year.}
	\label{sfig:capacity_other}
\end{figure}

\end{adjustwidth}
 % <--------------- uncomment to include sup. mat.

\end{document}